\documentclass[12pt]{iopart}

\expandafter\let\csname equation*\endcsname\relax

\expandafter\let\csname endequation*\endcsname\relax

\usepackage{iopams,graphicx,amsmath,amsfonts,relsize,epstopdf,upgreek,color,mathtools,bm,times,braket,rotating,tabularx,booktabs,setspace,makecell,algorithm,algpseudocode}
\usepackage[colorlinks=true,linkcolor=blue,citecolor=blue,urlcolor =blue]{hyperref}
\usepackage[hyphenbreaks]{breakurl}
\usepackage[T1]{fontenc}

\usepackage{footmisc}
\setfnsymbol{wiley}

\usepackage[normalem]{ulem}

\renewcommand{\ket}[1]{\left| #1\right>}
\renewcommand{\bra}[1]{\left< #1\right|}
\newcommand{\inner}[2]{\langle #1|#2\rangle}
\newcommand{\opinner}[3]{\langle #1|#2|#3\rangle}
\newcommand{\onevec}{\bm{\mathit{1}}}
\newcommand{\zerovec}{\bm{\mathit{0}}}
\newcommand{\rvec}[1]{\pmb{#1}}
\newcommand{\dyadic}[1]{{\bf#1}}
\renewcommand{\tr}[1]{\mathrm{tr}\!\left\{#1\right\}}

\newcommand{\D}{\mathrm{d}}
\newcommand{\I}{\mathrm{i}}

\newcommand{\E}[1]{\mathrm{e}^{\mbox{\footnotesize$#1$}}}
\newcommand{\erf}[1]{\mathrm{erf}\!\left(#1\right)}
\newcommand{\erfc}[1]{\mathrm{erfc}\!\left(#1\right)}

\newcommand{\RE}[1]{\mathrm{Re}\!\left\{#1\right\}}
\newcommand{\IM}[1]{\mathrm{Im}\!\left\{#1\right\}}

\newcommand{\VAR}[2]{\mathrm{Var}_{#1}\!\left[#2\right]}
\newcommand{\rhoMEAN}[2]{\left<{#1}\right>_{#2}}
\newcommand{\estrhoMEAN}[2]{\widehat{\left<{#1}\right>}_{#2}}
\newcommand{\AVG}[2]{\mathbb{E}_{#1}\!\left[#2\right]}

\renewcommand{\PR}{\mathrm{pr}}

\newcommand{\vacket}{\ket{\textsc{vac}}}
\newcommand{\vacbra}{\bra{\textsc{vac}}}
\newcommand{\appropto}{\mathrel{\vcenter{
			\offinterlineskip\halign{\hfil$##$\cr
				\propto\cr\noalign{\kern2pt}\sim\cr\noalign{\kern-2pt}}}}}
\newcommand{\GSP}[1]{P(#1,#1^*)}
\newcommand{\HQ}[1]{Q(#1,#1^*)}
\newcommand{\psuccVMZ}{p_\textsc{vmz}}
\newcommand{\psuccPSGPEC}{p_\textsc{psg-pec}}
\newcommand{\psuccHYBRID}{p_\textsc{psg-vmz}}
\newcommand{\CX}{\textsc{cx}}

\def\NoNumber#1{{\def\alglinenumber##1{**}\State #1}\addtocounter{ALG@line}{-1}}

\begin{document}
	
	\title[Linear-optical protocols for mitigating and suppressing noise in bosonic systems]{Linear-optical protocols for mitigating and suppressing noise in bosonic systems}
	
	\author{Y S Teo$^{1,*,\dag}$, S U Shringarpure$^{1,*}$,
		S Cho$^1$ and H Jeong$^{1,\ddag}$}
	\address{$^1$ IRC NextQuantum, Department of Physics \& Astronomy, 
		Seoul National University, 08826 Seoul, South Korea}
	\address{$^*$ These authors contributed equally to this work.}
	\address{$^\dag$ ys\_teo@snu.ac.kr}
	\address{$^\ddag$ h.jeong37@gmail.com}
	
	\begin{abstract}
		Quantum-information processing and computation with bosonic qubits are corruptible by noise~channels. Using interferometers and photon-subtraction gadgets~(PSGs) accompanied by linear amplification and attenuation, we establish linear-optical methods to mitigate and suppress bosonic noise channels. We first show that by employing amplifying and attenuating PSGs respectively at the input and output of either a thermal or random-displacement channel, probabilistic error cancellation~(PEC) can be carried out to mitigate errors in expectation-value~estimation. We also derive optimal physical estimators that are properly constrained to improve the sampling accuracy of PEC. Next, we prove that a purely-dephasing channel is \emph{coherently} suppressible using a multimode Mach--Zehnder interferometer and conditional vacuum measurements~(vacuum-based Mach--Zehnder scheme or the VMZ scheme). In the limit of infinitely-many ancillas, with nonvanishing success rates, VMZ using either Hadamard or two-design interferometers turns \emph{any} dephasing channel into a phase-space-rotated linear-attenuation channel that can subsequently be inverted with (rotated) linear amplification \emph{without} Kerr nonlinearity. Moreover, for weak central-Gaussian dephasing, the suppression fidelity increases monotonically with the number of ancillas and most optimally with Hadamard~interferometers. We demonstrate the performance of these linear-optical mitigation and suppression schemes on common noise channels (and their compositions) and popular bosonic~codes. While the theoretical formalism pertains to idling noise channels, we also provide numerical evidence supporting mitigation and suppression capabilities with respect to noise from universal gate~operations.
	\end{abstract}
	
	
	\maketitle
	
	\section{Introduction}
	
	Advancement in our understanding of qubits encoded in continuous-variable~(CV) bosonic modes (each governed by ladder operators $[a,a^\dag]=1$)~\cite{Cochrane:1999macroscopically,Gottesman:2001encoding,Michael:2016new,Albert:2018performance} in terms of their generation and control~\cite{Jacobs:2008energy,Vasconcelos:2010all-optical,Leghtas:2013hardware-efficient,Mirrahimi:2014dynamically,Eaton:2019non-gaussian,Ma:2021quantum,Hastrup:2021measurement-free,Fukui:2022generating,Dahan:2023creation,Konno:2024logical,Takase:2024generation} supplies ample evidence that these systems form vital resources in quantum-information processing~\cite{Lund:2004conditional,Jeong:2004generation,Ourjoumtsev:2007generation,Andersen:2010continuous-variable,Mari:2014quantum,Joshi:2021quantum,Anteneh:2023sample} and computation~\cite{Ralph:2003quantum,Aliferis:2009fault,Noh:2019quantum,Grimsmo:2020quantum,Omkar:2020resource-efficient,Madsen:2022quantum,Xu:2024fault-tolerant,Calcluth:2024sufficient,Lee:2024fault-tolerant,Lemonde:2024hardware-efficient,AghaeeRad:2025scaling,Jeong:2002efficient,Lee:2013near-deterministic}. In~the current noisy intermediate-scale quantum~(NISQ) era, unfortunately, such encoded CV systems are susceptible to bosonic noise channels~\cite{Caruso:2008multi-mode,Konig:2013classical,Suter:2016colloquim,Noh:2020enhanced,Leviant:2022quantum,Fukui:2023efficient,Huang:2024exact,Bose:2024long-distance}. Bosonic error correction~\cite{Ralph:2011quantum,Bergmann:2016quantum,Puri:2019stabilized,Noh:2020:fault-tolerant,Grimsmo:2021quantum,Cai:2021bosonic,Hastrup:2022all-optical,Brady:2024advances,Schmidt:2024error-corrected} can potentially reduce logical error rates dramatically for sufficiently-small physical errors, ensuring reliable execution of quantum protocols running on bosonic qubits and that they possess distinct advantages from their \emph{best-known} classical~counterparts~\cite{Gagatsos:2017bounding,Gao:2018efficient,Noh:2020efficient,Oh:2021classical,Oh:2022classical,Fontana:2023classical,Zhang:2023noisy,Bugalho:2023resource-efficient,Oh:2024classical,Bressanini:2024gaussian}. This, however, demands encodings corresponding to highly
	nonclassical states of (typically) multiple excitation modes, whose preparation remains a technological~challenge. 
	
	With lower resource overheads, error-suppression methods employing specialized ``cat'' codes~\cite{Ofek:2016extending,Hastrup:2020deterministic,Su:2022universal}, noiseless attenuation and amplification~\cite{Micuda:2012noiseless,Winnel:2020generalized,Nunn:2021heralding,Shringarpure:2022coherence,Nunn:2022modifying,Nunn:2023transforming,Shringarpure:2024error} could reduce the impact of photon~losses on bosonic systems. Techniques involving error filtration~\cite{Gisin:2005error,Lee:2023error} and virtual channel purification~\cite{Liu:2024virtual} were also proposed. By recognizing that observable expectation values are what is often sought after in a quantum task, error mitigation~\cite{Cai:2023quantum} also offers experimentally-feasible ways to further improve the quality of classical measurement outputs with postprocessing on symmetric bosonic~codes~\cite{Endo:2022quantum,Endo:2024projective,Li:2024performance}. 
	
	In this work, we introduce operationally feasible linear-optical protocols with classical postprocessing~\cite{Knill:2001scheme} to effectively mitigate and suppress common noise channels. Suppression of photon-loss channels using noiseless attenuation and amplification~(noiseless loss suppression or NLS) on multicomponent ``cat'' codes was discussed in~\cite{Shringarpure:2024error}. More recently, probabilistic error cancellation~(PEC)~\cite{vandenBerg:2023probabilistic,Gupta:2024probabilistic,Taylor:2024quantum} to invert pure photon-loss~channels were~proposed. 
	
	We extend the aforementioned finding by employing \emph{attenuating and amplifying photon-subtraction gadgets}~(PSGs), comprising both linear-attenuation (or amplification) and photon subtraction operations, to mitigate general thermal (encompassing the familiar photon-loss channels)~\cite{Anderson:2022fundamental} and random-displacement noise~(RDN) channels~\cite{Oh:2024entanglement-enabled}, a common type of which being Gaussian-displacement noise~(GDN) channels~\cite{Schafer:2011gaussian,Noh:2020encoding}, which cannot otherwise be suppressed using NLS in~general. We show that for arbitrary bosonic states and observables, these noise types can be asymptotically inverted by applying an \emph{amplifying PSG before} and an \emph{attenuating PSG after} the noise channel. As a way to circumvent the unphysical nature of amplifying PSGs that prevents any form of their deterministic implementation, we propose to carry out error mitigation using both PSGs and PEC on observable expectation~values that permits the sampling of measurement outcomes constituting these amplifying PSGs as their indirect realization. As the expectation values are generally constrained, we develop optimal estimators that are properly constrained to maximize the PEC's sampling~accuracy.

	Next, we discuss the \emph{coherent error suppression} of bosonic dephasing channels~\cite{Huang:2024exact,Mele:2024quantum,Arqand:2023energy-constrained,Rexiti:2022discrimination,Leviant:2022quantum,Zhuang:2021quantum-enabled,Fanizza:2021squeezing-enhanced,Arqand:2020quantum}, which, for instance, models phase randomization in optical fibers under temperature and coherence fluctuation~\cite{Wanser:1992fundamental,Derickson:1998fiber} and quantum memories~\cite{Terhal:2015quantum}. Mitigation attempts were recently known to be difficult in practice as they require nonlinear elements such as Kerr's effect~\cite{Taylor:2024quantum}. In what shall be coined the \emph{vacuum-based Mach--Zehnder}~(VMZ) suppression scheme in this work, the input state immediately enters an input of a multimode interferometer before all output modes being subjected to independent and identically-distributed~(i.i.d.) dephasing noise and another Hermitian-conjugated multimode interferometer afterwards. Recent progress in optical computing allows for large multimode interferometers that can accommodate up to 10 or more modes in the photonic integrated chip~\cite{Zhu:2022space,Fu:2024optical} making our propositions practically~feasible.
    Finally, vacuum measurements on all output ancillas are performed. In the limit of infinitely-many ancillas, we found that VMZ with either Hadamard or unitary-two-design~\cite{harrow_random_2009,Dankert:2009exact,Brandao:2016local,Haferkamp:2022random} interferometers turns any dephasing channel into a phase-space-rotated linear-attenuation channel with \emph{nonvanishing} success probability, which may be subsequently inverted with linear amplification at the cost of a lower success~rate. This presents a linear-optical dephasing-suppression protocol that requires neither feedforward nor nonlinear Kerr's~effects.
	
	We demonstrate the performance of error mitigation and suppression using the PSG and VMZ schemes (which are mutually commuting) for the common thermal, Gaussian-displacement and dephasing channels, as well as their combinations on well-established bosonic qubit encodings. We find that the output encoded-qubit fidelities improve significantly for reasonable noise rates, even in the presence of small measurement~errors. Despite the fact that all theoretical arguments in Secs.~\ref{sec:mit_TN_RDN} and~\ref{sec:supp_deph} apply to idling noise during the storage or propagation of encoded states, Sec.~\ref{sec:perf} also numerically showcases our protocols' capabilities in treating noise from universal gate operations, thereby presenting feasible linear-optical recipes for bosonic error mitigation and~suppression.
	
	\section{Mitigating thermal noise and RDN}
	\label{sec:mit_TN_RDN}
	
	\subsection{Thermal-noise and RDN channels}
	\label{subsec:TN_RDN_chn}
	
	We write the optical state \mbox{$\rho=\int\frac{(\D\alpha)}{\pi}\ket{\alpha}\GSP{\alpha}\bra{\alpha}$} in terms of its Glauber--Sudarshan P~function and coherent states~$\{\ket{\alpha}\bra{\alpha}\}$, where $(\D\alpha)/\pi$ is the phase-space volume~measure. Then, structurally, the thermal and random-displacement noise~(RDN) channels relate through the displacement-operator~$(D)$~mixture
	\begin{align}
		\mathcal{E}[\rho]=&\,\int\frac{(\D\alpha)}{\pi}\,\GSP{\alpha}\int\frac{(\D\beta)}{\pi}\,p(\beta,\beta^*)\,D(\nu\beta)\ket{\mu\alpha}\bra{\mu\alpha}D(\nu\beta)^\dag\,.
		\label{eq:gen_thm_RDN}
	\end{align}
	For~a thermal noise channel~$\mathcal{E}=\mathcal{E}^{(\eta,\bar{n})}_\text{therm}$ defined by the thermal photon number~$\bar{n}\geq0$ and noise rate $0\leq\eta\leq1$, $\nu=\sqrt{\eta}=\sqrt{1-\mu^2}$ and $p(\beta,\beta^*)=\E{-|\beta|^2/\bar{n}}/(\bar{n})$. The special case \mbox{$\bar{n}=0$} corresponds to the (vacuum-)loss~channel.
	
	On the other hand, a general RDN channel corresponds to $\mu=1=\nu$ and some normalized probability distribution~$p(\beta,\beta^*)$. For a special case of such a channel which we shall focus on, namely the Gaussian-displacement noise~(GDN) channel, we have $p(\beta,\beta^*)=\E{-|\beta|^2/\sigma^2}/(\sigma^2)$ for some $\sigma>0$ characterizing its strength. It can be shown that an application of the quantum-limited amplification operation on the thermal-noise channel results in the GDN channel~\cite{Noh:2019quantum}, namely $\mathcal{E}^{(\sigma)}_\text{GDN}[\rho]=\mathcal{A}_{1/(1-\eta)}\circ\mathcal{E}^{(\eta,\bar{n})}_\text{therm}[\rho]$, where $\sigma^2=\eta\,(1+\bar{n})/(1-\eta)$ and $\mathcal{A}_G$~is the quantum-limited amplification map~(details in \ref{app:TN_GDN_DPH}). The purpose of such a noise conversion (akin to Pauli twirling for discrete variables~\cite{vandenBerg:2023probabilistic}) was to facilitate the error correction of GDN using the Gottesmann--Kitaev--Preskill~(GKP) code, although it is now known that this code may also correct thermal noise directly in conjunction with properly-optimized decoding methods~\cite{Albert:2018performance}. Note also that $\mathcal{E}^{(\sigma')}_\text{GDN}[\rho]=\mathcal{E}^{(\eta,\bar{n})}_\text{therm}[\rho]\circ\mathcal{A}_{1/(1-\eta)}$, where $\sigma'^2=\eta\,(1+\bar{n})<\sigma^2$. Hence, as a conservative choice for testing error-mitigation capabilities, we shall consider~$\mathcal{E}^{(\sigma)}_\text{GDN}$, although for small $\eta$, both GDN channels are almost the~same.
	
	\subsection{Asymptotic channel-inverting capability in PSGs}
	\label{subsec:asymp_chn_invert_PSG}
	
	The amplifying or attenuating photon-subtraction gadget~(PSG), characterized by a parameter $g>0$, is a gadget consisting of either a noiseless linear amplifier~($g>1$) or attenuator~($g<1$), and photon subtraction. Its complete action may be described by the trace-preserving~(TP)~map
	\begin{equation}
		\mathcal{M}_{\textsc{psg},\,g}[\,\bm{\cdot}\,]=\sum^\infty_{k=0}\dfrac{(-1+g^{-2})^k}{k!}\,a^k\,g^{a^\dag a}\,\bm{\cdot}\,g^{a^\dag a}\,a^{\dag\,k}
		\label{eq:psg_map}
	\end{equation}
	for any~$g$. \emph{Only when} $g\leq1$ is the map $\mathcal{M}_{\textsc{psg},\,g}$ completely~positive~(CP) and physically implementable in a deterministic way with photon counting after a beam~splitter~(BS) of transmittance~$g^2$.
	
	Suppose that the quantum bosonic system traverses through an idling noise channel, which arises either during its storage or propagation, of the form in~\eqref{eq:gen_thm_RDN}. Then, the channel-inverting scheme involves an amplifying PSG of gain factor~\mbox{$g>1$} applied on the bosonic system before and another attenuating~PSG of loss factor~$g'<1$ after the noise~channel. Since
	\begin{align}
		\mathcal{M}_{\textsc{psg},\,g}[\ket{\alpha}\bra{\alpha}]=&\,\ket{g\alpha}\bra{g\alpha}\,,
		\label{eq:psg_props}
	\end{align}
	for \emph{any real}~$g$, when $g'g\mu=1$, the output state becomes
	\begin{align}
		\rho_\mathrm{out}=&\,\mathcal{M}_{\textsc{psg},\,g'=1/(g\mu)<1}[\mathcal{E}[\mathcal{M}_{\textsc{psg},\,g>1}[\rho_\mathrm{in}]]]\nonumber\\
		=&\,\int\dfrac{(\D\beta)}{\pi}\,p(\beta,\beta^*)\,D\!\left(\frac{\nu}{g\mu}\beta\right)\rho_\mathrm{in}\,D\!\left(\frac{\nu}{g\mu}\beta\right)^\dag,
		\label{eq:rho_out_mit}
	\end{align}
	which is always a physical state since it is just a scaled RDN channel on $\rho_\mathrm{in}$, despite the unphysical~$\mathcal{M}_{\textsc{psg},\,g>1}$. We then see that if~$g\gg1$, then $\rho_\mathrm{out}\rightarrow\rho_\mathrm{in}$ and noise-channel inversion is \emph{asymptotically}~achieved. A useful rule-of-thumb guide for setting appropriate $g$ and $g'$ would be 
	\begin{equation}
		\begin{array}{ll}
			g'=\frac{1}{g\sqrt{1-\eta}}\,,\,\,g\gg2\sqrt{(2\ln 2)\frac{\bar{n}\eta}{1-\eta}} & \text{(thermal noise)}\,,\\
			g'=\frac{1}{g}\,,\,\,g\gg2\sqrt{2\ln 2}\,\sigma & (\text{GDN})\,,
		\end{array}
	\end{equation}
	where the two criteria for $g$ relate to full-width-at-half-maxima of the respective Gaussian~distributions. 
	
	\begin{figure}[t]
		\centering
		\includegraphics[width=0.85\columnwidth]{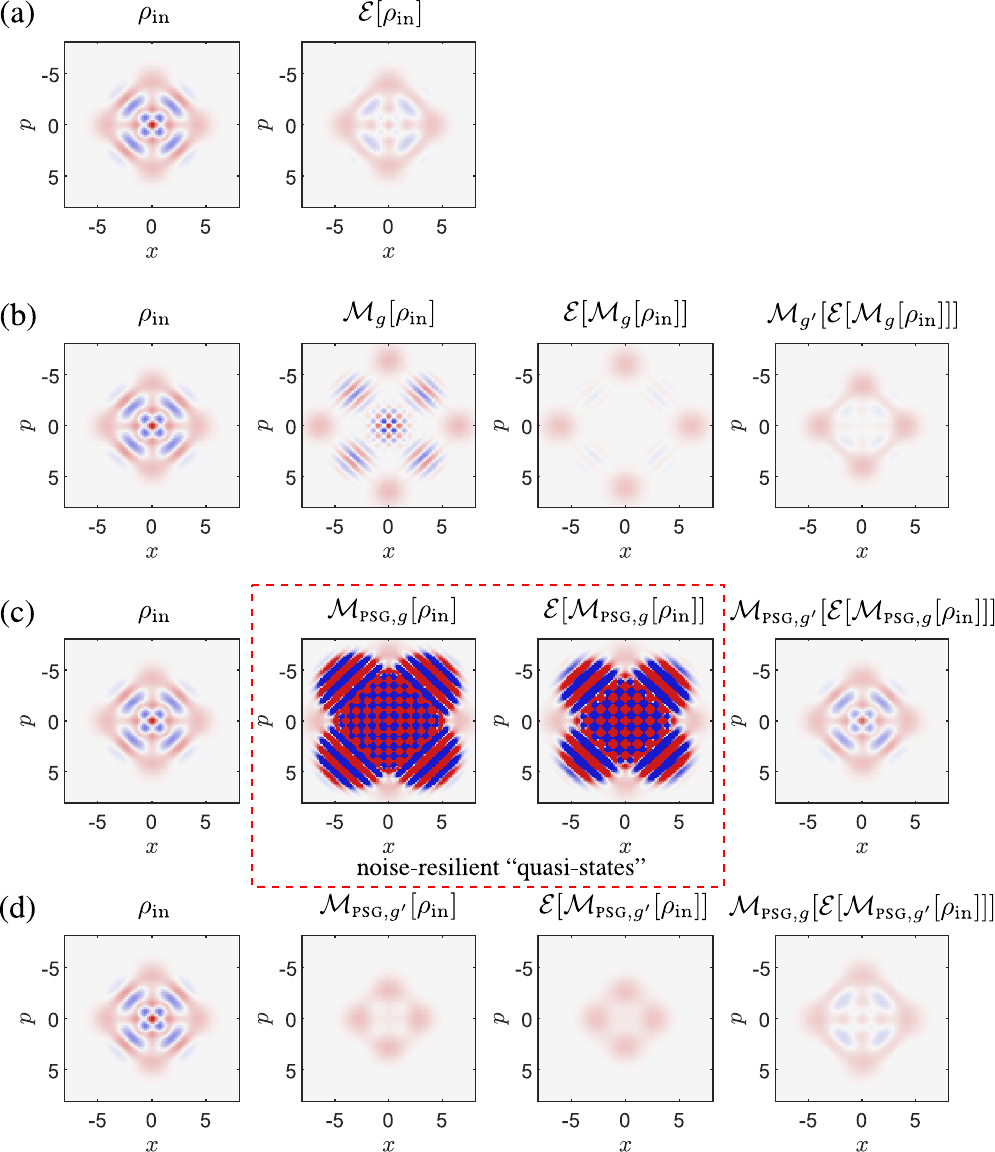}
		\caption{\label{fig:phys_expl}Evolution of the Wigner function under various transformations. A four-component ``cat'' state $(\propto\ket{\alpha}+\ket{-\alpha}+\ket{\I\alpha}+\ket{-\I\alpha})$ of amplitude $\alpha=2$ undergoing (a)~bare thermal-noise corruption~($\mathcal{F}=0.475$), (b) through an ordered sequence of linear amplification, thermal noise and linear attenuation~($\mathcal{F}=0.265$), and finally (c) through an ordered sequence of amplifying PSG, thermal noise and attenuating PSG~($\mathcal{F}=0.825$). Here, $\eta=0.1$~(a rate of~10\%), $\bar{n}=0.5$, $g=1.6$ and $g'=1/(1.6\sqrt{1-0.1})=0.659$. Although the Wigner functions in the two intermediate panels of~(c) represent unphysical ``quasi-states'' due to the non-CP nature of amplifying PSGs, they nonetheless reveal the vital role photon subtraction plays as an amplitude modifier that protects interference features and nullify state distortion at the final~output if such amplifying PSGs are realized through error mitigation (next~subsection). The fortified Wigner-function interference is clearly seen in~(c) and in striking contrast with~(b), where the absence of photon subtraction leads to an almost completely washed-out Wigner interference~fringes. (d) Note that reversing the PSG order results in an output state of low quality~($\mathcal{F}=0.424$).}
	\end{figure}
	
	In the special case where $\bar{n}=0$, the thermal-noise channel is simply the loss~channel. Then,~$p(\beta,\beta^*)=\delta(\RE{\beta})\delta(\IM{\beta})$ and the mixture in \eqref{eq:rho_out_mit} becomes $\rho_\mathrm{in}$, which implies that \emph{any} $g$ and $g'=1/(g\sqrt{1-\eta})$ \emph{exactly} inverts the~channel. Hence, the simplest setup that does this uses only one PSG~\cite{Taylor:2024quantum}---$g=1$ for instance. Otherwise, both thermal-noise and RDN channels can only be \emph{asymptotically inverted} with $g\rightarrow\infty$ in such PSG-based~setups.
	
	Before we elaborate on how the amplifying PSG map may be realized nondeterministically in the next subsection, one can intuitively understand the asymptotic channel-inversion properties of PSGs through the help of Wigner-function diagrams supposing that such a map is (indirectly) realized. Figure~\ref{fig:phys_expl} shows the possible fates of a nonclassical~\cite{Tan:2020negativity} input state undergoing different routes of operations, (b)~one starting with a linear amplification, that is $\mathcal{M}_{g>1}[\,\bm{\cdot}\,]=g^{a^\dag a}\,\bm{\cdot}\,g^{a^\dag a}/\tr{g^{a^\dag a}\,\bm{\cdot}\,g^{a^\dag a}}$, followed by thermal noise and ends with a linear attenuation~$\mathcal{M}_{g'<1}$, and (c)~another consisting of an amplifying PSG, then thermal noise and finally an attenuating PSG of the same~$g>1$ and~$g'<1$. Here, the important difference stems from the additional photon subtraction operations in~PSGs. Without them, linear amplification alone simply increases the average photon number of the encoded state, leaving it even more susceptible to~noise. The additional photon subtraction operations are crucial in strengthening the interference features of the state so that they become more noise-resilient---it is the \emph{combined efforts} of photon subtraction, linear amplification and linear attenuation that achieves asymptotic channel inversion. (d)~Reversing the PSG order results in no such noise-protective enhancement since an attenuating PSG on the input state dilutes the encoded quantum information. We emphasize that such interference fortification is only possible through nondeterministic implementation of the amplifying~PSG.

	\begin{figure}[t]
		\centering
		\includegraphics[width=0.7\columnwidth]{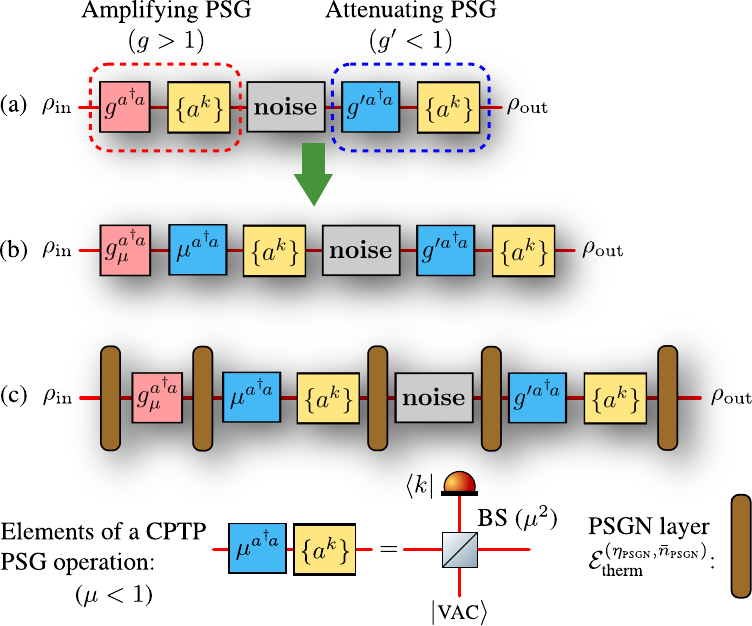}
		\caption{\label{fig:psg}The PSG-PEC mitigation setup. (a)~The first amplifying PSG of a gain factor~$g>0$ is not CP and so cannot be directly~realized. (b)~One may portion it into a probabilistic amplification of gain $g_\mu=g/\mu$ and an attenuating PSG of parameter~$0<\mu<1$. The~choice of $\mu$ should optimize the error-mitigation~performance. (c)~More realistically, each distinct operation of the setup is noisy and may be accompanied by a PSGN layer before and another after the operation. The first PSGN layer may also account for $\rho_\mathrm{in}$-preparation~errors. We model each PSGN layer as a thermal channel~$\mathcal{E}^{(\eta_\textsc{psgn},\bar{n}_\textsc{psgn})}_\text{therm}$ having identical thermal-photon number~$\bar{n}_\textsc{psgn}$ and $\eta_\textsc{psgn}=\eta_0/5$. Noise from observable measurements may also be incorporated into the PSGN~layers. The total PSGN is therefore thermal with error rate~$\eta_0$ by assuming a small~$\eta_0$, which follows from the approximate additivity of thermal-noise channels of identical thermal photon~number: $\mathcal{E}^{(\eta_2,\bar{n})}_\text{therm}\circ\,\mathcal{E}^{(\eta_1,\bar{n})}_\text{therm}\cong\mathcal{E}^{(\eta_1+\eta_2,\bar{n})}_\text{therm}$ for small $\eta_1$ and~$\eta_2$~(see \ref{app:additivity}).}
	\end{figure}

    Intuitively, when implemented indirectly through sampling, the first amplifying PSG map effectively spreads the encoded quantum information sufficiently far away from the phase-space origin, making space to introduce highly intensive interference reinforcements for the encoded qubit to sustain the effects of thermal noise or RDN. Thereafter, the second attenuating PSG undoes the spread and interference reinforcements to recover the encoded~state.
    
	\subsection{Error mitigation with PSGs and PEC}
	\label{subsec:err_mit_PSG-PEC}
	To circumvent the problem of a non-CP amplifying PSG that fundamentally prevents its direct experimental implementation, we can express ($g_\mu\equiv g/\mu$ and $0<\mu<1$)
	\begin{align}
		\mathcal{M}_{\textsc{psg},\,g>1}[\rho_\mathrm{in}]=&\,\sum^\infty_{k=0}\underbrace{\mathcal{N}_{g_\mu}\dfrac{(-1+g^{-2})^k}{(-1+\mu^{-2})^k}}_{\displaystyle\equiv\omega_k}\!K_{\mu,k}\,\rho_\mathrm{in}(g_\mu)K_{\mu,k}^\dag
		\label{eq:ampPSG_decomp}
	\end{align}
	where $\mathcal{N}_{g_\mu}=\tr{g_\mu^{a^\dag a}\rho_\mathrm{in}g_\mu^{a^\dag a}}$, $\rho_\mathrm{in}(g_\mu)=g_\mu^{a^\dag a}\rho_\mathrm{in}g_\mu^{a^\dag a}/\mathcal{N}_{g_\mu}$ and the Kraus operators $K_{\mu,k}=\sqrt{(-1+\mu^{-2})^{k}/k!}\,a^k\mu^{a^\dag a}$ constitute~$\mathcal{M}_{\textsc{psg},\,\mu<1}$. That is, the unphysical amplifying PSG is exactly equivalent to a linear combination of Kraus operations from a deterministic attenuating PSG on a linearly-amplified state~$\rho_\mathrm{in}(g_\mu)$~(see~Fig.~\ref{fig:psg}), where the coefficient~$\omega_k$ is known given some accurate estimate of~$\mathcal{N}_{g_\mu}$~(see ** in~Alg.~\ref{alg:psgpec}). Such a linear amplification is \emph{necessarily probabilistic}~\cite{Caves:1982quantum,Caves:2012quantum} and may be carried out using quantum scissors that requires only linear optics and Fock resource~states~\cite{Winnel:2020generalized,Guanzon:2024saturating}, although alternative schemes exist for Gaussian states and certain classes of their superpositions~\cite{Lund:2004conditional,Neergaard-Nielsen:2013quantum,Fiurasek:2022teleportation,Guanzon:2023noiseless,Zhao:2023enhancing,Fiurasek:2024analysis,Jeon:2024amplifying}.
	
	To proceed, we realize that a typical goal in quantum-information and computation tasks is to acquire expectation values of an observable (such as a desired pure target state, moments of any state property, POVM outcome, \emph{etc.}), $O=\sum_l\ket{o_l}o_l\bra{o_l}$, where $\ket{o_l}\bra{o_l}$ are states that can be measured through a feasible circuit-sampling~scheme. These, for instance, could be the eigenstates of~$O$, in which case $o_l$ correspond to its~eigenvalues, or those that span the operator subspace containing~$O$. In this perspective, based on~\eqref{eq:ampPSG_decomp}, we can apply an error-mitigation strategy known as \emph{probabilistic error cancellation}~(PEC) that would allow us to realize the map~$\mathcal{M}_{\textsc{psg},\,g'<1}\circ\mathcal{E}\circ\mathcal{M}_{\textsc{psg},\,g>1}$ through measurement-outcome~sampling and probabilistically invert both the thermal-noise and RDN channels. More precisely, PEC mitigates estimation errors of $\rhoMEAN{O}{\mathrm{out}}=\tr{\rho_\mathrm{out}O}$ for the noisy~$\rho_\mathrm{out}$ in~Eq.~\eqref{eq:rho_out_mit}:
	\begin{align}
		\rhoMEAN{O}{\mathrm{out}}=&\,\sum^\infty_{k=0}\sum^\infty_{l=0}\omega_k\,o_l\,p_{kl}\,,\,\,p_{kl}=\opinner{o_l}{\sigma_k}{o_l}\,,\nonumber\\
		\sigma_k=&\,\sum^\infty_{k'=0}K_{g',k'}\mathcal{E}[K_{\mu,k}\,\rho_{\mathrm{in}}(g_\mu)K_{\mu,k}^\dag]K_{g',k'}^\dag\,,
		\label{eq:Oout}
	\end{align}
	where $p_{kl}$ is the joint probability of measuring $K_{\mu,k}$ and $o_l$ ($\sum_{k,l}p_{kl}=1$). From~\eqref{eq:rho_out_mit}, the larger the~$g$, the more complete noise cancellation shall~be. 
	
	Thus, the employment of PEC turns the unfeasible problem of coherently obtaining $\rho_\mathrm{out}$ in~\eqref{eq:rho_out_mit} into a feasible, albeit indirect, solution of mitigating errors in the measured expectation values through circuit sampling and classical postprocessing. For a finite sampling copy number~$N$, the output expectation value~$\rhoMEAN{O}{\mathrm{out}}$ is estimated by some numerical estimator $\estrhoMEAN{O}{\mathrm{out}}$ that is random according to the collected measurement data, where the desired situation is that $\estrhoMEAN{O}{\mathrm{out}}$ is statistically close to the target expectation value~$\rhoMEAN{O}{\mathrm{targ}}=\tr{\rho_\mathrm{in}O}$, here evaluated with the noiseless target~$\rho_\mathrm{targ}=\rho_\mathrm{in}$. The figure of merit for statistical quality that we shall adopt is the \emph{mean squared-error}~(MSE): $\mathcal{D}=\AVG{}{\left(\estrhoMEAN{O}{\mathrm{out}}-\rhoMEAN{O}{\mathrm{targ}}\right)^2}$, where $\AVG{}{\,\bm{\cdot\,}}$ denotes the data average. In the large-$N$ limit, we would like this estimator to be \emph{consistent}, that is $\estrhoMEAN{O}{\mathrm{out}}\rightarrow\rhoMEAN{O}{\mathrm{targ}}$.
	
	We are left with deciding on the value of $\mu$ to execute the \emph{PSG-PEC~protocol}. In the infinite-$N$ limit, it is obvious that any choice of $\mu$ gives the same~$\rhoMEAN{O}{\mathrm{out}}$ as $\mathcal{M}_{\textsc{psg},\,g>1}[\rho]$ only depends on $g$ and not how one portions this amplifying PSG~map. For any finite~$N$, however, the value of~$\mu$ does affect the sampling error~$\mathcal{D}$.
	
	\subsection{Optimal PSG-PEC estimators}
	\label{subsec:opt_psgpec_est}
	
	We take $p_{kl}$ to be joint probabilities of a multinomial distribution for some fixed~$N$, which is a very common experimental scenario. One might then na{\"i}vely replace $p_{kl}$ in the first equality of~\eqref{eq:Oout} by~$\nu_{kl}$, which is the relative frequency of measuring $K_{\mu,k}$ and $o_l$~($\sum_{k,l}\nu_{kl}=1$) and regard the resulting sum to be the estimator. This is a fallacy because $\rhoMEAN{O}{\mathrm{out}}$ is generally a \emph{constrained} expectation~value. For instance, when $O=\rho_\mathrm{targ}$ is a pure target~state, we have the fidelity $\mathcal{F}=\tr{\rho_\mathrm{targ}\,\rho_\mathrm{out}}$, which is clearly constrained between~0 and~1.  However, it is evident that $\sum_{k,l}\omega_k\,o_l\,\nu_{kl}$ is a random number that obeys no such physical constraints and, thus, cannot be a meaningful estimator.
	
	Fortunately, for a constrained $a\leq\rhoMEAN{O}{\mathrm{out}}\leq b$, the immediate fix is to define
	\begin{align}
		\estrhoMEAN{O}{\mathrm{out}}=&\,\,\Theta\!\left(a-r\right)a+\Theta\!\left(r-b\right)b+\Theta\!\left(r-a\right)\Theta\!\left(b-r\right)r\,,\nonumber\\
		r=&\,\sum_{k,l}\omega_k\,o_l\,\nu_{kl}\equiv(\rvec{\omega}\otimes\rvec{o})\bm{\cdot}\rvec{\nu}\,,
		\label{eq:constr_est}
	\end{align}
	as its corresponding constrained PSG-PEC estimator, where $\Theta(x)$ is the Heaviside step~function that gives the desired constraining properties---$\estrhoMEAN{O}{\mathrm{out}}=a$ when $r<a$ and $\estrhoMEAN{O}{\mathrm{out}}=b$ if~$r>b$. The relevant MSE~reads
	\begin{align}
		\mathcal{D}=&\,\underbrace{\VAR{}{\estrhoMEAN{O}{\mathrm{out}}}}_{\displaystyle\text{variance}}+\underbrace{\left(\AVG{}{\estrhoMEAN{O}{\mathrm{out}}}-\rhoMEAN{O}{\mathrm{targ}}\right)^2}_{\displaystyle\text{noise and statistical bias}}\,,
		\label{eq:theory_D}
	\end{align}
	which, for values of~$N$ such that \eqref{eq:char_CLT} holds, is defined~by
	\begin{align}
		\AVG{}{\estrhoMEAN{O}{\mathrm{out}}}=&\,\frac{a+b}{2}+\frac{B-a}{2}\,\erf{\frac{B-a}{\sqrt{2A}}}-\frac{b-B}{2}\,\erf{\frac{b-B}{\sqrt{2A}}}\nonumber\\
		&+\sqrt{\frac{A}{2\pi}}\left[\E{-\frac{(B-a)^2}{2A}}-\E{-\frac{(b-B)^2}{2A}}\right],\nonumber\\
		\AVG{}{\estrhoMEAN{O}{\mathrm{out}}^2}=&\,\frac{a^2+b^2}{2}+\frac{A+B^2-a^2}{2}\,\erf{\frac{B-a}{\sqrt{2A}}}+\frac{A+B^2-b^2}{2}\,\erf{\frac{b-B}{\sqrt{2A}}}\nonumber\\
		&+\sqrt{\frac{A}{2\pi}}\left[\E{-\frac{(B-a)^2}{2A}}(B+a)-\E{-\frac{(b-B)^2}{2A}}(B+b)\right],
		\label{eq:theory_D2}
	\end{align}
	where $A=\rvec{\chi}\bm{\cdot}\left(\dyadic{Q}-\rvec{q}\,\rvec{q}\right)\bm{\cdot}\rvec{\chi}/N$, $B=\rvec{\chi}\bm{\cdot}\rvec{q}$, $\rvec{\chi}=\rvec{\omega}\otimes\rvec{o}$ and
	\begin{equation}
		\dyadic{Q}=\begin{pmatrix}
			p_{11} & 0 & \cdots & 0 & 0 &\cdots & 0\\
			0 & p_{12} & \cdots & 0 & 0 &\cdots & 0\\
			\vdots & \vdots & \ddots & \vdots & \vdots & \ddots & \vdots\\
			0 & 0 & \cdots & p_{21} & 0 &\cdots & 0\\
			0 & 0 & \cdots & 0 & p_{22} &\cdots & 0\\
			\vdots & \vdots & \vdots & \vdots & \vdots & \ddots & \vdots
		\end{pmatrix},\,\,\rvec{q}=\begin{pmatrix}
			p_{11}\\
			p_{12}\\
			\vdots\\
			p_{21}\\
			p_{22}\\
			\vdots
		\end{pmatrix}.
	\end{equation}

    \begin{algorithm}[b]
		\caption{\label{alg:psgpec}PSG-PEC}
		\begin{algorithmic}[1]
			\State Determine the noise type and calibrate parameters~$\eta$ and~$\bar{n}$.
			\State Set~$g>1$ that is reasonably~large.
			\State Decide on \mbox{$O=\sum_l\ket{o_l}o_l\bra{o_l}$} and feasible measurement outcomes \mbox{$\ket{o_l}\bra{o_l}$}.
			\State Fix the number of copies~$N$ to be sufficiently large.
			\State Use \eqref{eq:theory_D} and \eqref{eq:theory_D2} to obtain $\mu_\mathrm{opt}$, assuming $\rho_\mathrm{targ}$ and no~PSGN.
			\State Set up the PSG-mitigation scheme with~$g$, $\eta$, $\bar{n}$ and~$\mu_\mathrm{opt}$. 
			\State Sample the multinomial distribution defined by $N$ and $p_{kl}$s [see~\eqref{eq:Oout}], each being the probability of subtracting $k$~photons and measuring state $\ket{o_l}\bra{o_l}$, and collect data~$\nu_{kl}$.			
			\State Compute the error-mitigated  $\estrhoMEAN{O}{\mathrm{out}}$ defined in~\eqref{eq:constr_est}, where $\omega_k$ is defined in \eqref{eq:ampPSG_decomp}. The constraints of $\estrhoMEAN{O}{\mathrm{out}}$ follow that of~$\rhoMEAN{O}{\mathrm{out}}$.
			\NoNumber An accurate estimate of~$\mathcal{N}_{g_\mu}$ is needed for the given input~state. To this end, one may follow the steps above and sample any complete basis to find $\mathcal{N}_{g_\mu}$ such that~$\estrhoMEAN{1}{\mathrm{out}}\equiv1$. This estimate would be accurate if $N$ is sufficiently~large.
	\end{algorithmic} 
    \end{algorithm}
    
	As a sanity check, when the constraints are lifted ($a\rightarrow-\infty$ and $b\rightarrow\infty$), all error functions~(erf) approach unity and we have $\AVG{}{\estrhoMEAN{O}{\mathrm{out}}}=B=\sum_{k,l}\omega_k\,o_l\,p_{kl}=\rhoMEAN{O}{\mathrm{out}}$ and $\AVG{}{\estrhoMEAN{O}{\mathrm{out}}^2}=A+B^2=A+\rhoMEAN{O}{\mathrm{out}}^2$, which are precisely the first and second moments of the unconstrained~estimator for multinomial~data, so that $\mathcal{D}=A+(\rhoMEAN{O}{\mathrm{out}}-\rhoMEAN{O}{\mathrm{targ}})^2$. It~is straightforward to see that this expression also holds when $N\gg1$ based on the fact that $\erf{x/\kappa}\rightarrow\Theta(x)$ as \mbox{$\kappa\rightarrow0$}, which tells us that both the unconstrained and constrained estimation lead to the same estimator in the large-$N$~limit. For sufficiently large $g$, the second bias tends to~zero. 	
	With $\mathcal{D}$ in \eqref{eq:theory_D} (derived in \ref{app:constr_opt}), we now supply an answer to the optimal~$\mu$ for~PSG-PEC: this is the value that \emph{minimizes~$\mathcal{D}$}. We first realize that although there is $\mu$ dependence in both $\omega_k$ and $p_{kl}$, and thus~$\rvec{\chi}$, $B=\rvec{\chi}\bm{\cdot}\rvec{q}=\rhoMEAN{O}{\mathrm{out}}$ is independent of $\mu$ by~definition, so $A$ is the only function of~$\mu$. If $\mu=0$ or~$1$, $A=\infty=\mathcal{D}$, which means that the global minimum for $\mathcal{D}$ is within the interval $\mu\in(0,1)$. As $N$ increases, the landscape of $\mathcal{D}$ over $\mu$ becomes flatter over a larger~subinterval. Furthermore, the fact that $N$ \emph{only} appears in~$A$ as its reciprocal prefactor implies that the optimal~$\mu$ that minimizes~$\mathcal{D}$ is \emph{independent} of the sampling copy number~$N$.
    
	One can make use of \eqref{eq:theory_D} to obtain the global minimum $\mu_\mathrm{opt}$ in the following way: if one expects the input state prepared is the ideal target state~$\rho_\mathrm{targ}=\rho_\mathrm{in}$ (neglecting the existence of~PSGN and other confounding errors) and knows the noise-channel parameters, then $p_{kl}$ are the probabilities computed based on~$\rho_\mathrm{targ}$ and noise~channel. The corresponding $\mu_\mathrm{opt}$ that minimizes $\mathcal{D}$ in~\eqref{eq:theory_D} with this target state can then be used to collect actual experimental data that are almost always contaminated with additional PSG noise and observable-measurement errors~(collectively coined as PSGN) that would depend on the platform hardware. Here, these are modeled as thermal-noise layers sandwiching every key stages of the PSG-PEC mitigation scheme~[see Fig.~\ref{fig:psg}(c)]. This approximately accounts for realistic imperfections that are present in both the linear-amplification and attenuating-PSG stages, the implementation of both depends on actual~hardware~\cite{Scherer:2009quantum}. 
    
	\subsection{Complete procedure of PSG-PEC error mitigation}
	
	\begin{figure}
		\centering
		\includegraphics[width=.68\columnwidth]{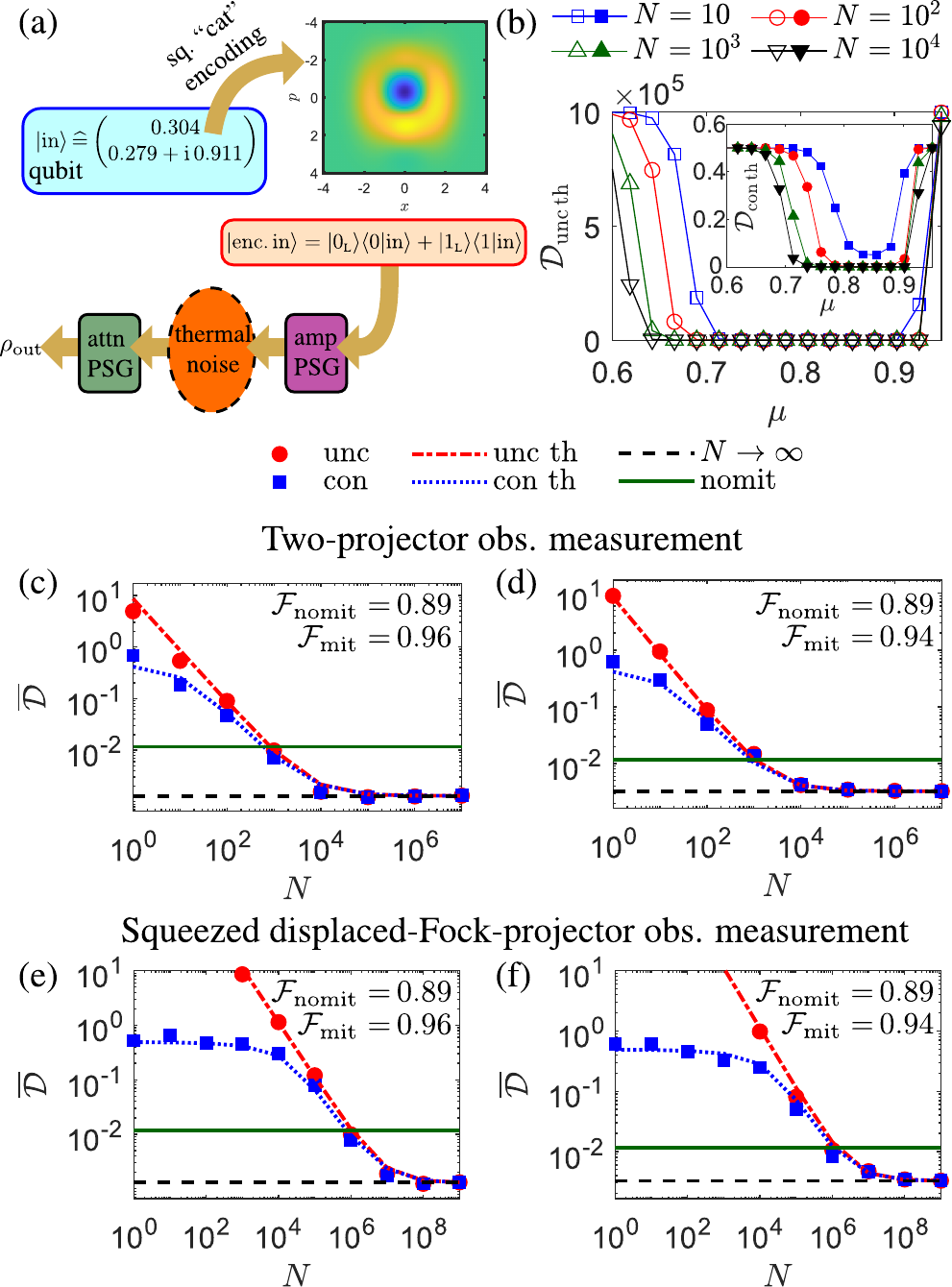}
		\caption{\label{fig:MSE_therm}(a) An example qubit state encoded in a 3dB-squeezed-``cat'' code~($r=0.345$, $\phi=0$, $\alpha=1$) is subjected to amplifying~(amp, $g=1.4$) and attenuating~(attn, $g'=0.855$)~PSGs for mitigating thermal noise of $\eta=0.05$ and~$\bar{n}=0.5$. (b)~The landscapes of the theoretical MSE from \eqref{eq:theory_D} and \eqref{eq:theory_D2} for both the unconstrained~(unc) and constrained~(con) estimators with respect to~$\mu$ eventually flatten out as $N$~increases (shown here for the two-projector measurement~case). The corresponding $\mu_\mathrm{opt}$s are used to formulate both unc and con estimators for $\rhoMEAN{\rho_\mathrm{targ}}{\mathrm{out}}=\tr{\rho_\mathrm{in}\,\rho_\mathrm{out}}$ ($\rho_\mathrm{targ}=\rho_\mathrm{in}$). Averaged MSE curves for the (c,d)~two-projector and (e,f) 64-squeezed-displaced-Fock-projector observable measurements are plotted. A total of 50 experiments are simulated to compute the average MSE~$\overline{\mathcal{D}}$, (d,f)~with and (c,e)~without PSGN, where in~(d,f), PSGN is distributed into five layers as in Fig.~\ref{fig:psg} such that $\eta_0=0.02$ and~$\bar{n}_{\textsc{psgn}}=0.1$. All fitted curves are derived from the analytical expressions in~\eqref{eq:theory_D} and~\eqref{eq:theory_D2}. The~fidelities in unmitigated~($\mathcal{F}_\mathrm{nomit}$) and mitigated~($\mathcal{F}_\mathrm{mit}$) scenarios, where the latter refers to the fidelity in the $N\rightarrow\infty$ limit, are~reported. The black dashed and green solid lines respectively represent the bias~$(\rhoMEAN{\rho_\mathrm{in}}{\mathrm{out}}-1)^2$ with and without PSG~mitigation. The larger number of projectors used in (e,f) consequently requires a larger~$N$ to approach the asymptotic~$\mathcal{F}_\mathrm{mit}$ relative to~(c,d).}
	\end{figure}
	
	\begin{figure}
		\centering
		\includegraphics[width=.68\columnwidth]{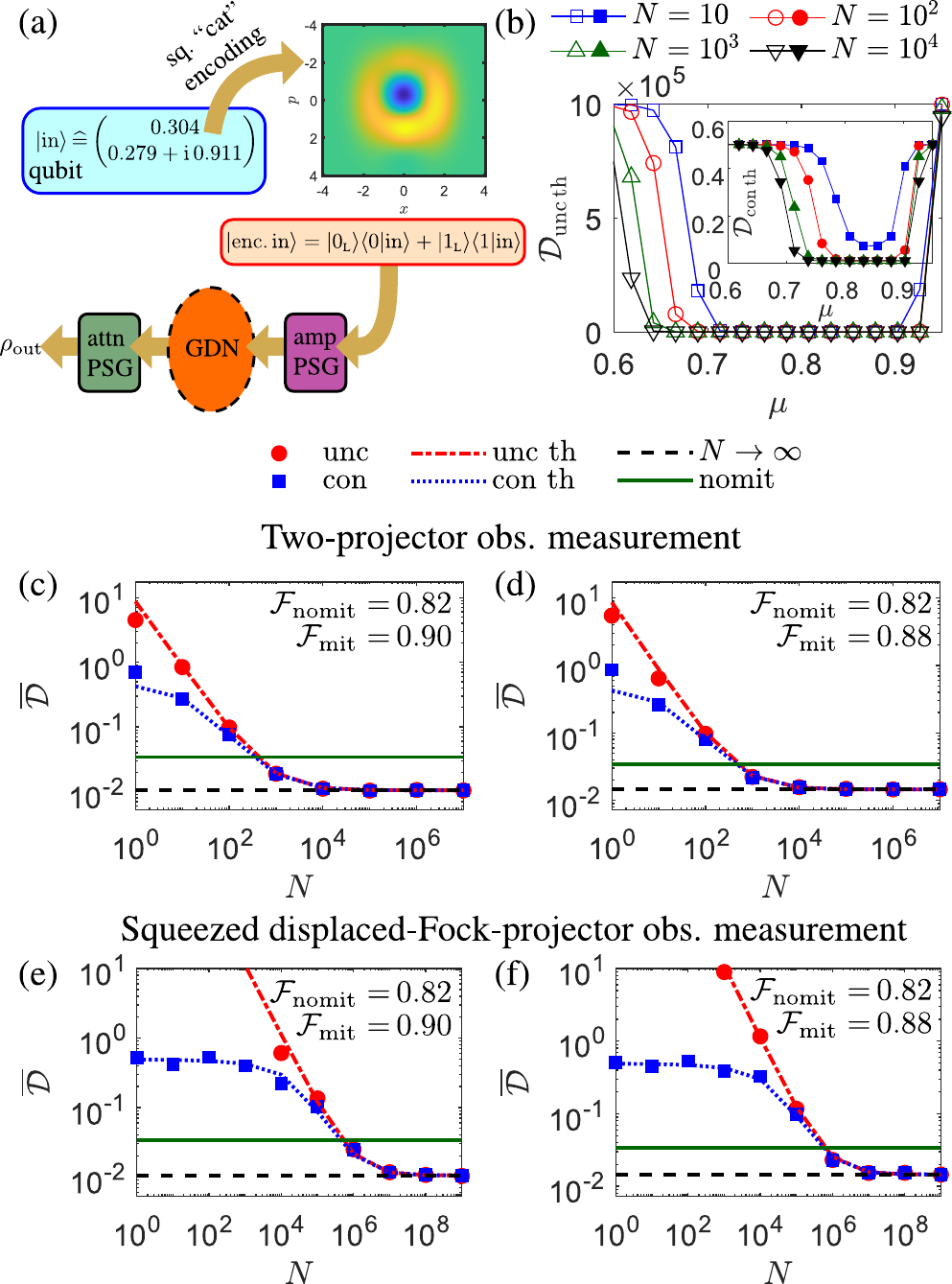}
		\caption{\label{fig:MSE_GDN}PSG-PEC mitigation of GDN for a qubit state encoded in the same 3dB-squeezed-``cat'' code presented in Fig.~\ref{fig:MSE_therm}. To align 
 with the parameters specifying the thermal-noise channel in Fig.~\ref{fig:MSE_therm}, we set $\sigma=0.281$ in accordance with the link between thermal noise and GDN through quantum-limited amplification described in Sec.~\ref{subsec:TN_RDN_chn}. For this channel, $g=1.4$ and $g'=0.733$, while all other specifications follow those of Fig.~\ref{fig:MSE_therm}.}
	\end{figure}
	
	In summary, PSG-PEC follows Algorithm~\ref{alg:psgpec}. Figures~\ref{fig:MSE_therm} and~\ref{fig:MSE_GDN} present the respective performances of \mbox{Alg.~\ref{alg:psgpec}} in mitigating thermal noise and GDN on an example qubit encoded on squeezed-``cat'' logical states [$S\ket{0}_\mathrm{L}\propto S(\ket{\alpha}+\ket{-\alpha})$, $S\ket{1}_\mathrm{L}\propto S(\ket{\alpha}-\ket{-\alpha})$], where $S=S(z)$ is the squeeze operator of amplitude \mbox{$z=r\E{\I\phi}$}. The~observable~$O$ is taken to be $\rho_\mathrm{targ}$~(fidelity estimation), with the ket~$\ket{\text{enc. in}}$ characterizing~$\rho_\mathrm{targ}$ stated in Figs.~\ref{fig:MSE_therm}(a) and~\ref{fig:MSE_GDN}(a). Ideally, one may measure just the two projectors $\ket{o_0}\bra{o_0}=\ket{\text{enc. in}}\bra{\text{enc. in}}$ and $1-\ket{o_0}\bra{o_0}$. For a more practical means to estimate~$\rhoMEAN{\rho_\mathrm{targ}}{\mathrm{out}}$ using squeezing, displacements and photon~counting~\cite{Sendonaris:2024ultrafast}, we write
	\begin{equation}
		O=\rho_\mathrm{targ}=\sum^K_{k=1}\sum^\infty_{n=0}SD(\beta_k)\ket{n}c_{kn}\bra{n}D(\beta_k)^\dag S^\dag
		\label{eq:dispfock_O}
	\end{equation}
	as a linear combination of squeezed displaced-Fock projectors, the ``$\ket{o_l}\bra{o_l}$''s (after normalization in $p_{kl}$), with varying weights $c_{kn}$, which are the ``$o_l$''s. \ref{app:dispfock} lists the required $\{\beta_k\}$ and $\{c_{kn}\}$ that constitute 64~squeezed displaced-Fock projectors ($K=8$) for $z=0.345=3\text{dB}$ and~$\alpha=1$. Note that in the limit $N\rightarrow\infty$, \emph{any} observable measurement leads to the same asymptotic expectation-value estimator, as correctly confirmed by the~figures.
	
	\section{Suppressing dephasing-noise channels}
	\label{sec:supp_deph}

    The PSG-PEC mitigation scheme presented in Sec.~\ref{sec:mit_TN_RDN} is incapable of treating purely-dephasing channels. Instead, here we propose an interferometer-based scheme that can coherently suppress such~channels.
    
	\subsection{Dephasing channel}
	
	A purely-dephasing channel on a state~$\rho$ is modeled by
	\begin{equation}
		\mathcal{E}^{(\gamma)}_\text{deph}[\rho]=\int\D\phi'\,p_\gamma(\phi')\,\E{\I\phi' a^\dag a}\rho\,\E{-\I\phi' a^\dag a}
		\label{eq:deph_chn}
	\end{equation}
	describing a CPTP phase-randomization operation, where $p_\gamma(\phi')$ is the distribution of this mixture and $\gamma\geq0$~quantifies the dephasing~strength. If $\gamma=0$, $p_0(\phi')=\delta(\phi')$. A special class of dephasing channels, namely the central-Gaussian dephasing channels, are characterized by the central Gaussian distribution $p_\gamma(\phi')=\E{-\phi'^2/(2\gamma)}/\sqrt{2\pi\gamma}$.
	
	Since in \ref{app:TN_GDN_DPH}~(also in~\cite{Gagatsos:2017bounding}, for instance), we know that the thermal-noise map is a composition of the loss and quantum-limited amplification channels, the fact that $a^k\E{\I\phi' a^\dag a}=\E{\I\phi'a^\dag a}a^k\E{\I k\phi'}$ implies that
	\begin{align}
		\mathcal{E}^{(\gamma)}_\text{deph}\circ\mathcal{E}^{(\eta,\bar{n})}_\text{therm}=&\,\,\mathcal{E}^{(\eta,\bar{n})}_\text{therm}\circ\mathcal{E}^{(\gamma)}_\text{deph}\,,\nonumber\\
		\mathcal{E}^{(\gamma)}_\text{deph}\circ\mathcal{E}^{(\sigma)}_\text{GDN}=&\,\,\mathcal{E}^{(\sigma)}_\text{GDN}\circ\mathcal{E}^{(\gamma)}_\text{deph}\,;
		\label{eq:chn_commute}
	\end{align}
	that is, thermal-noise channels and GDN channels independently commute with dephasing-noise channels for all channel~parameters. For the same reason, the PSG~map in Eq.~\eqref{eq:psg_map} commutes with $\E{\I\phi' a^\dag a}$, such that PSG-PEC fails for \emph{all} dephasing~channels. Fortunately, there is a linear-optical strategy that suppresses such~channels. 
	
	\begin{figure}
		\centering
		\includegraphics[width=0.7\columnwidth]{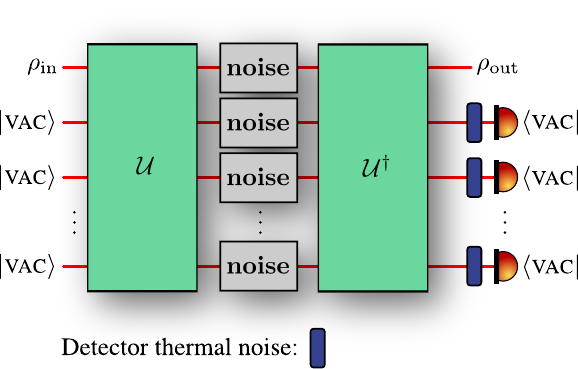}
		\caption{\label{fig:VMZ_supp}The VMZ suppression scheme involving linear-optical elements and conditional vacuum measurements (implementable with on-off photodetectors in~principle). The complete Mach--Zehnder interferometer is described by $M$-mode unitary transformations $\mathcal{U}$ and $\mathcal{U}^\dag$ that both sandwich i.i.d. dephasing noise channels. More realistic considerations would involve additional thermal-noise channels before all photodetectors to account for measurement~imperfections.}
	\end{figure}
	
	\subsection{VMZ suppression scheme}
	
	We propose the \emph{vacuum-based Mach--Zehnder}~(VMZ) suppression scheme to coherently treat dephasing channels, as schematically laid out in Fig.~\ref{fig:VMZ_supp}. This comprises an initial $M$-mode beam~splitter described by the unitary matrix~$\mathcal{U}$ that maps mode ladder operators~$a_j$ to their linear combinations: $a_j\mapsto a'_j=\sum^M_{k=1}\mathcal{U}_{jk}\,a_k$. This is not to be confused with the corresponding $M$-mode unitary evolution operator~$U_M$ acting on a single-mode state $\rho$ and vacuum ancillas---$\rho_M=U_M[\rho\otimes(\vacket\vacbra)^{\otimes M-1}]U_M^\dag$. This unitary map acts on the initially-prepared single-mode quantum state~$\rho_\mathrm{in}$, resulting in an $M$-mode output state~$\rho_M$ that next undergoes i.i.d. dephasing~channels. At the end of the channels, another $M$-mode beam~splitter described by~$\mathcal{U}^\dag$ to complete the mode interference. This Mach--Zehnder interferometric setup is then followed by vacuum measurements on $M-1$ output ancillary~modes. As these vacuum measurements are heralding, they collectively succeed with a non-unit~probability. For~any~$\gamma$ and~$M$, this success probability~is
	\begin{align}
		\psuccVMZ=&\,\mathrm{tr}\Big\{U^\dag_M\,\mathcal{E}^{(\gamma)}_\text{deph}[U_M[\rho_\mathrm{in}\otimes \left(\vacket\vacbra\right)^{\otimes M-1}]U^\dag_M]U_M\left[1\otimes\left(\vacket\vacbra\right)^{\otimes M-1}\right]\Big\}.
	\end{align}
	
	The $M$-mode unitary transformation~$\mathcal{U}$ (or equivalently operator~$U_M$) can either be physically decomposed into a sequence of two-mode BSs and phase~shifters~(PSs)~\cite{Reck:1994experimental} or implemented in other more sophisticated architectures that are feasible in the laboratory~\cite{Clements:2016optimal,Arends:2024decomposing,Hanamura:2024implementing}. The vacuum measurements may also be carried out with on-off detectors~\cite{Nunn:2021heralding}, or with superconducting-nanowire single-photon detectors~(SNSPD) for higher efficiencies~\cite{Hummel:2023nanosecond,Schapeler:2024electrical}. Thus far, no specification regarding $\mathcal{U}$ has been made for VMZ. We shall first show that there exists (at least) two classes of $\mathcal{U}$, namely the Hadamard transformations ($|\mathcal{U}_{jk}|=1/\sqrt{M}$) and unitary two-design transformations~($\mathcal{U}$ with Haar-random~\cite{Collins:2006integration,Mezzadri:2007qr,Puchala_Z._Symbolic_2017,Mele:2024introduction} first and second moments), that can facilitate the inversion of dephasing channels of \emph{any}~$p_\gamma$.
	
	\subsection{Inverting dephasing channels with infinitely many ancillas}
	\label{subsec:vmz_inf_anc}
	
	To understand the channel-inversion capability of VMZ, it is instructive to analyze its action on an arbitrary coherent state~$\ket{\alpha}\bra{\alpha}$ and generalize the result to any state with the P~function~formalism. After the first beam~splitter~($\mathcal{U}$), the coherent~ket~$\ket{\alpha}\vacket^{\otimes M-1}\mapsto\ket{\mathcal{U}_{11}\alpha,\mathcal{U}_{21}\alpha,\ldots,\mathcal{U}_{M1}\alpha}$ is mapped into a product of coherent~kets. The resultant state then undergoes i.i.d. dephasing, which from~Eq.~\eqref{eq:deph_chn} is equivalent to an average over all possible phase-space rotations on each mode according to some distribution~$p_\gamma$.
	
	Let us suppose that the $j$th mode is subjected to a single phase-space rotation by an angle~$\phi_j$. Then
	\begin{align}
		&\,\E{\I\phi_1 a_1^\dag a_1}\E{\I\phi_2 a_2^\dag a_2}\ldots\E{\I\phi_M a_M^\dag a_M}\ket{\mathcal{U}_{11}\alpha,\mathcal{U}_{21}\alpha,\ldots,\mathcal{U}_{M1}\alpha}\nonumber\\
		=&\,\ket{\mathcal{U}_{11}\alpha\,\E{\I\phi_1},\mathcal{U}_{21}\alpha\,\E{\I\phi_2},\ldots,\mathcal{U}_{M1}\alpha\,\E{\I\phi_M}}.
		\label{eq:rand_rot}
	\end{align}
	The second beam~splitter~($\mathcal{U}^\dag$) then turns the output ket in~\eqref{eq:rand_rot} into $\ket{s_1\alpha,s_2\alpha,\ldots,s_M\alpha}$, where the random complex number $s_l=\sum^M_{j=1}\E{\I\phi_j}\,\mathcal{U}^\dag_{lj}\,\mathcal{U}_{j1}$ and $\sum^M_{l=1}|s_l|^2=1$ owing to unitarity of~$\mathcal{U}$ or $\sum^M_{j=1}\mathcal{U}^\dag_{lj}\,\mathcal{U}_{jl'}=\delta_{l,l'}$. The probabilistic conditional vacuum measurements lead to the final~ket
	\begin{align}
		\ket{\alpha}_{\rvec{\phi}}\propto&\,\ket{s_1\alpha}\E{-\frac{1}{2}|\alpha|^2(1-|s_1|^2)}=s_1^{a^\dag a}\ket{\alpha}\bra{\alpha}s_1^{*\,a^\dag a}\,.
		\label{eq:output_coh_ket}
	\end{align}
	
	Equation~\eqref{eq:output_coh_ket} highlights the underlying mechanism behind~VMZ. The interferometric nature of VMZ sets up a mutual Mach--Zehnder~interference on every independently-dephased optical modes between the beam~splitters. Together with the $M-1$ ancillary vacuum measurements, the entire system essentially functions as a quantum adder that computes the sum ($s_1$) of squared magnitudes of random amplitudes that is encoded in a linear-attenuation operation as in Eq.~\eqref{eq:output_coh_ket} (since $|s_1|\leq1$).
	
	We are now ready to investigate the actual output state
	\begin{align}
		\rho_\mathrm{out,\textsc{vmz}}=&\,\dfrac{\overline{s_1^{a^\dag a}\rho_\mathrm{in}\, s_1^{*\,a^\dag a}}}{\tr{\overline{s_1^{a^\dag a}\rho_\mathrm{in}\, s_1^{*\,a^\dag a}}}}\,,
		\label{eq:mixture_linatt}
	\end{align}
	where $\overline{f(\phi_1\ldots\phi_M)}=\int\D\phi_1 p_\gamma(\phi_1)\ldots\int\D\phi_M p_\gamma(\phi_M) f(\phi_1\ldots\phi_M)$. For \emph{any finite}~$M$, Eq.~\eqref{eq:mixture_linatt} tells us that VMZ turns $\rho_\mathrm{out,\textsc{vmz}}$ into a heralded mixture of linearly-attenuated states. To~proceed, we invoke Chebyshev's inequality, which states that~(\ref{app:chebyshev})
	\begin{align}
		\PR(|s_1-\overline{s_1}|\geq\epsilon)\leq&\,\dfrac{\overline{|s_1-\overline{s_1}|^2}}{\epsilon^2}=\dfrac{1-|\lambda_\gamma|^2}{\epsilon^2}\sum^M_{j=1}|\mathcal{U}_{j1}|^4,
		\label{eq:Markov_Chebyshev}
	\end{align}
	where $\lambda_\gamma=\overline{\E{\I\phi_j}}$ is constant for all~$j$ and~$|\lambda_\gamma|\leq1$ just by the triangle~inequality. Owing to the unitarity of~$\mathcal{U}$, we note, also, that $\overline{s_l}=\lambda_\gamma\,\delta_{l,1}$. It turns out that there exist at least two classes of $\mathcal{U}$ for which \mbox{$Y\equiv\sum^M_{j=1}|\mathcal{U}_{j1}|^4\rightarrow0$} as~\mbox{$M\gg1$}. The first is the set of Hadamard transformations, where \mbox{$|\mathcal{U}_{jk}|^2=1/M$} directly gives $Y=1/M$. The second is the set of unitary two-design transformations, where it is shown in \ref{app:HaarU} that the two-design average $\overline{Y}\strut^{\,\textsc{td}}=2/(M+1)\sim1/M$, so that $Y\rightarrow0$ \emph{typically} as~$M\rightarrow\infty$.
	
	Therefore, in the asymptotic limit $M\rightarrow\infty$, for Hadamard and two-design interferometers, \eqref{eq:Markov_Chebyshev} implies that \mbox{$s_l\rightarrow\lambda_\gamma\,\delta_{l,1}$} (the weak law of large numbers), which gives us the key result
	\begin{equation}
		\rho_\mathrm{out,\textsc{vmz}}\xlongrightarrow[M\rightarrow\infty]{}\dfrac{\lambda_\gamma^{a^\dag a}\rho_\mathrm{in}\,\lambda_\gamma^{*\,a^\dag a}}{\tr{\lambda_\gamma^{a^\dag a}\rho_\mathrm{in}\,\lambda_\gamma^{*\,a^\dag a}}}\,,
		\label{eq:deph2linatt}
	\end{equation}
	where the success probability of VMZ is given by $\psuccVMZ=\tr{\lambda_\gamma^{a^\dag a}\rho_\mathrm{in}\,\lambda_\gamma^{*\,a^\dag a}}>0$ and is typically not~small even in the infinite-$M$ regime. Finally, if $\lambda_\gamma=|\lambda_\gamma|\E{\I\,\theta_\gamma}$, the rotated linear-amplification operation~$\lambda_\gamma^{-a^\dag a}=|\lambda_\gamma|^{-a^\dag a}\E{-\I\,\theta_\gamma a^\dag a}$ may be further applied to probabilistically invert this~channel.
	
	In order to better appreciate~\eqref{eq:deph2linatt}, we focus on the central-Gaussian dephasing channel, for which~$\lambda=\E{-\frac{\gamma}{2}}$. Such a channel also possesses a simple Kraus representation~[Eq.~\eqref{eq:deph_Kraus}], where each Kraus operator is accompanied with an exponential operator that is quadratic in~$a^\dag a$. Under this context, we remark that~\eqref{eq:deph2linatt} may also be understood as an experimentally-feasible linearization of these quadratic exponents that permits successful inversion \emph{without} requiring Kerr nonlinearities, where the latter is currently known to be technologically challenging. It has been shown that linear optics suffices to approach the maximum success rate for linear amplification~\cite{Guanzon:2024saturating}. We shall next show that even with a finite~$M$, VMZ already possesses dephasing-suppression qualities for weak dephasing strengths~(small~$\gamma$).
	
	\subsection{Suppression capabilities with finite number of ancillas}
	
	Let us analyze the circumstance with finite~$M$ by focusing on the popular \emph{central-Gaussian} dephasing channel. For this channel, it is more expedient to express the \emph{pure} input state~$\rho_\mathrm{in}=\sum^\infty_{m,n=0}\ket{m}\rho_{mn}\bra{n}$ in the Fock basis to acquire the VMZ-suppressed output~state
	
	\begin{align}
		\rho_\mathrm{out,\textsc{vmz}}=\dfrac{\sum^\infty_{m,n=0}\ket{m}\rho_{mn}\,S_{mn}\bra{n}}{\sum^\infty_{n=0}\rho_{nn}\,S_{nn}}\,,
		\label{eq:rhooutvmz}
	\end{align}
	where $S_{mn}=\overline{s_1^ms_1^{*n}}$ and $\psuccVMZ=\sum^\infty_{n=0}\rho_{nn}\,S_{nn}$. In~the limit of small~$\gamma$, with~$w=1-\sum^M_{j=1}|\mathcal{U}_{j1}|^4$, the Kraus representation in \eqref{eq:deph_Kraus} and calculations in \ref{app:smsn} give
	
	\begin{equation}
		S_{mn}\cong1-\dfrac{\gamma}{2}\!\left[(m+n)w+(m-n)^2\!\left(1-w\right)\right].
		\label{eq:Smn_small_gamma}
	\end{equation}
	
	To compare the quality of the suppressed output state relative to the unsuppressed one, we shall calculate the fidelities $\mathcal{F}_\mathrm{supp}=\tr{\rho_\mathrm{in}\,\rho_\mathrm{out,\textsc{vmz}}}$ and $\mathcal{F}_\mathrm{nosupp}=\tr{\rho_\mathrm{in}\,\rho_\mathrm{out,no\,\textsc{vmz}}}$, where $\rho_\mathrm{out,no\,\textsc{vmz}}=\sum^\infty_{m,n=0}\ket{m}\rho_{mn}\,\E{-\frac{\gamma}{2}(m-n)^2}\bra{n}$. After invoking Taylor's expansion on the numerator and denominator of Eq.~\eqref{eq:rhooutvmz} \emph{via} the use of~\eqref{eq:Smn_small_gamma}, their difference reads
	\begin{equation}
		\Delta\mathcal{F}\equiv \mathcal{F}_{\rm supp}-\mathcal{F}_{\rm nosupp}=\dfrac{\gamma\, w}{2} \sum^\infty_{m,n=0}|\rho_{mn}|^2(m-n)^2
		\label{eq:Delta_F}
	\end{equation}
	up to first order in~$\gamma$, whence we find that $\mathcal{F}_{\rm supp}\geq\mathcal{F}_{\rm nosupp}$ for small~$\gamma$. This~states that for \emph{any} beam~splitter~$\mathcal{U}$, dephasing suppression with VMZ is \emph{always}~successful in the weak-dephasing~regime.
	
	\begin{figure}[t]
		\centering
		\includegraphics[width=.7\columnwidth]{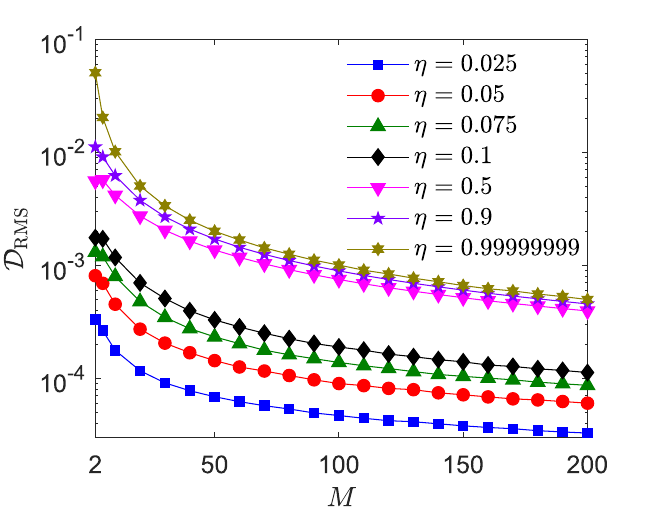}
		\caption{\label{fig:Had_smsn}The root-mean-square distance~$\mathcal{D}_{\mathrm{RMS}}=\sqrt{\sum^{9}_{m,n=0}(S_{\mathrm{Had},mn}-\widetilde{S}_{mn})^2/10^2}$ against~$M$ for different central-Gaussian dephasing channels of dephasing rate $\eta=1-\E{-\gamma}$, where a cutoff dimension of $10$ was chosen for illustration. The largest $\mathcal{D}_{\mathrm{RMS}}$ occurs at $\gamma=\infty$ and $M=2$, corresponding to $S_{\mathrm{Had},mn}=\delta_{m,n}4^{-m}\binom{2m}{m}$~[obtained from Eq.~\eqref{eq:smsn} with $p_j=1/M$] and $\widetilde{S}_{mn}=\delta_{m,0}\,\delta_{n,0}$. Hence, $\max\mathcal{D}_{\mathrm{RMS}}\cong0.089$.}
	\end{figure}
	
	Equation~\eqref{eq:Delta_F} also provides us the opportunity to search for the optimal beam~splitter~$\mathcal{U}$ that maximizes~$\Delta\mathcal{F}$ in this~regime. As $w$ is the only function of~$\mathcal{U}$, under the notation~$p_j\equiv|\mathcal{U}_{j1}|^2$, this amounts to a simple minimization of $w$ over $\mathcal{U}$ under the unitarity~constraint, which is equivalent to maximizing $1-\sum^M_{j=1}p^2_j$ over $p_j$ such that~$\sum^M_{j=1}p_j=1$. Thus, the optimal solution is $p_j=1/M$; that is, the optimal~$\mathcal{U}$ is~Hadamard. Moreover, \emph{any} Hadamard interferometer, not just a specific choice, can optimally suppress all central-Gaussian dephasing channels if $\gamma$ is sufficiently~small. The corresponding optimal infidelity \emph{in this context} is given by
	\begin{equation}
		1-\mathcal{F}^{\rm opt}_{\rm supp}=\dfrac{\gamma}{2M} \sum^\infty_{m,n=0}|\rho_{mn}|^2(m-n)^2+\{\text{higher-order terms in $\gamma$}\}\,.
	\end{equation}
    The small-$\gamma$ condition implies that this optimal infidelity asymptotically goes to zero \emph{up to the leading order in $\gamma$~only}.
	
	Interestingly, for \emph{any}~$\gamma$, one can also arrive at an approximate analytical formula for $S_{mn}$ with $p_j=1/M$ (denoted as $S_{\mathrm{Had},mn}$). A very good approximation to $S_{\mathrm{Had},mn}$ is in fact
	\begin{equation}
		S_{\mathrm{Had},mn}\cong \widetilde{S}_{mn}=\dfrac{\E{-\frac{\gamma}{2M}(m-n)^2}}{\left[1+\frac{\gamma}{M}(m+n)\right]^{\frac{M-1}{2}}}\,.
		\label{eq:S_Hadmn}
	\end{equation}
	Strictly speaking, the central limit theorem is applied to multinomial distributions~\cite{Severini:2005aa} in order to derive \eqref{eq:S_Hadmn}, which normally requires~$m,n\gg1$. In this regime and the limit~\mbox{$M\rightarrow\infty$}, \eqref{eq:S_Hadmn} approaches~$\E{-\frac{\gamma}{2}(m+n)}$ in accordance with~Eq.~\eqref{eq:deph2linatt}, so that the discrepancy between~$S_{\mathrm{Had},mn}$ and~$\widetilde{S}_{mn}$ tends to zero with increasing~$M$. Details are found in \ref{app:smsn} and~\ref{app:multinomial}. Rather surprisingly, Fig.~\ref{fig:Had_smsn} demonstrates remarkable accuracy of~\eqref{eq:S_Hadmn} even for small~$m$ and~$n$.
	
	\section{Performance on known bosonic codes}
	\label{sec:perf}
	
	We showcase the performance of error mitigation and suppression of noise channels for bosonic codes that are of general interest to quantum~computation. For a given code, a single-qubit ket~$\ket{\mathrm{qbit}}=\ket{0}c_0+\ket{1}c_1$ is encoded onto bosonic logical kets~$\ket{0}_\textsc{l}$ and~$\ket{1}_\textsc{l}$ to generate the input state $\rho_\mathrm{in}=\ket{\text{enc.~in}}\bra{\text{enc.~in}}$, with $\ket{\text{enc.~in}}=\ket{0}_\textsc{l}c_0+\ket{1}_\textsc{l}c_1$. All mitigation and suppression protocols serve to protect this input state against thermal~noise, GDN, dephasing~noise and their~compositions. State fidelity is chosen as the figure of merit to quantify the mitigation and suppression qualities, since a high fidelity would naturally entail a good quality in either the coherent output state or any observable expectation-value estimation, be it at the logical or bosonic qubit~level.
	
	For a binary number $b$, in the Fock and position~$\{\ket{x}_q\}$ bases, the four bosonic codes considered here that are often of interest in quantum information and computation~are
	\begin{itemize}
		\item \textbf{Binomial}~\cite{Michael:2016new,Soule:2023concatenating}, $\texttt{bin}(n,\kappa)$: 
		$$\ket{b}_\textsc{l} =2^{-\frac{\kappa-1}{2}}\sum_{k=0}^{\lfloor \frac{\kappa-b}{2}\rfloor}\ket{(2k+b)n}\sqrt{\binom{\kappa}{2k+b}}.$$
		\item \textbf{$n$-component ``cat''}~\cite{Shringarpure:2024error,Mirrahimi:2014dynamically,Hastrup:2020deterministic,Su:2022universal,Hastrup:2022all-optical}, $\texttt{cat}(n,\alpha)$: 
		$$\ket{b}_\textsc{l} \propto\sum_{k=0}^{\infty} \ket{n(k+b/2)}\frac{\alpha^{n(k+b/2)}}{\sqrt{[n(k+b/2)]!}}\quad(\text{even }n).$$
		\item \textbf{Squeezed two-component ``cat''}~\cite{Gao:2018programmable,Teh:2020overcoming,Schlegel:2022quantum,Provazník:2024adapting}, $\texttt{sqcat}(z,\alpha)$.
		\item \textbf{Finite-energy GKP}~\cite{Gottesman:2001encoding,Royer:2020stabilization,Tzitrin:2020progress,Grimsmo:2021quantum,Rojkov:2024two-qubit,Marek:2024ground}, $\texttt{gkp}(\Delta)$:
		$$\ket{b}_\textsc{l} \propto \sum^\infty_{k=-\infty}\E{-\Delta^{2}\,a^\dag a}\ket{(2k+b)\sqrt{\pi}}_q.$$
	\end{itemize}
	Subsequently reported squeezing strength $|z|=r$ and~$\Delta$ are in decibels~(dB)--- $r=r_\mathrm{dB}(\ln 10)/20$ and~$\Delta=10^{-\Delta_\mathrm{dB}/20}$~\cite{Tzitrin:2020progress}. Numerical simulations are carried out with the help of \texttt{Strawberry Fields}~\cite{Killoran:2019strawberry,Bromley:2020applications} and \texttt{Mr~Mustard}~\cite{Yao:2024riemannian}~packages.

    Finally, all PSG and VMZ schemes shall be accompanied by measurement errors in the forms of PSGN layers and detection noise after the interferometer as described previously, which we take to be~thermal. In general, state-preparation errors are also an important consideration for evaluating the performance of error-mitigation and suppression schemes. These errors are generally not thermal in~character, with state-encoding errors being a typical example. Unfortunately, since the modeling of state-preparation errors heavily depends on the specific procedure considered in the generation of the encoded input state, a general inclusion of these errors into the simulations is not~possible. Nonetheless, in \ref{app:stateprep}, we carry out an investigation on the quality of ``cat'' logical states undergoing state-preparation noise that originates from the approximate preparation and the photon-lossy~heralding. There, we demonstrate the robustness of our error mitigation and suppression schemes to \emph{both} state-preparation and measurement~errors.
	
	\begin{figure}[t]
		\centering
		\includegraphics[width=.7\columnwidth]{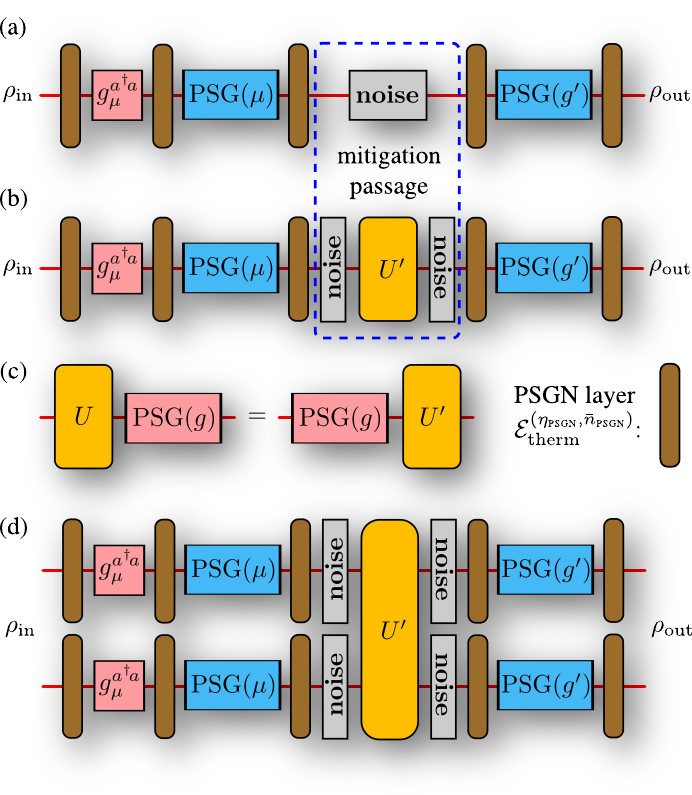}
		\caption{\label{fig:PSG_schematic_nogate_gate}Noise mitigation with PSG-PEC in two scenarios: a PSG-treated state that is either (a)~idling or (b)~subjected to a gate~operation. (c)~If the intended unitary-gate operation on~$\rho_\mathrm{in}$ is~$U$, then a proper gate modification~($U'$) should be carried out to minimize, if not eradicate, gate distortion coming from the commutation through $g^{a^\dag a}_\mu$ and~$\mathrm{PSG}(\mu)$, which would otherwise introduce unwanted excess errors in addition to gate~noise. All diagrams refer to single-mode gate operations, which are easily generalized to multimode cases by employing independent PSG-mitigation setups for each~mode. (d)~Shown here is an example for a two-mode~system.}
	\end{figure}
		
	\subsection{Mitigating thermal noise and GDN on bosonic codes}
	\label{subsec:mit_codes}
	
	We measure the performance of PSG-PEC on all four bosonic codes by how they fare in estimating the fidelity between the target state~$O=\rho_\mathrm{targ}$ with respect to the mitigated output state in comparison to the unmitigated~(noisy) output~state. We shall compare asymptotic fidelities in the limit of large~$N$. For thermal-noise and GDN channels, we will consider two scenarios in which noise is mitigated. The first scenario is one in which a quantum system, after being operated on by an amplifying~PSG, undergoes idle~noise as considered in Secs.~\ref{subsec:asymp_chn_invert_PSG} and~\ref{subsec:err_mit_PSG-PEC}, either during storage and/or propagation stages of its journey in a quantum-information~task. The target state $\rho_\mathrm{targ}=\rho_\mathrm{in}$ is the ideal input~state.
		
	In the second scenario~[Fig.~\ref{fig:PSG_schematic_nogate_gate}(b)], we would like to perform PSG-PEC on noise generated from an additional unitary gate~$U$ applied to the system \emph{within} the passage that is in between the amplifying and attenuating PSGs~(mitigation passage). This is an important application of PSG-PEC to quantum~computation. The target state is then~$\rho_\mathrm{targ}=U\rho_\mathrm{in}U^\dag$. However, we caution that care must be taken to ensure that the gate~$U'$ inside the mitigation passage is properly adjusted so that it corresponds to the intended gate operation~$U$. The two are generally not the same as $\mathcal{M}_{\textsc{psg},g}[U\,\bm{\cdot}\,U^\dag]=U'\mathcal{M}_{\textsc{psg},g}[\,\bm{\cdot}\,]U'^\dag\neq U\mathcal{M}_{\textsc{psg},g}[\,\bm{\cdot}\,]U^\dag$ and a wrong choice of $U'$ would result in low mitigated quality due to additional gate~distortion by the~PSG~[Fig.~\ref{fig:PSG_schematic_nogate_gate}(c)]. Single-mode gate noise is then modeled as two identical (thermal-noise or GDN) layers of noise rate~$\eta/2$ before and after~$U'$, where $\eta$ is also the noise rate of the idle noise~channel. Noise of a two-mode gate is modeled as two layers before~$U'$ and two after~(all identical), each possessing a noise rate of~$\eta/2$. For the given encoded $\rho_\mathrm{in}$, fair performance comparisons between idle- and gate-noise scenarios can thus be made under the same total noise rate per mode when $\eta$ is small~(see \ref{app:additivity}).
	
	\begin{figure}
		\centering
		\includegraphics[width=.57\columnwidth]{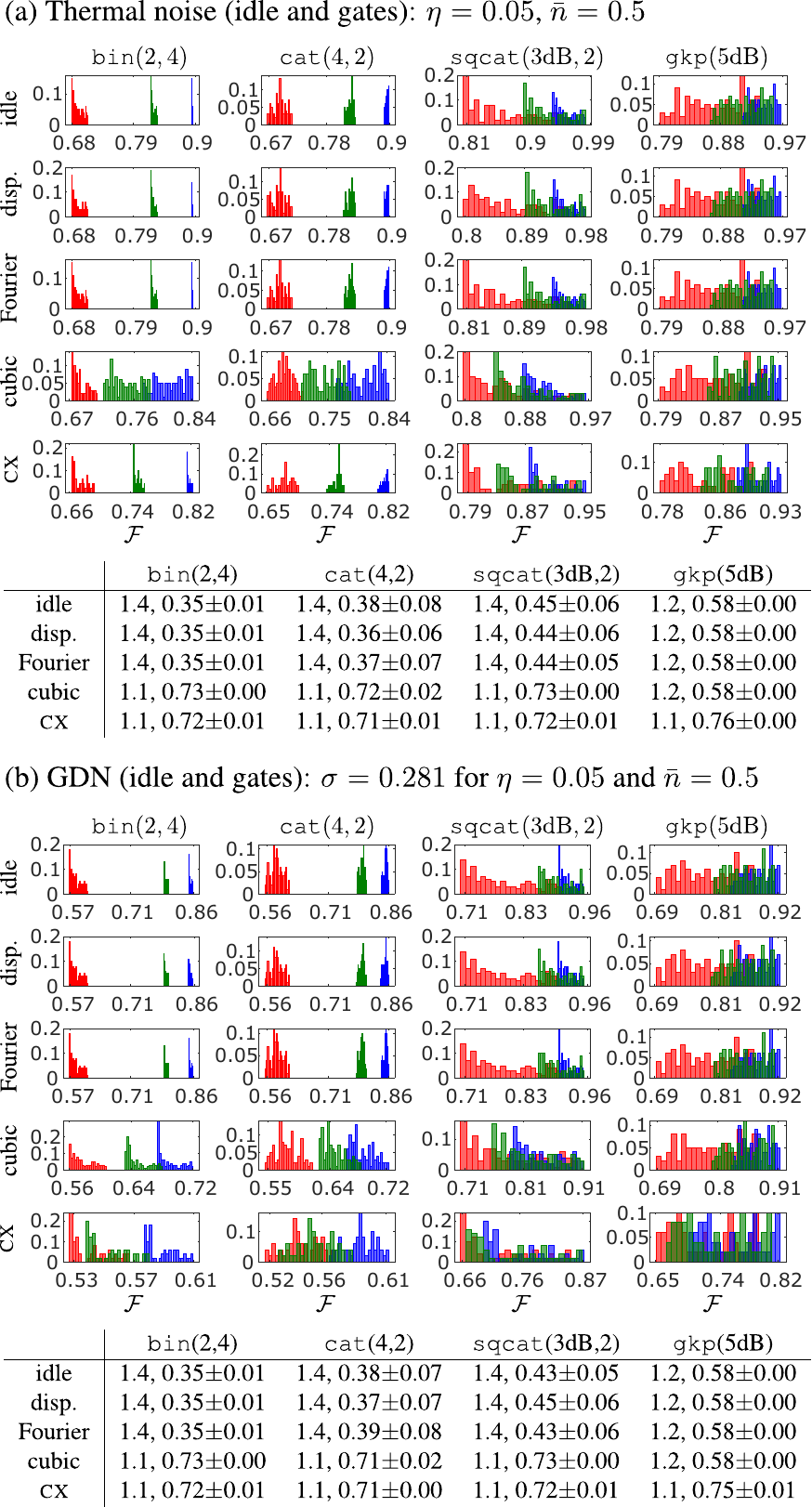}
		\caption{\label{fig:PSG-PEC_therm_GDN_allgates}The large-$N$ estimated fidelities with PSG-PEC mitigation in the absence~(blue) and presence~(green) of PSGN ($\eta_\textsc{psgn}=0.02,\bar{n}_\textsc{psgn}=0.1$), and without~(red) are shown here for both (a)~thermal noise and (b)~GDN. All histograms are normalized in~probability. A total of 100 random single-mode~$\rho_\mathrm{in}$s for each bosonic code are generated by encoding qubit states chosen uniformly from the Bloch sphere to simulate the PSG-PEC performance with the idle, $D(0.27+0.55\I)$, $\I^{a^\dag a}$ and $\E{\I\,0.05\,q^3}$~operations. For the $\CX$~gate~$\E{\I\,0.1\,q_1p_2}$, 50 random two-mode~$\rho_\mathrm{in}$s per code are generated from product states, where one input mode ($\CX$~target) is the vacuum state and the other ($\CX$~control) a random encoded-qubit~state. Both~$g$ and \emph{state-averaged} $\max\psuccPSGPEC$ from Eq.~\eqref{eq:best-case-psuccPSGPEC}, along with its standard deviation, are given, where smaller $g$ values are applied to both cubic and \CX~gates to avoid significant gate~distortion. All red and blue histograms for both idle and Fourier-gate operations are \emph{almost} identical since all channels are additive for small $\eta$ (such as those considered here) and $\I^{a^\dag a}$ commutes with the PSG and both the thermal-noise and GDN channels based on the commutation relation~$a^k\,\I^{a^\dag a}=\I^{a^\dag a} (\I a)^k$.}
	\end{figure}
	
	We will focus on the universal gate-set $\mathcal{S}=\left\{D(\alpha),\I^{a^\dag a},\E{\I \chi q^3},\E{\I\chi' q\otimes p}\right\}$~\cite{Hillmann:2020universal,Calcluth:2024sufficient,Budinger:2024all-optical} consisting of the displacement, Fourier, cubic and \CX~gates in this~order. The modded gate~$U'$ for PSG-PEC mitigation is straightforward regarding the first two gates; that is, the pair $U=D(\alpha)$, $U'=D(g\alpha)$ for the displacement gate and $U=\I^{a^\dag a}=U'$ for the Fourier gate, both valid for a PSG of gain/loss factor~$g$, are respectively a result of Eq.~\eqref{eq:psg_props} and the commutativity between a rotation gate and the PSG~operation. The cubic phase~gate and \CX~gate, on the other hand, do not exhibit easy forms of~$U'$ to our~knowledge. The search for such a~$U'$ for these gates that either exactly nullifies or partially cancel PSG~gate~distortion is an interesting future~work. Nevertheless, when $\chi$ and $\chi'$ are not too large, (multimode) PSG-PEC is still successful even with realistic imperfect operations as gate distortion by the amplifying PSG is not~major.
	
	In both idling and gate-operated scenarios, for either thermal noise or GDN, the general procedure for PSG-PEC is given in Alg.~\ref{alg:psgpec}. Furthermore, the linear-amplification operation~$g^{a^\dag a}_{\mu_\mathrm{opt}}$, being nondeterministic, succeeds with a certain probability that determines the success rate of our PSG-PEC mitigation~protocol. We adopt the linear-optical recipe proposed in~Ref.~\cite{Guanzon:2024saturating} that saturates the \emph{maximum achievable} success probability, which was found to be
	\begin{equation}
		\max\psuccPSGPEC=g_{\mu_\mathrm{opt}}^{-2}=(g/\mu_\mathrm{opt})^{-2}
		\label{eq:best-case-psuccPSGPEC}
	\end{equation}
	in that~reference~$(g>1)$ and is also the one that we take to benchmark the protocol for all bosonic~codes. Finally, noisy gates are modeled as the gate $U'$ sandwiched between two identical noise channels~(be it thermal or GD) of rate~$\eta/2$. In this way, the \emph{total gate-noise channel}~(composition of the two sandwiching channels) is the same as the idle noise~channel of rate~$\eta$ for small~$\eta$ due to channel additivity~(see \ref{app:additivity}), which ensures an equal-footing comparison of PSG-PEC performance during idling and gate~operations.	
	
	\begin{figure}[t]
		\centering
		\includegraphics[width=.7\columnwidth]{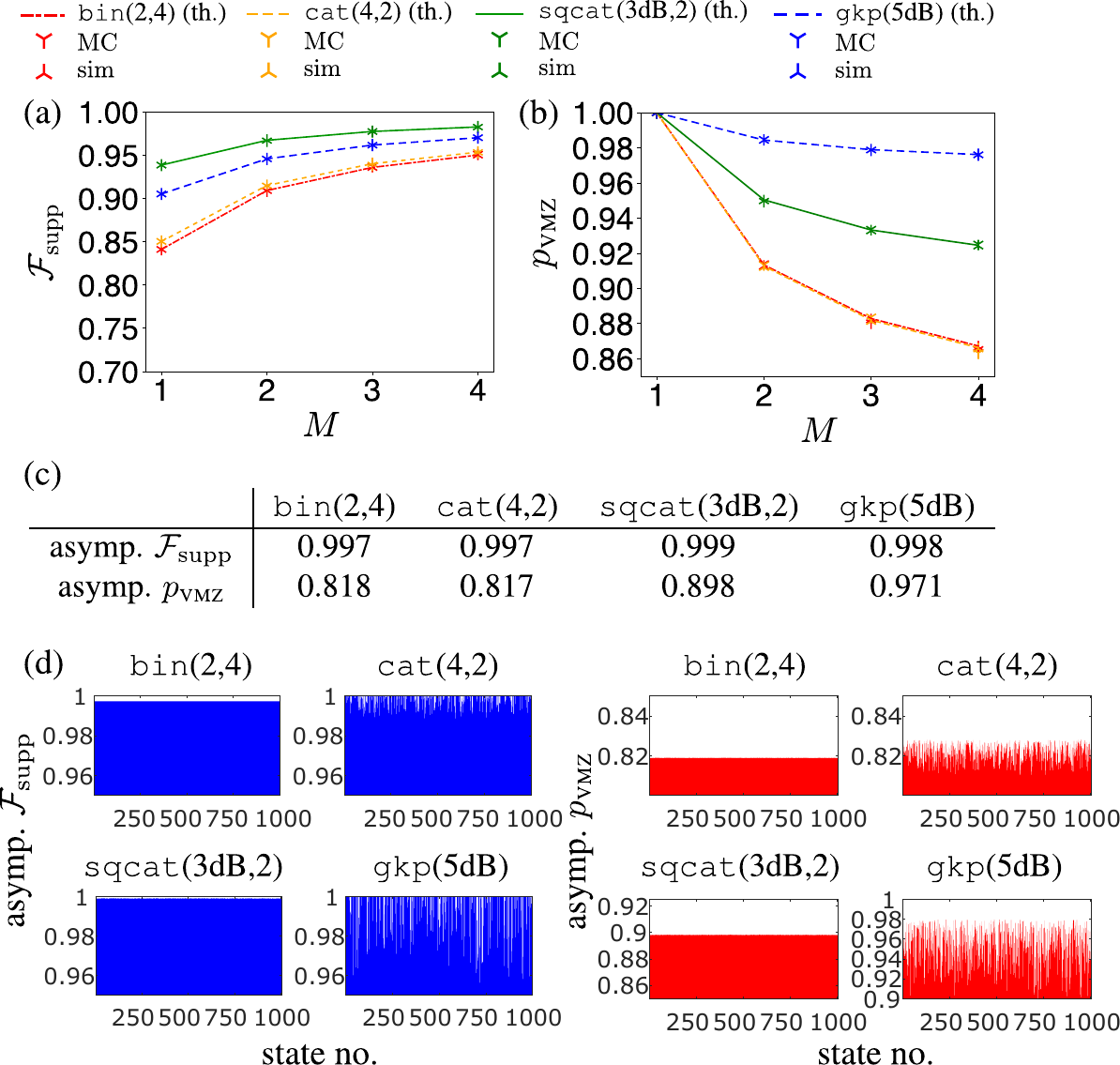}
		\caption{\label{fig:VMZ_fid_psucc_th_fit}VMZ~suppression of an idling central-Gaussian dephasing channel \mbox{($\eta=1-\E{-\gamma}=0.05$)} for bosonic codes encoding a randomly-chosen qubit ket $\ket{\mathrm{qbit}}=\ket{0}0.765+\ket{1}(0.641+0.058\I)$. using a Hadamard~$\mathcal{U}$ of $M$~modes ($M-1$ ancillary~modes). The (a)~suppression fidelity~$\mathcal{F}_\mathrm{supp}$ and (b)~success probability~$\psuccVMZ$ are plotted, where $M=1$ signifies the absence of dephasing suppression (no~VMZ). The analytical approximation~(th.) in~\eqref{eq:S_Hadmn} and success-probability formula are compared with both numerical computation of the summations in Eq.~\eqref{eq:smsn} that explicitly defines~$S_{mn}$ through Monte~Carlo multinomial sampling~(MC) and simulations using \texttt{Strawberry Fields} and \texttt{Mr~Mustard}~(sim). (c)~Up to the first decimal place, these finite-$M$ values are very close to the asymptotic ones~($M\rightarrow\infty$) already at~$M=2$. (d)~Moreover, for 1000 randomly-chosen qubits (uniform on the Bloch sphere) encoded in these bosonic codes, asymp.~$\mathcal{F}_\mathrm{supp}$ are near-unity and asymp.~$\psuccVMZ$ are relatively~high. Here, all vacuum measurements suffer no additional thermal noise for~simplicity.}
	\end{figure}
	
	Figure~\ref{fig:PSG-PEC_therm_GDN_allgates}(a) presents the performance of PSG-PEC ($g>1$) in mitigating thermal noise for idle and universal gate operations on all four bosonic~codes. For all tested encoded-qubit states, \emph{all} unmitigated estimated fidelities are lower than mitigated ones, with or without~PSGN. In particular, a smaller~$g$ is applied to \texttt{gkp} in order to simulate this code with the same Fock-space dimension as other codes for the same computation~efficiency~($d=20$ per mode for $\CX$ and $d=60$ for single-mode~gates). The fact that with such low values of~$g$, PSG-PEC still permits significant improvements in fidelity-estimation~accuracy is testament to the practical utility of~PSG-PEC in mitigating thermal-noise and GDN~channels. The relatively high best-case~$\psuccPSGPEC$, achievable with linear-optical methods~\cite{Guanzon:2024saturating}, motivates the search for evermore efficient ways to perform linear-amplification~operations.
	
	Mitigating GDN with $\sigma=\sqrt{\eta(1+\bar{n})/(1-\eta)}$ using \mbox{$\eta=0.05$} and $\bar{n}=0.5$ set in Fig.~\ref{fig:PSG-PEC_therm_GDN_allgates}(a), as shown in Fig.~\ref{fig:PSG-PEC_therm_GDN_allgates}(b), proves to be more difficult with the same values of~$g$. The~overall estimated fidelities (mitigated or not) are lower than when noise channels are thermal for the same~$\eta$ and~$\bar{n}$. The performance on the cubic phase gate and $\CX$~gate tell us that PSG-PEC is especially prone to even slight PSG imperfections in the case of~GDN.
	
	\begin{table}[b]
		\caption{\label{tab:modU_VMZ}Gate modifications for VMZ of~$M=2$, where $\mathrm{BS}_{1,2}(\vartheta,\varphi)=\exp(\vartheta\,(\E{\I\varphi}a_1a^\dag_2-\E{-\I\varphi}a^\dag_1a_2))$ describes a real~BS, here taken to be balanced, $\widetilde{\mathrm{BS}}_{1,2}\equiv\mathrm{BS}_{1,2}(\pi/4,0)$ for simplicity, $R_j(\varphi)=\exp(\I\,\varphi\,a^\dag_j a_j)$ the rotation operator and all subscripts refer to the mode numbers acted~upon. For single-mode gates, the second and third columns respectively list the modified gates according to~$\widetilde{\mathrm{BS}}_{1,2}\,U=V'\,\widetilde{\mathrm{BS}}_{1,2}$ and~$R_1(\varphi)\,U=V'\,R_1(\varphi)$. For two-mode gates, they coincide with~$\widetilde{\mathrm{BS}}_{1,2}\widetilde{\mathrm{BS}}_{3,4}\,U=V''\,\widetilde{\mathrm{BS}}_{1,2}\widetilde{\mathrm{BS}}_{3,4}$ and~$R_1(\varphi)R_2(\varphi')\,U=V''\,R_1(\varphi)R_2(\varphi')$. Regarding the CX~gate, modes~$1$ and~$3$ refer to the inputs, and~$2$ and~$4$ the BS~ancillas.}
		\vspace{2ex}
		\centering
		\begin{tabular}{c|c|c}
			\hline\hline
			$U$ & $U'=V'$ & $U'=V''$\\
			\hline
			$D_1(\alpha)$ & $D_1\!\left(\alpha/\sqrt{2}\right)D_2\!\left(\alpha\sqrt{2}\right)$ & $D_1(\alpha\,\E{\I\varphi})$\\
			$\I^{a_1^\dag a_1}$ & $\I^{\frac{1}{2}\left(a_1^\dag+a_2^\dag\right)(a_1+a_2)}$ & $\I^{a_1^\dag a_1}$\\
			$\E{\I\chi q_1^n}$ & $\E{\I\frac{\chi}{2^{n/2}} (q_1+q_2)^n}$ & $\E{\I\chi (q_1\!\cos\varphi+p_1\!\sin\varphi)^n}$\\
			$\E{\I\chi'q_1p_3}$ & \!\!\!$\E{\I\frac{\chi'}{2}(q_1+q_2)(p_3+p_4)}$ &
			\!$\begin{array}{c}\exp(\I\chi'(q_1\!\cos\varphi+p_1\!\sin\varphi)\\	\,\times(p_3\cos\varphi'-q_3\sin\varphi'))\end{array}$\!\!\\
			\hline\hline
		\end{tabular}		
	\end{table}
	
	\begin{figure}[t]
		\centering
		\includegraphics[width=.6\columnwidth]{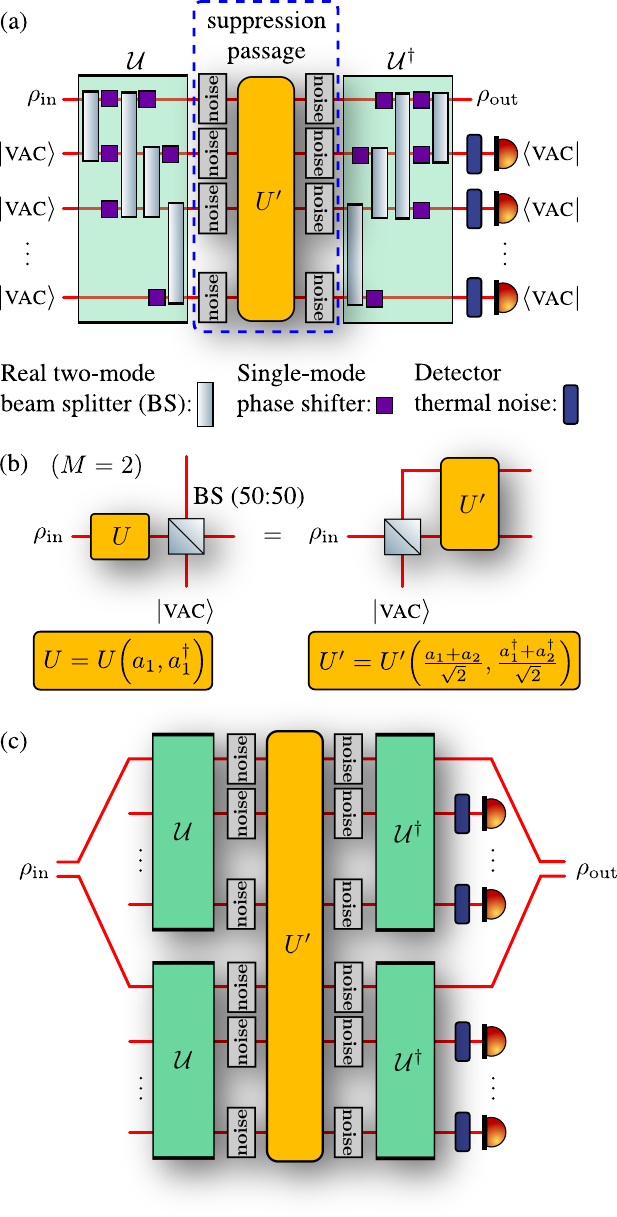}
		\caption{\label{fig:VMZ_schematic_gate}Single-mode gate modification in VMZ suppression to avoid interferometric gate~distortion. (a)~From~\cite{Clements:2016optimal}, every $\mathcal{U}$ can be implemented with real, two-mode BSs and PSs (phase shifters). (b)~For $M=2$, a typical Hadamard $\mathcal{U}$ corresponds to the balanced two-mode~BS. The~$U'$ of the (noiseless) gate operation~$U$ on the input state~$\rho_\mathrm{in}$ is, thus, a function of linear combinations of ladder operators on different~modes. For larger~$M$, $U'$ is further altered by~PSs. (c)~Multimode gate-noise suppression (shown here for two input modes) is carried out using independent VMZ~setups.}
	\end{figure}	
	
	\subsection{Suppressing dephasing noise on bosonic codes}
	\label{subsec:supp_codes}
	
	We first look at scenario~one in which the encoded qubit is subjected to an idling central-Gaussian dephasing~channel. Figure~\ref{fig:VMZ_fid_psucc_th_fit} plots the convergence of both the VMZ-suppression fidelity~$\mathcal{F}_\mathrm{supp}$ and success probability~$\psuccVMZ$ corresponding to a randomly-chosen qubit that is encoded on all four bosonic~codes. This encoded qubit is one of the many other random ones tested that yields an $\mathcal{F}_\mathrm{supp}$ and $\psuccVMZ$ which are already close to the asymptotic values~($M\rightarrow\infty$) even when $M$ is small, up to the first decimal place. Moreover, over the qubit Bloch sphere, the distributions of $\mathcal{F}_\mathrm{supp}$ and $\psuccVMZ$ are both generally high for these tested bosonic~codes. From numerical observations, typically, the increase in~$\mathcal{F}_\mathrm{supp}$ is largest in going from $M=1$~(no VMZ suppression) to~$M=2$.	
    
    \begin{figure}[t]
    	\centering
    	\includegraphics[width=.7\columnwidth]{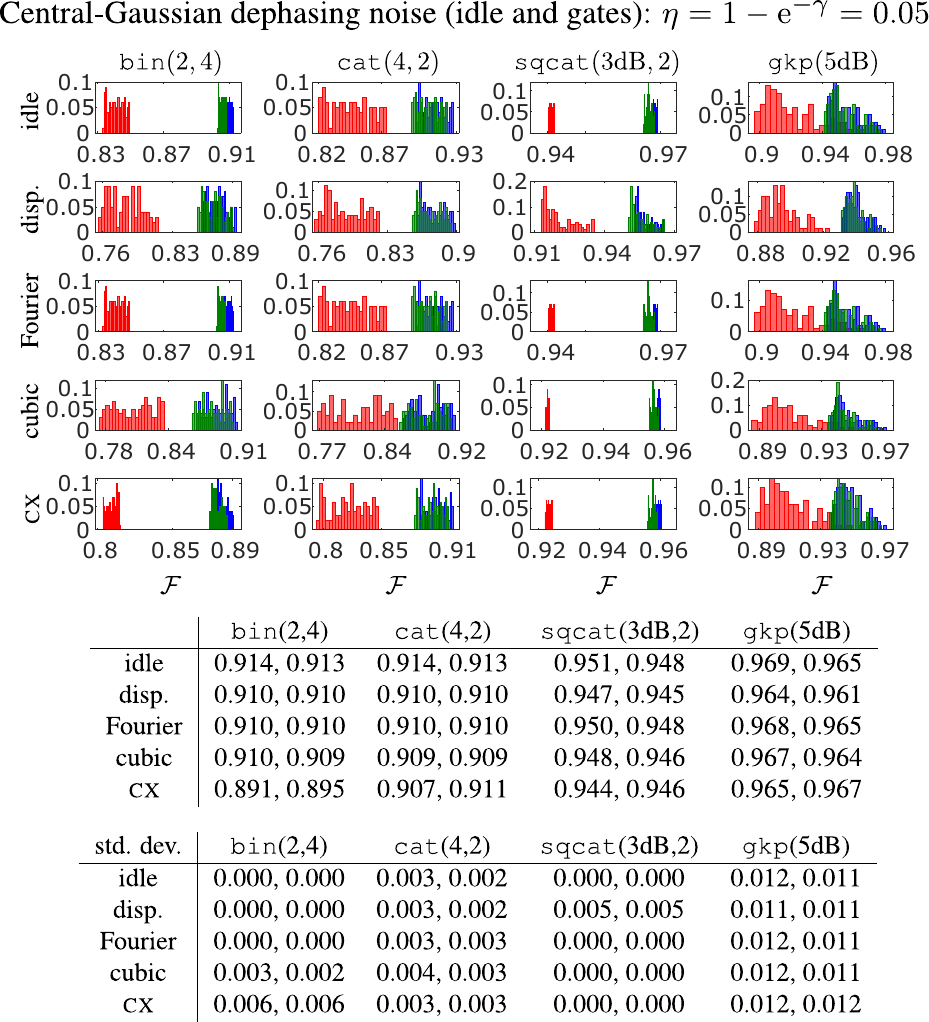}
    	\caption{\label{fig:VMZ_deph_allgates}Output fidelities with Hadamard-VMZ dephasing suppression~\mbox{($M=2$)} in the absence~(blue) and presence~(green) of detector thermal~noise ($\eta_\textsc{det}=0.05$, $\bar{n}_\textsc{det}=0$ for the \CX~gate and 0.1 for all other gates), and without~(red). The~operations considered are idle, $D(0.27+0.55\I)$, $\I^{a^\dag a}$, $\E{\I\,0.05\,q^3}$ and~$\E{\I\,0.5\,q_1p_2}$, where $\rho_\mathrm{in}$s are generated in the same way as in Fig.~\ref{fig:PSG-PEC_therm_GDN_allgates}. VMZ~success probabilities in the absence~($\psuccVMZ$, left) and presence~($\widetilde{p}_\textsc{vmz}$, right) of detector thermal noise are~listed. Just like PSG-PEC, all red and blue histograms for both idle and Fourier-gate operations are \emph{almost} identical since the dephasing channel is additive for small $\eta$ and $\I^{a^\dag a}$ clearly commutes with this noise~channel.}
    \end{figure}
    
	When a single-mode gate~$U$ are applied to the encoded qubit~(scenario~two), then the attempt to suppress purely-dephasing channels coming from the gate operation by the VMZ scheme would also result in gate distortion. From Ref.~\cite{Clements:2016optimal}, the beam~splitter~$\mathcal{U}$ may be decomposed into real two-mode~BSs and single-mode phase~shifters~[Fig.~\ref{fig:VMZ_schematic_gate}(a,b)]. The gate distortion is therefore coming from commuting $U$ through these linear-optical passive components. Fortuitously, there exist convenient algebraic forms of~$U'$ needed inside the suppression~passage. Such prescriptions may be generalized to multimode gate operations~[Fig.~\ref{fig:VMZ_schematic_gate}(c)]. Table~\ref{tab:modU_VMZ} lists the corresponding~$U'$s for all $U$s in the universal gate-set $\mathcal{S}$ when~$M=2$. Noisy $U'$s are, likewise, modeled in exactly the same way as in Sec.~\ref{subsec:mit_codes}. Here, the dephasing-noise layers sandwiching the noisy~$U'$, be it a single- or two-mode gate, also each possesses a rate $\eta/2$ per mode such that the total gate-dephasing channel is that of the idle-dephasing channel of rate~$\eta$ for small~$\eta$ as per \ref{app:additivity}.

	Figure~\ref{fig:VMZ_deph_allgates} lists the performance of the VMZ suppression scheme on all bosonic codes. Remarkably, just using a simple balanced BS~($M=2$ or single ancillary mode) with no additional phases, significant improvements in output-state fidelities is still achievable. For reasonably-small detector imperfections, the suppressed fidelities remain almost~unchanged.
	
	\subsection{Suppressive mitigation of noise-channel compositions}
	
	We take this moment to reiterate some crucial notes on the~PSG and VMZ~schemes. Recall that the PSG is unable to offer any error mitigation on dephasing~channels. At~the same time, we show in \ref{app:thm-GDN-VMZ} that the VMZ protocol cannot suppress \emph{both} thermal noise and~GDN. More generally, however, noise channels may be modeled as a composition of different kinds. An example is the composition of a thermal-noise channel and a dephasing channel, which is also known to be degradable~\cite{Leviant:2022quantum}. The commutativity between thermal-noise and dephasing-noise channels, and between GDN and dephasing-noise channels, as in~\eqref{eq:chn_commute}, tells us that any other complicated compositions of thermal-noise~(or GDN) and dephasing channels may be simplified to such a two-layered~composition. Our error-mitigation and suppression strategies would prove useful if they can be utilized to suppressively mitigate these noise-channel~compositions.
	
	\begin{figure}[b]
		\centering
		\includegraphics[width=.6\columnwidth]{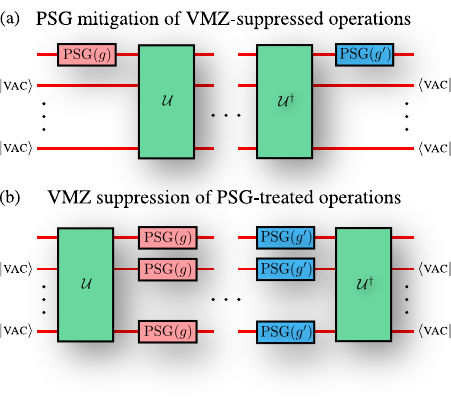}
		\caption{\label{fig:VMZ_PSG_commute}Compatibility between PSG-mitigation and VMZ-suppression schemes: (a) and~(b) are equivalent for \emph{any} $\mathcal{U}$, $g$ and~$g'$.}
	\end{figure}
	
	Indeed, it is possible to simultaneously implement PSG-PEC and VMZ-suppression protocols to treat such noise compositions. According to Fig.~\ref{fig:VMZ_PSG_commute}, one may either (a)~sandwich the multimode VMZ~interferometer with PSGs or, instead, (b)~sandwich multimodal PSGs with the VMZ~interferometer. It turns out (see \ref{app:PSG_VMZ_exclu}) that both configurations are identical, which tells us that PSG-mitigation and VMZ-suppression schemes are \emph{mutually~compatible}, so that we can conveniently carry out both schemes without paying attention to their~ordering. Advice on the forms of~$U'$ depending on the gate operation~$U$ that are given in Secs.~\ref{subsec:mit_codes} and~\ref{subsec:supp_codes} may be heeded to cope with gate~distortions.

    \begin{figure}[t]
		\centering
		\includegraphics[width=1\columnwidth]{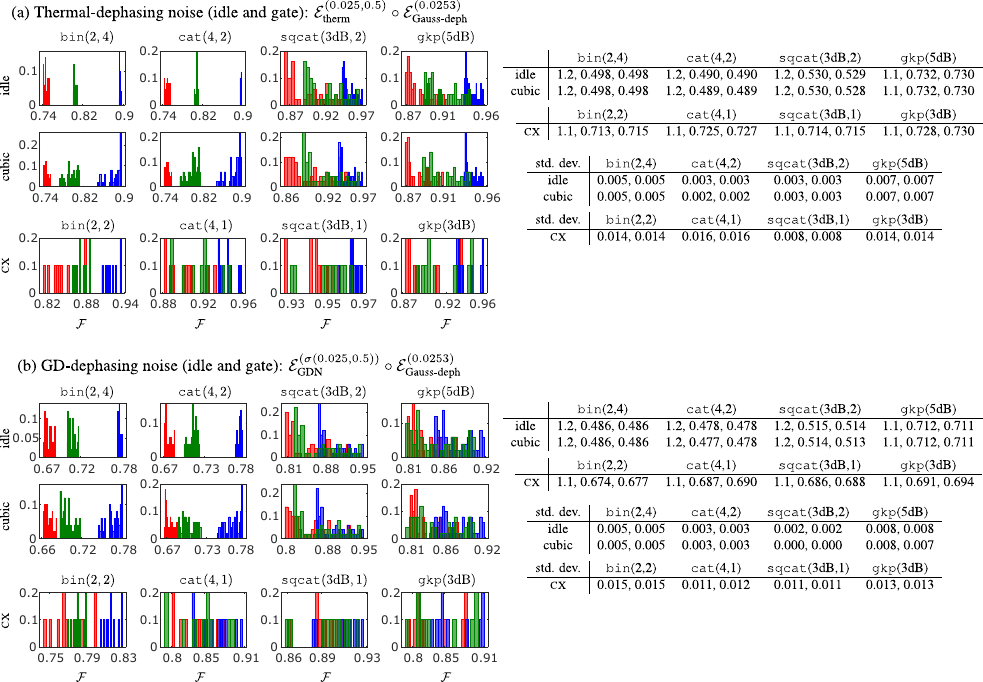}
		\caption{\label{fig:hybridPSGVMZ}Suppressive mitigation of (a)~thermal-dephasing noise and (b)~GD-dephasing noise for the idle operation and the cubic phase gate~$\E{\I\,0.02\,q^3}$. Here the two channel-noise layers are equally distributed in $\eta$ so that the \emph{total noise rate} remains as~0.05, with $\gamma=0.0253$ corresponding to the dephasing rate of~0.025. The triplet of values $g$, $\max\psuccHYBRID$ and $\max\widetilde{p}_\textsc{psg-vmz}$ (accounting for the vacuum-measurement thermal noise in the VMZ~part), together with the standard deviations of both types of success probabilities over 50 states are quoted~(10 for~\CX). In contrast to the unmitigated fidelities~(red), all PEC mitigated ones, be it in the ideal case~(blue) or in the presence of PSGN \emph{and} vacuum-detection thermal noise~(green), are always higher for thermal-dephasing~noise. On the other hand, the same sort of difficulty in mitigating GD-dephasing noise as exhibited in Fig.~\ref{fig:PSG-PEC_therm_GDN_allgates}(b)~exists. The parameters characterizing PSGN and VMZ-detector noise are $\eta_\textsc{psgn}=0.02$, $\bar{n}_\textsc{psgn}=0.1$, $\eta_\textsc{det}=0.05$, $\bar{n}_\textsc{det}=0$ for the \CX~gate and 0.1 for all other~gates. [\textsc{note}]~For \CX, owing to extended computation time on multimode Hilbert spaces, slightly smaller bosonic codes are considered.}
    \end{figure}
	
	In short, the \emph{hybrid PSG-VMZ scheme} is equivalent to PSG-PEC mitigation carried out on the VMZ-suppressed passage as shown in Fig.~\ref{fig:VMZ_PSG_commute}(a), where the final mitigation procedure is once again described in Alg.~\ref{alg:psgpec}. As a key demonstration of the hybrid PSG-VMZ protocol for suppressive mitigation of composite noise channels, Fig.~\ref{fig:hybridPSGVMZ} presents the PEC error-mitigation results pertaining to the idle, cubic-phase- and \CX-gate operations for thermal-dephasing and GD-dephasing~noise. Here, all dephasing noise are assumed to be central-Gaussian. The best-case success probability for this hybrid scheme is: 
	\begin{equation}
		\max\psuccHYBRID=(g/\mu_\mathrm{opt})^{-2}\sum^\infty_{k=0}p_k\,p_{\textsc{vmz},k}\,,
	\end{equation}
	where $p_k$ is the probability of subtracting $k$~photons and $p_{\textsc{vmz},k}$ is the VMZ success rate when this~happens. We see that when PSGN and thermally-noisy vacuum detectors~\cite{Scherer:2009quantum} are considered in the hybrid protocol, the difficulty in mitigating GD-dephasing noise manifests itself as possible lower mitigated fidelities in comparison to the unmitigated~ones, which originates from the GDN part of the composite noise~channel.
	
	\section{Discussion}
	
	In the age of NISQ devices, where information-processing and computation tasks are carried out in relatively noisy environments, it is of paramount importance to be able to protect quantum information from the influence of noise using feasible resources and methodologies. In this work, we have successfully made use of linear-optical techniques to mitigate and suppress common noise that plague qubits encoded onto quantum bosonic~codes. 
	
	In particular, using gadgets involving photon subtraction, linear attenuation and amplification, all realizable with passive linear components, photodetectors and manageable nonclassical resource states, we showed that thermal noise and general random displacement noise can be asymptotically mitigated from measured expectation values. We further developed proper constrained statistics theory to improve the error-mitigation~quality. Even more interestingly, we proved that just by employing a straightforward multimode Mach--Zehnder interferometer and vacuum-state measurements, any purely-dephasing noise can be turned into a linear-attenuation channel, with an, often, high success probability when the number of modes is very~large. This can then be, in principle, completely inverted with a linear~amplification operation. Such an inversion, previously thought to be accomplishable only with nonlinear optical~elements, is now known to be doable linear~optically. Moreover, for any finite number of modes and dephasing rate that is not too large, we show that such an interferometric scheme always suppresses the commonly-considered central-Gaussian dephasing channel and that a uniform multimode beam~splitter is optimal for this~purpose. In the simplest case of a single ancillary mode, a balanced beam-splitter-based interferometer suffices and can lead to significant state-fidelity improvement for reasonable noise~parameters. Last, but not the least, we found that both the mitigation and suppression schemes are mutually compatible with each other, so that they may be implemented simultaneously in any order to mitigate more realistic noise channels that are compositions of the aforementioned types, namely thermal-dephasing and Gaussian-displacement-dephasing. 
	
	For common bosonic encodings, we numerically verified our established theoretical formalism and confirmed, indeed, that our linear-optical mitigation and suppression schemes can reduce errors due to idling noise~channels. More importantly, we tested the schemes on universal gate-set operations and found that they are also effective in mitigating and suppressing gate~noise. Furthermore, all results demonstrated that these protocols are robust to slight detector~imperfections.
	
	Owing to the recent advancement in experimental implementation of linear amplification at higher success rates and multimode interferometers, the realization of mitigation and suppression protocols presented here is becoming increasingly more feasible in practice. Therefore, this work finds immediate practical applications of noise mitigation and suppression on current quantum~technologies.\\
	
	\noindent
	\emph{Special Note.---}After publishing this work, we were informed of the existence of Ref.~\cite{Swain:2024improving} that employed the same setup as in our VMZ scheme~(with the technique coined ``unitary averaging''~\cite{Marshman:2018passive}) to reduce the effect of dephasing errors on Gaussian~states using Hadamard~interferometers. In Ref.~\cite{Swain:2024improving}, quality measures of Gaussian-state purity, squeezing and entanglement were shown to improve with the~suppression. In our work, we analytically showed that (1)~the VMZ setup can completely invert dephasing noise with the help of linear amplification in the $M\rightarrow\infty$~limit, that (2)~the setup applies to general input states beyond Gaussian~ones, and that (3)~Hadamard interferometers indeed maximize suppression infidelity under central-Gaussian dephasing errors for finite~$M$ and small~$\gamma$.
	
	\section{Acknowledgments}
	
	We thank A.~Taylor, H.~Kwon, K.~Park, \mbox{N.~Lo~Piparo} and O.~Solodovnikova for fruitful discussions. This work is supported by the National Research Foundation of Korea (NRF) grants funded by the Korean government (MSIT) (Grant~Nos.~NRF-2023R1A2C1006115, RS-2023-00237959, RS-2024-00413957, RS-2024-00437191, and RS-2024-00438415) via the Institute of Applied Physics at Seoul National University, the Institute of Information \& Communications Technology Planning \& Evaluation (IITP) grant funded by the Korea government (MSIT) (IITP-2024-2020-0-01606), and the Brain Korea 21 FOUR Project grant funded by the Korean Ministry of Education.
	
	\appendix
	\section{Properties of thermal, GD and dephasing channels}
	\label{app:TN_GDN_DPH}
	
	\subsection{Definitions}
	
	Subsequent discussion relies on the identity~\cite{Teo:2015qs,Perina:1991aa}
	\begin{align}
		&\,\int\dfrac{(\D\beta)}{\pi}\E{-a|\beta|^2+b_1\beta+b_2\beta^*+c_1\beta^2+c_2\beta^{*2}}=\dfrac{1}{\sqrt{a^2-4\,c_1c_2}}\exp\!\left(\dfrac{a\,b_1b_2+c_1b_2^2+c_2\,b_1^2}{a^2-4\,c_1c_2}\right)
		\label{eq:cmplx_gauss}
	\end{align}
	for $\RE{a}\geq|c_1+c_2^*|$, offering an interesting route through the use of the coherent-state basis in contrast to the usual covariance-matrix comparison of Gaussian channels~\cite{Serafini:2017aa}.
	
	Both the loss and quantum-limited amplification channels may be defined in terms of Kraus operators inasmuch as	
	\begin{align}
		\mathcal{L}_{\mu}[\rho]=&\,\sum^\infty_{k=0}\dfrac{\mu^k}{k!}\,(1-\mu)^{a^\dag a/2}\,a^k\,\rho\,a^{\dag\,k}\,(1-\mu)^{a^\dag a/2}\,,\nonumber\\
		\mathcal{A}_G[\rho]=&\,\dfrac{1}{G}\sum^\infty_{k=0}\dfrac{(1-G^{-1})^k}{k!}\,a^{\dag\,k}\,G^{-a^\dag a/2}\,\rho\,G^{-a^\dag a/2}\,a^{k}\,.
	\end{align}
	If $\mu=1-(1-\eta)/G$ and $G=1+\eta\,\bar{n}$, then it can be shown that $\mathcal{A}_{G}\circ\mathcal{L}_{1-(1-\eta)/G}=\mathcal{E}^{(\eta,\bar{n})}_\text{therm}$, where $\mathcal{L}_{\eta}=\mathcal{E}^{(\eta,0)}_\text{therm}$. To see this, we observe what happens to an arbitrary coherent state under these CPTP~maps. While it is simple to arrive at $\mathcal{L}_\mu[\ket{\alpha}\bra{\alpha}]=\ket{\sqrt{1-\mu}\,\alpha}\bra{\sqrt{1-\mu}\,\alpha}$, we look at the Husimi Q~function of the final state:
	\begin{align}
		&\,\HQ{\beta}=\bra{\beta}\mathcal{A}_G\!\left[\ket{\alpha\sqrt{1-\mu}}\bra{\alpha\sqrt{1-\mu}}\right]\ket{\beta}\nonumber\\
		=&\,\dfrac{1}{G}\sum^\infty_{k=0}\dfrac{(1-G^{-1})^k}{k!}|\beta|^{2k}\,\E{(G^{-1}-1)(1-\mu)|\alpha|^2}\,\E{-|\beta-\alpha\sqrt{(1-\mu)\,G^{-1}}|^2}\nonumber\\
		=&\,\dfrac{1}{G}\,\E{-\frac{1}{G}|\beta-\alpha\sqrt{1-\eta}|^2}\,.
	\end{align}
	From the connection
	\begin{equation}
		\HQ{\beta}=\int\dfrac{(\D\beta')}{\pi}\,\GSP{\beta'}\,\E{-|\beta'-\beta|^2}
	\end{equation}
	between the Q and P~functions~\cite{Cahill:1969qd}, and the fact that $\HQ{\beta}$ is a Gaussian function, it is clear that $\GSP{\beta}$ is also Gaussian. Indeed, after recognizing~that
	\begin{equation}
		\mathcal{E}^{(\eta,\bar{n})}_\text{therm}[\ket{\alpha}\bra{\alpha}]=\int\dfrac{(\D\beta)}{\pi}\ket{\beta}\dfrac{\E{-\frac{1}{\eta\,\bar{n}}|\beta-\alpha\sqrt{1-\eta}|^2}}{\eta\,\bar{n}}\bra{\beta}
	\end{equation}
	from \eqref{eq:gen_thm_RDN}, upon inserting the P~function
	\begin{equation}
		\GSP{\beta}=\dfrac{\E{-\frac{1}{\eta\,\bar{n}}|\beta-\alpha\sqrt{1-\eta}|^2}}{\eta\,\bar{n}}
	\end{equation}
	of $\mathcal{E}^{(\eta,\bar{n})}_\text{therm}[\ket{\alpha}\bra{\alpha}]$, we see that
	\begin{align}
		\HQ{\beta}=&\,\dfrac{\E{-\frac{1-\eta}{\eta\,\bar{n}}|\alpha|^2-|\beta|^2}}{\eta\,\bar{n}}\int\dfrac{(\D\beta')}{\pi}\,\E{-\frac{1+\eta\,\bar{n}}{\eta\,\bar{n}}|\beta'|^2}\E{\left(\frac{\sqrt{1-\eta}}{\eta\,\bar{n}}\alpha^*+\beta^*\right)\!\beta'+\mathrm{c.c.}}\nonumber\\
		=&\,\dfrac{\E{-\frac{1-\eta}{\eta\,\bar{n}}|\alpha|^2-|\beta|^2}}{1+\eta\,\bar{n}}\,\E{\frac{1}{\eta\,\bar{n}\,(1+\eta\,\bar{n})}|\alpha\sqrt{1-\eta}+\beta\,\eta\,\bar{n}|^2}\nonumber\\
		=&\,\dfrac{1}{G}\,\E{-\frac{1}{G}|\beta-\alpha\sqrt{1-\eta}|^2}\,.
	\end{align}
	
	By the same token, using Eq.~\eqref{eq:cmplx_gauss}, we may verify that $\mathcal{A}_{1/(1-\eta)}[\mathcal{E}^{(\eta,\bar{n})}_\text{therm}[\ket{\alpha}\bra{\alpha}]]=\mathcal{E}^{(\sigma)}_{\text{GDN}}[\ket{\alpha}\bra{\alpha}]$ through comparison of state Q~functions. After some steps of calculations, the Q~function of the left-hand side is
	\begin{equation}
		\opinner{\beta}{\mathcal{A}_{1/(1-\eta)}[\mathcal{E}^{(\eta,\bar{n})}_\text{therm}[\ket{\alpha}\bra{\alpha}]]}{\beta}=\dfrac{1-\eta}{1+\eta\,\bar{n}}\,\E{-\frac{1-\eta}{1+\eta\,\bar{n}}|\alpha-\alpha'|^2}\,,
	\end{equation}
	whereas that of the right-hand side is
	\begin{equation}
		\opinner{\beta}{\mathcal{E}^{(\sigma)}_{\text{GDN}}[\ket{\alpha}\bra{\alpha}]}{\beta}=\dfrac{1}{1+\sigma^2}\,\E{-\frac{1}{1+\sigma^2}|\alpha-\alpha'|^2}\,,
	\end{equation}
	so that $\sigma^2=\eta\,(1+\bar{n})/(1-\eta)$.
	
	Lastly, the central-Gaussian dephasing channel leads to
	\begin{align}
		\int\D\phi\,\dfrac{\E{-\frac{\phi^2}{2\gamma}}}{\sqrt{2\pi\gamma}}\,\E{\I\phi\,a^\dag a}\ket{m}\bra{n}\E{-\I\phi\,a^\dag a}=\ket{m}\E{-\frac{\gamma}{2}(m-n)^2}\bra{n}
	\end{align}
	on the operator element~$\ket{m}\bra{n}$. It is easy to see that the Taylor expansion $\E{-\frac{\gamma}{2}(m-n)^2}=\E{-\frac{\gamma}{2}m^2}\sum^\infty_{k=0}\frac{\gamma^k}{k!}(mn)^k\E{-\frac{\gamma}{2}n^2}$ straightforwardly supplies
	
	\begin{equation}
		\mathcal{E}^{(\gamma)}_\text{Gauss-deph}[\rho]=\sum^\infty_{k=0}\dfrac{\gamma^k}{k!}\,\E{-\frac{\gamma}{2}(a^\dag a)^2}\,(a^\dag a)^k\rho\,(a^\dag a)^k\,\E{-\frac{\gamma}{2}(a^\dag a)^2}.
		\label{eq:deph_Kraus}
	\end{equation}
	
	\subsection{Approximate channel additivity in noise parameters}
	\label{app:additivity}
	
	We study the result of composing two channels, beginning with thermal channels, where an application of~\eqref{eq:cmplx_gauss} yields
	\begin{align}
		&\,\mathcal{E}^{(\eta_2,\bar{n})}_\text{therm}\circ\mathcal{E}^{(\eta_1,\bar{n})}_\text{therm}[\ket{\alpha}\bra{\alpha}]\nonumber\\
		=&\,\int\dfrac{(\D\beta)}{\pi}\ket{\beta}\dfrac{\E{-\frac{1-\eta_1}{\eta_1\bar{n}}|\alpha|^2-\frac{1}{\eta_2\bar{n}}|\beta|^2}}{(\eta_1+\eta_2-\eta_1\eta_2)\bar{n}}\,\E{\frac{\eta_1\eta_2}{(\eta_1+\eta_2-\eta_1\eta_2)\,\bar{n}}\left|\frac{\sqrt{1-\eta_1}}{\eta_1}\alpha+\frac{\sqrt{1-\eta_2}}{\eta_2}\beta\right|^2}\bra{\beta}.
	\end{align}
	Now, for small~$\eta_1$ and~$\eta_2$, since
	\begin{align}
		\eta_1+\eta_2-\eta_1\eta_2\cong&\,\,\eta_1+\eta_2\,,\nonumber\\
		\dfrac{1-\eta_1}{\eta_1}\left(\frac{\eta_2}{\eta_1+\eta_2-\eta_1\eta_2}-1\right)\cong&\,-\dfrac{1-\eta_1-\eta_2}{\eta_1+\eta_2}\,,\nonumber\\
		\frac{1}{\eta_2}\left[(1-\eta_2)\frac{\eta_1}{\eta_1+\eta_2-\eta_1\eta_2}-1\right]\cong&\,-\dfrac{1}{\eta_1+\eta_2}\,,
	\end{align}
	it is clear that for any $\bar{n}\geq0$
	\begin{equation}
		\mathcal{E}^{(\eta_2,\bar{n})}_\text{therm}\circ\mathcal{E}^{(\eta_1,\bar{n})}_\text{therm}\cong\mathcal{E}^{(\eta_1+\eta_2,\bar{n})}_\text{therm}\,\,\text{for small~$\eta_1$, $\eta_2$.}
	\end{equation}
	
	Next, for the GDN channels, the sleight-of-hand substitution~$u=\beta+\beta'$ and $u'=\beta-\beta'$ corresponding to a Jacobian of determinant equal to \emph{four} followed by the employment of~\eqref{eq:cmplx_gauss} on the $u'$ integration~yields,
	\begin{align}
		&\,\mathcal{E}^{(\sigma_2)}_\text{GDN}\circ\mathcal{E}^{(\sigma_1)}_\text{GDN}[\ket{\alpha}\bra{\alpha}]\nonumber\\
		=&\,\int\!\dfrac{(\D\beta)}{\pi\sigma_2^2}\int\!\dfrac{(\D\beta')}{\pi\sigma_1^2}\ket{\alpha+\beta+\beta'}\E{-\frac{|\beta'|^2}{\sigma_1^2}-\frac{|\beta|^2}{\sigma_2^2}}\bra{\alpha+\beta+\beta'}\nonumber\\
		=&\,\int\dfrac{(\D u)}{\pi(\sigma_1^2+\sigma_2^2)}\ket{\alpha+u}\E{-\frac{|u|^2}{\sigma_1^2+\sigma_2^2}}\bra{\alpha+u},
	\end{align}
	which clearly means that
	\begin{equation}
		\mathcal{E}^{(\sigma_2)}_\text{GDN}\circ\mathcal{E}^{(\sigma_1)}_\text{GDN}=\mathcal{E}^{\left(\sqrt{\sigma_1^2+\sigma_2^2}\right)}_\text{GDN}\,\,\text{for any~$\sigma_1$ and~$\sigma_2$},
	\end{equation}
	the standard consequence of convolving two Gaussian~distributions. If $\sigma^2_j=\eta_j(1+\bar{n})/(1-\eta_j)$, then for small $\eta_1$ and~$\eta_2$,
	\begin{align}
		\sigma^2_1+\sigma^2_2=&\,(1+\bar{n})\,\frac{\eta_1+\eta_2-2\eta_1\eta_2}{1-\eta_1-\eta_2-\eta_1\eta_2}\nonumber\\
		\cong&\,(1+\bar{n})\,\dfrac{\eta_1+\eta_2}{1-\eta_1-\eta_2}\,.
	\end{align}
	In other words,
	\begin{equation}
		\mathcal{E}^{(\sigma(\eta_2,\bar{n}))}_\text{GDN}\circ\mathcal{E}^{(\sigma(\eta_1,\bar{n}))}_\text{GDN}\cong\mathcal{E}^{\left(\sigma(\eta_1+\eta_2,\bar{n})\right)}_\text{GDN}\,\,\text{for small~$\eta_1$, $\eta_2$.}
	\end{equation}
	
	Finally, the statement
	\begin{equation}
		\mathcal{E}^{(\gamma_2)}_\text{Gauss-deph}\circ\mathcal{E}^{(\gamma_1)}_\text{Gauss-deph}=\mathcal{E}^{(\gamma_1+\gamma_2)}_\text{Gauss-deph}\,\,\text{for any~$\gamma_1$ and~$\gamma_2$}
	\end{equation}
	follows immediately from the actions
	\begin{align}
		&\,\mathcal{E}^{(\gamma_2)}_\text{Gauss-deph}\circ\mathcal{E}^{(\gamma_1)}_\text{Gauss-deph}[\ket{m}\bra{n}]\nonumber\\
		=&\,\mathcal{E}^{(\gamma_2)}_\text{Gauss-deph}\!\left[\ket{m}\E{-\frac{\gamma_1}{2}(m-n)^2}\bra{n}\right]\nonumber\\
		=&\,\ket{m}\E{-\frac{\gamma_1+\gamma_2}{2}(m-n)^2}\bra{n}
	\end{align}
	on the arbitrary Fock element~$\ket{m}\bra{n}$. In terms of the dephasing rate $\eta=1-\E{-\gamma}$, the Reader is invited to verify that
	\begin{equation}
		\mathcal{E}^{(\eta_2)}_\text{Gauss-deph}\circ\mathcal{E}^{(\eta_1)}_\text{Gauss-deph}=\mathcal{E}^{(\eta_1+\eta_2)}_\text{Gauss-deph}\,\,\text{for small~$\eta_1$, $\eta_2$.}
	\end{equation}
	
	To conclude, all these noise channels are additive in the regime of small noise rates~$\eta$.

	\section{Constrained parameter estimation}
	\label{app:constr_opt}
	
	Let $x=\sum^M_{j=1}w_j\nu_j=\rvec{w}\bm{\cdot}\rvec{\nu}$, where $w_j$s are just weights of the sum over all multinomial relative frequencies $\nu_j$. If one defines a new random variable $y=\Theta(x-a)\Theta(b-x)x+\Theta(a-x)a+\Theta(x-b)b$ that is limited to the range $[a,b]$, where $\Theta$ is the Heaviside step function, then the task of interest here is the calculation of~$\AVG{}{y^n}$, where
	\begin{equation}
		y^n=\Theta(x-a)\Theta(b-x)x^n+\Theta(a-x)a^n+\Theta(x-b)b^n\,.
		\label{eq:ymom}
	\end{equation} 
	Because $\Theta(0)=1/2$ by convention, we may consider $a$ and $b>a$ to respectively be ever so slightly (up to numerical precision) smaller and larger than the lower and upper limits of the actual range of $\sum^M_{j=1}w_jp_j=\rvec{w}\bm{\cdot}\rvec{p}$. Using the famous integral representation
	\begin{equation}
		\Theta(x)=\lim_{\epsilon\rightarrow0^+}\int\dfrac{\D \tau}{2\pi \I}\,\dfrac{\E{\I x\tau}}{\tau-\I\epsilon}\,,
	\end{equation}
	and taking a sufficiently large copy number $N$ and invoking the central limit theorem [Eq.~(4.20) of~\cite{Rehacek:2004minimal}],
	\begin{equation}
		\AVG{}{\E{\I\lambda\, \rvec{w}\bm{\cdot}\rvec{\nu}}}\cong\E{-\frac{A}{2}\,\lambda^2+\I B\,\lambda}\,,
		\label{eq:char_CLT}
	\end{equation}
	where $A=\rvec{w}\bm{\cdot}(\dyadic{P}-\rvec{p}\,\rvec{p})\bm{\cdot}\rvec{w}/N$, $B=\rvec{w}\bm{\cdot}\rvec{p}$ (remembering that $a< B< b$) and both $\dyadic{P}=\mathrm{diag}(\rvec{p})$ and $\rvec{p}\,\rvec{p}$ are matrices characterizing the multinomial distribution. From hereon, we will incessantly exploit the deceptively simple~result
	\begin{equation}
		\dfrac{1}{\I(a-\I\epsilon)}=\int^\infty_0\D u\,\E{-\I u(a-\I\epsilon)}\quad(\epsilon>0).
	\end{equation}
	
	We first settle the second and third terms in \eqref{eq:ymom}:
	\begin{align}
		\AVG{}{\Theta(a-x)a^n}=&\,a^n\lim_{\epsilon\rightarrow0^+}\int\dfrac{\D \tau}{2\pi}\,\dfrac{\E{\I a\tau}}{\tau-\I\epsilon}\,\AVG{}{\E{-\I \tau\,\rvec{w}\bm{\cdot}\rvec{\nu}}}\nonumber\\
		=&\,a^n\int^\infty_0\D u\int\dfrac{\D \tau}{2\pi}\,\E{-\frac{A}{2}\,\tau^2+\I\,(a-B-u)\,\tau}\,\nonumber\\
		=&\,a^n\int^\infty_0\dfrac{\D u}{\sqrt{2\pi A}}\,\E{-\frac{1}{2A}\left(u-a+B\right)^2}\nonumber\\
		=&\,\dfrac{a^n}{2}\,\erfc{\dfrac{B-a}{\sqrt{2A}}}\,,\\
		\AVG{}{\Theta(x-b)b^n}=&\,\dfrac{b^n}{2}\,\erfc{\dfrac{b-B}{\sqrt{2A}}}.
	\end{align}
	It, then, follows that
	\begin{align}
		\AVG{}{y^n}=&\,\dfrac{a^n+b^n}{2}-\dfrac{a^n}{2}\,\erf{\dfrac{B-a}{\sqrt{2A}}}-\dfrac{b^n}{2}\,\erf{\dfrac{b-B}{\sqrt{2A}}}\nonumber\\
		&+\lim_{\epsilon,\epsilon'\rightarrow0^+}\int\dfrac{\D \tau}{2\pi \I}\,\dfrac{\E{-\I a\tau}}{\tau-\I\epsilon}\int\dfrac{\D \tau'}{2\pi \I}\dfrac{\E{\I b\tau'}}{\tau'-\I\epsilon'}\left.\left(-\I\dfrac{\D}{\D\lambda}\right)^n\AVG{}{\E{\I\lambda\, \rvec{w}\bm{\cdot}\rvec{\nu}}}\right|_{\lambda=\tau-\tau'}.
	\end{align}
	
	Specifically when $n=1$ and $2$, for instance,
	\begin{align}
		-\I\dfrac{\D}{\D\lambda}\AVG{}{\E{\I\lambda\, \rvec{w}\bm{\cdot}\rvec{\nu}}}\cong&\,\left(B+\I\,A\,\lambda\right)\E{-\frac{A}{2}\,\lambda^2+\I B\,\lambda}\,,\nonumber\\
		\left(-\I\dfrac{\D}{\D\lambda}\right)^2\AVG{}{\E{\I\lambda\, \rvec{w}\bm{\cdot}\rvec{\nu}}}\cong&\,\left(A+B^2+2\,\I\,A\,B\,\lambda-A^2\,\lambda^2\right)\E{-\frac{A}{2}\,\lambda^2+\I B\,\lambda}\,.
	\end{align}
	Since the exponential terms are analytic functions of $\tau$ and $\tau'$, the answers to $\AVG{}{y}$ and $\AVG{}{y^2}$ may be subsequently obtained by handling the double integrals inasmuch as
	\begin{align}
		\AVG{}{y}=&\,\dfrac{a+b}{2}-\dfrac{a}{2}\,\erf{\dfrac{B-a}{\sqrt{2A}}}-\dfrac{b}{2}\,\erf{\dfrac{b-B}{\sqrt{2A}}}\nonumber\\
		&+B\lim_{\epsilon,\epsilon'\rightarrow0^+}\int\dfrac{\D \tau}{2\pi \I}\,\dfrac{\E{-\I a\tau}}{\tau-\I\epsilon}\int\dfrac{\D \tau'}{2\pi \I}\,\dfrac{\E{\I b\tau'}}{\tau'-\I\epsilon'}\,\E{-\frac{A}{2}\,(\tau-\tau')^2+\I B\,(\tau-\tau')}\nonumber\\
		&+\I\,A\lim_{\epsilon,\epsilon'\rightarrow0^+}\int\dfrac{\D \tau}{2\pi \I}\,\dfrac{\E{-\I a\tau}}{\tau-\I\epsilon}\int\dfrac{\D \tau'}{2\pi \I}\,\dfrac{\E{\I b\tau'}}{\tau'-\I\epsilon'}\,(\tau-\tau')\,\E{-\frac{A}{2}\,(\tau-\tau')^2+\I B\,(\tau-\tau')}\,.
	\end{align}
	The first double integral simplifies to
	\begin{align}
		&\,\int^\infty_0\!\D u\!\int^\infty_0\!\D u'\!\int\dfrac{\D \tau}{2\pi}\,\E{\I (B-a-u)\tau}\int\dfrac{\D \tau'}{2\pi}\,\E{-\frac{A}{2}\,(\tau-\tau')^2+\I (b-B-u')\,\tau'}\nonumber\\
		=&\,\int^\infty_0\!\dfrac{\D u}{\sqrt{2\pi A}}\int^\infty_0\!\D u'\,\delta(b-a-u-u')\,\E{-\frac{1}{2A}(u'+B-b)^2},
	\end{align}
	where we point out that the delta function yields a nonzero answer only when $u\leq b-a$, which gives rise to
	\begin{align}
		&\,\int^{b-a}_0\!\!\!\!\dfrac{\D u}{\sqrt{2\pi A}}\E{-\frac{(u+a-B)^2}{2A}}\!=\!\dfrac{1}{2}\!\left[\erf{\dfrac{B-a}{\sqrt{2A}}}\!+\erf{\dfrac{b-B}{\sqrt{2A}}}\!\right]\!.
	\end{align}
	Similarly, the second double integral turns out to be
	\begin{align}
		&\,\int^\infty_0\!\D u'\!\int\dfrac{\D\tau\,\D\tau'}{4\pi^2\I}\,\E{\I (B-a)\tau}\E{-\frac{A}{2}\,(\tau-\tau')^2}\E{\I (b-B-u')\,\tau'}\nonumber\\
		&-\int^\infty_0\!\D u\!\int\dfrac{\D \tau\,\D\tau'}{4\pi^2\I}\,\E{\I (B-a-u)\tau}\E{-\frac{A}{2}\,(\tau-\tau')^2}\E{\I (b-B)\,\tau'}\nonumber\\
		=&\,\int^\infty_0\D u'\int\dfrac{\D \tau}{2\pi\I}\,\E{-\I (u'+a-b)\tau}\,\dfrac{1}{\sqrt{2\pi A}}\,\E{-\frac{1}{2A}(u'+B-b)^2}\nonumber\\
		&\,-\int^\infty_0\D u\int\dfrac{\D \tau}{2\pi\I}\,\E{-\I (u+a-b)\tau}\,\dfrac{1}{\sqrt{2\pi A}}\,\E{-\frac{1}{2A}(b-B)^2}\nonumber\\
		=&\,\dfrac{1}{\I\sqrt{2\pi A}}\left[\E{-\frac{1}{2A}(B-a)^2}-\E{-\frac{1}{2A}(b-B)^2}\right]\,.
	\end{align}
	So,
	\begin{align}
		\!\AVG{}{y}=&\,\dfrac{a+b}{2}+\dfrac{B-a}{2}\,\erf{\dfrac{B-a}{\sqrt{2A}}}-\dfrac{b-B}{2}\,\erf{\dfrac{b-B}{\sqrt{2A}}}\nonumber\\
		&+\sqrt{\dfrac{A}{2\pi}}\left[\E{-\frac{1}{2A}(B-a)^2}-\E{-\frac{1}{2A}(b-B)^2}\right].
	\end{align}
	If $a\rightarrow-\infty$ and $b\rightarrow\infty$, then $\AVG{}{y}\rightarrow B$. Also, since $\erf{x/k}\rightarrow 2\,\Theta(x)-1$ when $k\rightarrow0$, as \mbox{$N\rightarrow\infty$}, $A\rightarrow0$ and $\AVG{}{y}\rightarrow a\,\Theta\!\left(a-B\right)+b\,\Theta\!\left(B-b\right)+B\,\Theta\!\left(b-B\right)\Theta\!\left(B-a\right)=B$ as it should be.
	
	To obtain $\AVG{}{y^2}$, we need to evaluate
	\begin{align}
		&\,\lim_{\epsilon,\epsilon'\rightarrow0^+}\int\dfrac{\D \tau}{2\pi \I}\,\dfrac{\E{-\I a\tau}}{\tau-\I\epsilon}\int\dfrac{\D \tau'}{2\pi \I}\,\dfrac{\E{\I b\tau'}}{\tau'-\I\epsilon'}\,(\tau-\tau')^2\,\E{-\frac{A}{2}\,(\tau-\tau')^2+\I B\,(\tau-\tau')}\nonumber\\
		=&\,\lim_{\epsilon'\rightarrow0^+}\int\dfrac{\D \tau'}{2\pi \I}\,\dfrac{\E{\I (b-B)\tau'}}{\tau'-\I\epsilon'}\!\!\int\dfrac{\D \tau\tau}{2\pi \I}\,\E{-\frac{A}{2}\,(\tau-\tau')^2+\I (B-a)\,\tau}\nonumber\\
		&+\lim_{\epsilon\rightarrow0^+}\!\!\int\dfrac{\D \tau}{2\pi \I}\dfrac{\E{\I (B-a)\tau}}{\tau-\I\epsilon}\!\!\int\dfrac{\D \tau'\tau'}{2\pi \I}\E{-\frac{A}{2}\,(\tau-\tau')^2+\I (b-B)\,\tau'}\nonumber\\
		&-2\int\dfrac{\D \tau}{2\pi \I}\,\E{-\I a\tau}\int\dfrac{\D \tau'}{2\pi \I}\,\E{\I b\tau'}\,\E{-\frac{A}{2}\,(\tau-\tau')^2+\I B\,(\tau-\tau')}\,,
	\end{align}
	where the first integral gives	
	\begin{align}
		&\,\dfrac{\E{-\frac{1}{2A}(B-a)^2}}{A^{3/2}\sqrt{2\pi}}\lim_{\epsilon'\rightarrow0^+}\int\dfrac{\D \tau'}{2\pi \I}\,\dfrac{\E{\I (b-a)\tau'}}{\tau'-\I\epsilon'}(B-a-\I A\tau')\nonumber\\
		=&\,\dfrac{\E{-\frac{1}{2A}(B-a)^2}}{A^{3/2}\sqrt{2\pi}}\left[\Theta(b-a)(B-a)-A\,\delta(b-a)\right]\nonumber\\
		=&\,\dfrac{\E{-\frac{1}{2A}(B-a)^2}}{A^{3/2}\sqrt{2\pi}}(B-a)\,,
	\end{align}
	the second integral is simply the first upon replacing $B-a$ with $b-B$,
	and, finally, the third~being
	\begin{align}
		&\,-2\int\dfrac{\D \tau}{2\pi \I}\,\E{-\I a\tau}\int\dfrac{\D \tau'}{2\pi \I}\,\E{\I b\tau'}\,\E{-\frac{A}{2}\,(\tau'-\tau)^2-\I B\,(\tau'-\tau)}\nonumber\\
		=&\,\sqrt{\dfrac{2}{\pi A}}\,\E{-\frac{1}{2A}(b-B)^2}\delta(b-a)=0\,.
	\end{align}
	After all is said and done,
	\begin{align}
		\AVG{}{y^2}=&\,\dfrac{a^2+b^2}{2}+\dfrac{A+B^2-a^2}{2}\,\erf{\dfrac{B-a}{\sqrt{2A}}}+\dfrac{A+B^2-b^2}{2}\,\erf{\dfrac{b-B}{\sqrt{2A}}}\nonumber\\
		&+\sqrt{\dfrac{A}{2\pi}}\Big[\E{-\frac{(B-a)^2}{2A}}\left(B+a\right)-\E{-\frac{(b-B)^2}{2A}}\left(B+b\right)\Big].
	\end{align}
	Again, when $a\rightarrow-\infty$ and $b\rightarrow\infty$, $\AVG{}{y^2}\rightarrow A+B^2$. For very large $N$, we have that $\AVG{}{y^2}\rightarrow a^2\,\Theta\!\left(a-B\right)+b^2\,\Theta\!\left(B-b\right)+(A+B^2)\,\Theta\!\left(b-B\right)\Theta\!\left(B-a\right)=A+B^2$.
	
	\section{Thermal-noise- and GDN-suppression incapability with VMZ}
	
	\label{app:thm-GDN-VMZ}
	
	We shall analyze the action of VMZ on coherent states and generalize it to arbitrary states under the P~function~formalism. Let us first apply the $M$-mode unitary transformation described by the unitary matrix $\mathcal{U}$ on the coherent state $\ket{\alpha}\bra{\alpha}$:
	\begin{align}
		\ket{\alpha}\bra{\alpha}\mapsto&\,\E{\alpha\sum^M_{j=1}\mathcal{U}_{1j}a^\dag}\vacket\E{-|\alpha|^2}\!\!\vacbra\E{\alpha^*\sum^{M}_{j=1}\mathcal{U}^*_{1j}a_j}\nonumber\\
		=&\,\ket{\alpha\,\mathcal{U}_{11},\ldots,\alpha\,\mathcal{U}_{1M}}\bra{\alpha\,\mathcal{U}_{11},\ldots,\alpha\,\mathcal{U}_{1M}}\,,
		\label{eq:coh2splitcoh}
	\end{align}
	where $\sum^M_{j=1}\mathcal{U}_{1j}\mathcal{U}_{jk}^\dag=\updelta_{1,k}$. After the thermal-noise map,
	\begin{align}
		&\,\ket{\alpha\,\mathcal{U}_{11},\ldots,\alpha\,\mathcal{U}_{1M}}\bra{\alpha\,\mathcal{U}_{11},\ldots,\alpha\,\mathcal{U}_{1M}}\nonumber\\
		\mapsto&\,\int\dfrac{(\D\rvec{\beta})}{\bar{n}^M\,\pi^M}\,\E{\sum^M_{j=1}(\alpha\,\mathcal{U}_{1j}\sqrt{1-\eta}+\beta_j\sqrt{\eta})\,a^\dag_j}\vacket\nonumber\\
		&\qquad\quad\times\E{-\frac{1}{\bar{n}}\sum^M_{j=1}|\beta_j|^2-\sum^M_{j=1}\left|\alpha^*\,\mathcal{U}^*_{1j}\sqrt{1-\eta}+\beta^*_j\sqrt{\eta}\right|^2}\nonumber\\
		&\qquad\qquad\qquad\times\vacbra\E{\sum^M_{j=1}(\alpha^*\,\mathcal{U}^*_{1j}\sqrt{1-\eta}+\beta^*_j\sqrt{\eta})\,a^\dag_j}\,.
	\end{align}
	
	The final step is to apply the inverse unitary transformation followed by ($M-1$)-mode vacuum measurements on the resulting state. The former leads to the operator~exponent
	\begin{align}
		&\,\sum^M_{j=1}(\alpha\,\mathcal{U}_{1j}\sqrt{1-\eta}+\beta_j\sqrt{\eta})\,\sum^M_{k=1}\mathcal{U}^\dag_{jk}a^\dag_k\nonumber\\
		=&\,\!\left(\alpha\sqrt{1-\eta}+\sqrt{\eta}\sum^M_{j=1}\beta_j\,\mathcal{U}^\dag_{j1}\right)a^\dag_1+\sqrt{\eta}\sum^M_{j=1}\sum^M_{k=2}\beta_j\,\mathcal{U}^\dag_{jk}a^\dag_k\,.
	\end{align}
	After the vacuum measurements, we arrive at the output~state
	\begin{align}
		&\,(\ket{\alpha}\bra{\alpha})_\mathrm{supp}		\propto\int\dfrac{(\D\rvec{\beta})}{\bar{n}^M\,\pi^M}\ket{\alpha\sqrt{1-\eta}+\sqrt{\eta}\sum^M_{j=1}\beta_j\,\mathcal{U}^\dag_{j1}}\nonumber\\
		&\,\times\E{-\frac{1}{\bar{n}}\sum^M_{j=1}|\beta_j|^2}\,f(\alpha,\rvec{\beta};\eta,\mathcal{U})\bra{\alpha\sqrt{1-\eta}+\sqrt{\eta}\sum^M_{j=1}\beta_j\,\mathcal{U}^\dag_{j1}}\,,
	\end{align}
	where in terms of the computational basis,
	\begin{align}
		\log f(\alpha,\rvec{\beta};\eta,\mathcal{U})=&\,-\sum^M_{j=1}\left|\alpha\,\mathcal{U}_{1j}\sqrt{1-\eta}+\beta_j\sqrt{\eta}\right|^2+\left|\alpha\sqrt{1-\eta}+\sqrt{\eta}\sum^M_{j=1}\beta_j\,\mathcal{U}^\dag_{j1}\right|^2\nonumber\\
		=&\,-\eta\sum^M_{j=1}|\beta_j|^2+\eta\sum^M_{j=1}\sum^M_{k=1}\beta_j\beta^*_k\,\mathcal{U}^\dag_{j1}\,\mathcal{U}_{1k}\nonumber\\
		=&\,-\eta\,\rvec{\beta}^\dag\left(\dyadic{1}-\mathcal{U}^\top\rvec{e}\,\rvec{e}^\top\mathcal{U}^*\right)\rvec{\beta}
		\label{eq:vmzi_thm_loss_misc}
	\end{align}
	and $\rvec{e}=(1\,\,0\,\,\ldots\,\,0)^\top$.
	
	To proceed, one can perform the variable change $\rvec{\beta}'\equiv\mathcal{U}^*\rvec{\beta}$ of unit Jacobian, so that we may rewrite
	\begin{align}
		&\,(\ket{\alpha}\bra{\alpha})_\mathrm{supp}\propto\int\dfrac{(\D\rvec{\beta})}{\bar{n}^M\,\pi^M}\ket{\alpha\sqrt{1-\eta}+\sqrt{\eta}\,\beta'_1}\nonumber\\
		&\times\E{-\frac{1}{\bar{n}}\sum^M_{j=1}|\beta'_j|^2-\eta\,\rvec{\beta}'^\dag(\dyadic{1}-\rvec{e}\,\rvec{e}^\top)\rvec{\beta}'}\bra{\alpha\sqrt{1-\eta}+\sqrt{\eta}\,\beta'_1}\,,
	\end{align}
	where we now see that 
	\begin{align}
		\eta\,\rvec{\beta}'^\dag(\dyadic{1}-\rvec{e}\,\rvec{e}^\top)\rvec{\beta}'=\eta\sum^{M}_{j=2}|\beta'|^2.
	\end{align}
	This observation helps take care of $M-1$ integrals, which themselves give the vacuum-measurement success rate
	\begin{equation}
		p_{\mathrm{succ}}=\left(\int\dfrac{(\D\beta')}{\bar{n}\,\pi}\E{-\left(\frac{1}{\bar{n}}+\eta\right)|\beta'|^2}\right)^{M-1}=\left(\dfrac{1}{1+\bar{n}\,\eta}\right)^{M-1}\!\!,
	\end{equation}
	and produces the suppressed state as
	\begin{align}
		(\ket{\alpha}\bra{\alpha})_\mathrm{supp}=&\,\int\dfrac{(\D\beta'_1)}{\bar{n}\,\pi}\ket{\alpha\sqrt{1-\eta}+\sqrt{\eta}\,\beta'_1}\E{-\frac{1}{\bar{n}}|\beta'_1|^2}\bra{\alpha\sqrt{1-\eta}+\sqrt{\eta}\,\beta'_1}\,.
	\end{align}
	Therefore, the $M$-mode VMZ protocol leaves the thermally-noisy state intact. As this fails for the coherent state, we conclude likewise for \emph{any} state~$\rho$.
	
	A similar calculation with RDN yields the suppressed coherent state of a (expectedly) very similar~form:
	\begin{align}
		(\ket{\alpha}\bra{\alpha})_\mathrm{supp}		\propto&\,\int(\D\rvec{\beta})\,p(\beta_1,\beta^*_1)\ldots p(\beta_M,\beta^*_M)\,\ket{\alpha+\sum^M_{j=1}\beta_j\,\mathcal{U}^\dag_{j1}}\nonumber\\
		&\,\qquad\times\E{-\rvec{\beta}^\dag\left(\dyadic{1}-\mathcal{U}^\top\rvec{e}\,\rvec{e}^\top\mathcal{U}^*\right)\rvec{\beta}}\bra{\alpha+\sum^M_{j=1}\beta_j\,\mathcal{U}^\dag_{j1}}\,.
	\end{align}
	Hence, the same variable change would result in
	\begin{align}
		(\ket{\alpha}\bra{\alpha})_\mathrm{supp}	\propto&\,\int(\D\rvec{\beta}')\,p((\mathcal{U}^\top\rvec{\beta}')_1,(\mathcal{U}^\top\rvec{\beta}')^*_1)\ldots p((\mathcal{U}^\top\rvec{\beta}')_M,(\mathcal{U}^\top\rvec{\beta}')^*_M)\nonumber\\
		&\,\qquad\quad\times\ket{\alpha+\beta'_1}\E{-\sum^{M}_{j=2}|\beta'_j|^2}\bra{\alpha+\beta'_1}\,.
	\end{align}
	If $p(\beta,\beta^*)=\E{-|\beta|^2/\sigma^2}/(\pi\sigma^2)$, that is for a GDN channel, we realize that the output after the VMZ protocol is
	\begin{equation}
		(\ket{\alpha}\bra{\alpha})_\mathrm{supp}=\int(\D\beta'_1)\ket{\alpha+\beta'_1}\frac{\E{-\frac{|\beta'_1|^2}{\sigma^2}}}{\pi\sigma^2}\bra{\alpha+\beta'_1},
	\end{equation}
	which is just the original GDN-corrupted~state.
	
	\section{Mutual compatibility between PSG-PEC mitigation and VMZ-suppression schemes}
	\label{app:PSG_VMZ_exclu}
	
	The purpose is to show the equivalence between Figs.~\ref{fig:VMZ_PSG_commute}(a) and~(b), that is, the order in which an input state is treated with the PSG error-mitigation and $M$-mode VMZ error-suppression protocols is irrelevant. It is enough to look at the combined operations before the noise channel. Those after behave exactly the same way by symmetry. We adopt the usual trick of inspecting the action of these operations on the coherent state and generalize the result to all quantum states.
	
	In Fig.~\ref{fig:VMZ_PSG_commute}(a), we swiftly find that the $M$-mode state right after $U_M$, the evolution operator of~$\mathcal{U}$,
	\begin{align}
		\rho_{\text{(a)}}=&\,U_M\ket{g\,\alpha}\bra{g\,\alpha}\otimes(\vacket\vacbra)^{\otimes M-1}\,U_M^\dag\nonumber\\
		=&\,\E{g\,\alpha\sum^M_{j=1} \mathcal{U}_{1j}a_j^\dag}\vacket\E{-|g\,\alpha|^2}\vacbra\E{g\,\alpha^*\sum^M_{j'=1}\mathcal{U}^*_{1j'}a_{j'}}\nonumber\\
		=&\,\ket{g\,\alpha\,\mathcal{U}_{11},\ldots,g\,\alpha\,\mathcal{U}_{1M}}\bra{g\,\alpha\,\mathcal{U}_{11},\ldots,g\,\alpha\,\mathcal{U}_{1M}}\,.
	\end{align}	
	For the circuit in Fig.~\ref{fig:VMZ_PSG_commute}(b), because $g^{a^\dag_1 a_1+\ldots+a^\dag_M a_M}$ commutes with any BS, the $M$-mode state right after the PSGs~is
	\begin{align}
		\rho_{\text{(b)}}=&\sum^\infty_{k_1,\ldots,k_M=0}\!\!\!\!\!\!\!\dfrac{(-1+g^{-2})^{k_1+\ldots+k_M}}{k_1!\ldots k_M!}\prod^M_{j=1}a_j^{k_j}\,W\prod^M_{j'=1}\!\!a_{j'}^{\dag\,k_{j'}}\!,\nonumber\\
		W=&\,\,U_M\!\ket{g\,\alpha}\E{(g^2-1)|\alpha|^2}\!\!\bra{g\,\alpha}\otimes(\vacket\!\vacbra)^{\otimes M-1}U_M^\dag\nonumber\\
		=&\,\, U_M\,\E{g\,\alpha\,a_1^\dag}\vacket\E{-|\alpha|^2}\vacbra\E{g\,\alpha^*\,a_1}\,U_M^\dag\nonumber\\
		=&\,\,\E{g\,\alpha\sum^M_{j=1}\mathcal{U}_{1j}\,a_j^\dag}\vacket\E{-|\alpha|^2}\vacbra\E{g\,\alpha^*\sum^M_{j'=1}\mathcal{U}_{1j'}\,a_{j'}},
	\end{align}
	such that
	\begin{align}
		&\,\prod^M_{j=1}a_j^{k_j}\,W\prod^M_{j'=1}a_{j'}^{\dag\,k_{j'}}=\ket{g\,\alpha\,\mathcal{U}_{11},\ldots,g\,\alpha\,\mathcal{U}_{1M}}\E{(g^2-1)|\alpha|^2}\nonumber\\
		&\quad\times|g\,\alpha\,\mathcal{U}_{11}|^{2k_1}\ldots|g\,\alpha\,\mathcal{U}_{1M}|^{2k_M}\bra{g\,\alpha\,\mathcal{U}_{11},\ldots,g\,\alpha\,\mathcal{U}_{1M}}\,.
	\end{align}
	As the multiple sums simplify to
	\begin{align}
		&\,\prod^M_{j=1}\sum^\infty_{k_j=0}\dfrac{(-1+g^{-2})^{k_j}}{k_j!}\,|g\,\alpha\,\mathcal{U}_{1j}|^{2k_j}=\prod^M_{j=1}\E{|g\,\alpha|^2|\mathcal{U}_{1j}|^2(-1+g^{-2})}=\E{-|\alpha|^2(g^2-1)}\,,
	\end{align}
	we find that $\rho_{\text{(a)}}=\rho_{\text{(b)}}$. 
	
	\section{Miscellaneous}
	
	\subsection{Squeezed displaced-Fock representation for the squeezed-``cat'' code in Sec.~\ref{subsec:opt_psgpec_est}}
	\label{app:dispfock}
	
	Since displaced Fock states span the entire phase space (for example, in the manner of Wigner operators~\cite{Englert:1989operator}), we can write any operator as their linear combination inasmuch as~Eq.~\eqref{eq:dispfock_O}. As the bosonic code of interest here is the squeezed-``cat'' code, we can apply the same amount of squeezing to these displaced Fock projectors. Given that $O=\ket{\text{enc. in}}\bra{\text{enc. in}}$, the objective is to find~$\{\beta_k\}$ and~$\{c_{kn}\}$ by
	\begin{algorithm}[H]
		\caption{\label{alg:dispfock}Optimizing squeezed displaced-Fock meas.}
		\begin{algorithmic}[1]
			\State Specify the dimension~$d$ of the truncated Hilbert~space, number of squeezed displaced-Fock bases~$K$ and a small~$\epsilon>0$.
			\State Set termination flag $\texttt{flag}$ to~0.
			\State Define $\mathcal{F_{\rvec{\beta},\rvec{c}}}=\sum^K_{k=1}\sum^{d-1}_{n=0}c_{kn}\left|\bra{\text{enc. in}}S(z)D(\beta_k)\ket{n}\right|^2$.
			\State Start with a random~$\rvec{\beta}$ and $c_{kn}=1$.
			\While{$\texttt{flag}=0$}
			\State Maximize $\mathcal{F_{\rvec{\beta},\rvec{c}}}$ over~$\rvec{\beta}$ whilst fixing~$\dyadic{c}$.
			\State Set $\rvec{\beta}$ as the new optimized~column.
			\State\label{alg:sdp_start} Maximize $\mathcal{F_{\rvec{\beta},\rvec{c}}}$ over~$\dyadic{c}$ whilst fixing~$\rvec{\beta}$, 
			subject to 
			\State $\sum^K_{k=1}\sum^{d-1}_{n=0}c_{kn}=1$ and 
			\State\label{alg:sdp_end} $\sum^K_{k=1}\sum^{d-1}_{n=0}S(z)D(\beta_k)\ket{n}c_{kn}\bra{n}D(\beta_k)^\dag S(z)^\dag\geq0$.
			\State Set $\dyadic{c}$ as the new optimized~matrix.
			\If{$\mathcal{F_{\rvec{\beta},\rvec{c}}}>1-\epsilon$}
			\State Set $\texttt{flag}=1$.
			\EndIf
			\EndWhile
		\end{algorithmic} 
	\end{algorithm}
	\noindent
	Lines~\ref{alg:sdp_start} to~\ref{alg:sdp_end} form a semidefinite program that can be efficiently executed using packages such as \texttt{CVX}~\cite{Boyd:2009dv,cvx,gb08}. Algorithm~\ref{alg:dispfock} generalizes a procedure in~\cite{Izumi:2018projective}.	For \mbox{$z=3\text{dB}$} and $\alpha=1$, Alg.~\ref{alg:dispfock} yields the set of $K=8$ amplitudes $\{-0.1721 + 0.3930\I,0.4398 + 0.1135\I,0.3325 - 0.0789\I,0.2340 + 0.3229\I,0.0777 - 0.2548\I,0.2100 + 0.0366\I,0.4174 + 0.2144\I,0.1963 + 0.1448\I\}$ and the $c_{kn}$s that gave an optimized $\mathcal{F}_{\rvec{\beta},\dyadic{c}}>0.99$ are given in Tab.~\ref{tab:dispfock}. The 64~squeezed displaced-Fock projectors that specify $\ket{\text{enc. in}}\bra{\text{enc. in}}$ consist of 56~pure projectors
	$S(z)D(\beta_k)\ket{n}\bra{n}D(\beta_k)^\dag S(z)^\dag$ for $0\leq n\leq 6$, $1\leq k\leq8$ and 8~mixed projectors $1-\sum^6_{n=0}S(z)D(\beta_k)\ket{n}\bra{n}D(\beta_k)^\dag S(z)^\dag$ for~$1\leq k\leq8$.
	
	\begin{table}[t]
		\caption{\label{tab:dispfock}Values of $c_{kn}$ \emph{rounded off} to two decimal~places, with the vanishingly-small values for $n>6$~neglected. Actual values all sum to~unity as they~should.}
		\vspace{2ex}
		\centering
		\begin{tabular}{c|r|r|r|r|r|r|r}
			\hline\hline
			$c_{kn}$ & $n=0$ & $n=1$ & $n=2$ & $n=3$ & $n=4$ & $n=5$ & $n=6$\\
			\hline
			$k=1$ & $-1.30$ & $-0.44$ & $0.05$ & $-0.00$ & $-0.00$ & $0.00$ & $0.00$\\
			$k=2$ & $6.93$ & $1.43$ & $-0.01$ & $-0.00$ & $-0.00$ & $-0.00$ & $0.00$\\
			$k=3$ & $-2.87$ & $0.59$ & $0.18$ & $-0.05$ & $0.01$ & $0.00$ & $0.00$\\
			$k=4$ & $-0.25$ & $0.22$ & $0.06$ & $-0.02$ & $0.00$ &  $0.00$ & $0.00$\\
			$k=5$ & $-1.19$ & $-0.01$ & $0.09$ & $-0.02$ & $0.00$ & $-0.00$ & $0.00$\\
			$k=6$ & $2.43$ & $-1.51$ & $-0.68$ & $0.09$ & $-0.01$ & $-0.01$ & $-0.00$\\
			$k=7$ & $-5.57$ & $-0.29$ & $0.04$ & $-0.01$ & $0.00$ & $-0.00$ & $0.00$\\
			$k=8$ & $1.36$ & $1.74$ & $-0.02$ & $0.03$ & $-0.00$ &   $0.00$ & $-0.00$\\
			\hline\hline
		\end{tabular}		
	\end{table}
	
	\subsection{Short proof of inequality~\eqref{eq:Markov_Chebyshev}}
	\label{app:chebyshev}
	
	Since $s_l=\lambda_\gamma\delta_{l,1}$ and $\overline{\E{\I(\phi_j-\phi_{j'})}}=|\lambda_\gamma|^2$ if $j\neq j'$, 
	\begin{align}
		\overline{|s_l-\overline{s_l}|^2}=\overline{|s_l|^2}-|\lambda_\gamma|^2\delta_{l,1}
		=&\,\sum^M_{j=1}|\mathcal{U}_{j1}|^2|\mathcal{U}_{jl}|^2+|\lambda_\gamma|^2\underbrace{\sum_{j\neq j'}\,\mathcal{U}^\dag_{lj}\,\mathcal{U}_{j1}\,\mathcal{U}^\dag_{1j'}\,\mathcal{U}_{j'l}}_{\mathclap{\displaystyle{=\delta_{l,1}-\sum^M_{j=1}|\mathcal{U}_{j1}|^2|\mathcal{U}_{jl}|^2}}}-|\lambda_\gamma|^2\delta_{l,1}\nonumber\\
		=&\,\left(1-|\lambda_\gamma|^2\right)\sum^M_{j=1}|\mathcal{U}_{j1}|^2|\mathcal{U}_{jl}|^2\,,
	\end{align}
	and all else follows from the basic Chebyshev~inequality. 
	
	\subsection{Averaging over unitary two-designs}
	\label{app:HaarU}
	
	If $\mathcal{U}$ is an $M$-dimensional unitary two-design~(TD), which encompasses the Haar-random unitary operators, given $M$-dimensional~$\mathcal{A}$, $\mathcal{B}$, $\mathcal{C}$ and $\mathcal{Y}$, and $M^2$-dimensional~$\mathcal{Z}$, the following identities are well-known~\cite{Collins:2006integration,Puchala_Z._Symbolic_2017,Holmes:2022connecting,Mele:2024introduction,Teo:2023virtual}:
	\begin{align}
		\overline{\mathcal{U}\,\mathcal{Y}\, \mathcal{U}^\dag}\strut^{\,\textsc{td}}=&\,\dfrac{\tr{\mathcal{Y}}}{M}\,,\label{eq:useful_2design_1}\\
		\overline{\mathcal{U}^{\otimes2}\,\mathcal{Z}\, \mathcal{U}^{\dag\,\otimes2}}\strut^{\,\textsc{td}}=&\,\left[\dfrac{\tr{\mathcal{Z}}}{M^2-1}-\dfrac{\tr{\mathcal{Z}\tau}}{M(M^2-1)}\right]1\nonumber\\
		&\,+\left[\dfrac{\tr{\mathcal{Z}\tau}}{M^2-1}-\dfrac{\tr{\mathcal{Z}}}{M(M^2-1)}\right]\tau\,,\label{eq:useful_2design_2}\\
		\overline{\mathcal{U}\,\mathcal{A}\,\mathcal{U}^\dag \mathcal{B}\,\mathcal{U}\,\mathcal{C}\mathcal{U}^\dag}\strut^{\,\textsc{td}}=&\,\dfrac{\tr{\mathcal{A}}\tr{\mathcal{C}}\mathcal{B}+\tr{\mathcal{B}}\tr{\mathcal{A}\,\mathcal{C}}1}{M^2-1}\nonumber\\
		-&\,\dfrac{\tr{\mathcal{A}}\tr{\mathcal{B}}\tr{\mathcal{C}}1+\tr{\mathcal{A}\,\mathcal{C}}\mathcal{B}}{M(M^2-1)}\,,\label{eq:useful_2design_3}
	\end{align}
	where $\tau$ is the swap~operator. Using~\eqref{eq:useful_2design_3} and the substitutions $\mathcal{A}=\ket{k}\bra{k}$, $\mathcal{B}=\ket{j}\bra{j}$ and $\mathcal{C}=\ket{l}\bra{l}$, the two-design average
	\begin{align}
		&\,\sum^M_{j=1}\overline{\mathcal{U}\ket{k}\bra{k}\mathcal{U}^\dag\ket{j}\bra{j}\mathcal{U}\ket{l}\bra{l}\mathcal{U}^\dag}\strut^{\,\textsc{td}}\nonumber\\
		=&\,\frac{1}{M^2-1}\left(\ket{j}\bra{j}+\delta_{k,l}1\right)-\frac{1}{M(M^2-1)}\left(1+\ket{j}\delta_{k,l}\bra{j}\right)
	\end{align}
	straightforwardly yields
	\begin{equation}
		\sum^M_{j=1}|\mathcal{U}_{jk}|^2|\mathcal{U}_{jl}|^2=\dfrac{1+\delta_{k,l}}{M+1}\,.
	\end{equation}
	
	\subsection{Approximating $S_{mn}$ for central-Gaussian dephasing channels}
	\label{app:smsn}
	
	If $p_j=|\mathcal{U}_{j1}|^2$, where clearly $\sum^M_{j=1}p_j=1$, then $s_1=\sum^M_{j=1}p_j\E{\I\phi_j}$ and the average over the central-Gaussian distribution \emph{exactly}~gives
	\begin{align}
		S_{mn}=&\,\overline{s_1^ms_1^{*n}}\nonumber\\
		=&\,\sum_{\rvec{k},\rvec{k}'}\dfrac{m!}{k_1!\ldots k_M!}\dfrac{n!}{k'_1!\ldots k'_M!}\,p^{k_1+k'_1}_1\ldots p^{k_M+k'_M}_M\,\E{-\frac{\gamma}{2}(k_1-k'_1)^2-\ldots-\frac{\gamma}{2}(k_M-k'_M)^2},
		\label{eq:smsn}
	\end{align}
	where it is understood that the summations respect the constraints $\sum^M_{j=1}k_j=m$ and $\sum^M_{j=1}k'_j=n$. This is followed by a Taylor expansion of the exponential function $\E{-\frac{\gamma}{2}(k_j-k'_j)^2}\cong1-\frac{\gamma}{2}(k_j-k'_j)^2$ up to first order in~$\gamma$. Since the right-hand side of \eqref{eq:smsn} is just averages over two multinomial distributions that are both defined by~$p_j$, up to~$\mathcal{O}(\gamma)$,
	\begin{align}
		S_{mn}\cong&\,1-\frac{\gamma}{2}\sum^M_{j=1}\AVG{}{(k_j-k'_j)^2}\nonumber\\
		=&\,1-\frac{\gamma}{2}\sum^M_{j=1}\Big[(m+n)(p_j-p_j^2)+(mp_j)^2+(np_j)^2-2mnp_j^2\Big]\nonumber\\
		=&\,1-\frac{\gamma}{2}[(m+n)\,w+(m-n)^2(1-w)]\,.
	\end{align}
	
	For \emph{any} $\gamma$, if $m,n\gg1$, then one can apply the central limit theorem to multinomial distributions~\cite{Severini:2005aa}, which is summarized in \ref{app:multinomial} for completeness, such that 
	\begin{align}
		\dfrac{m!}{k_1!\ldots k_M!}\,\delta_{k_1+\ldots+k_M,m}\,p^{k_1}_1\,p^{k_2}_2\,\ldots\, p^{k_M}_M
		\cong&\,\dfrac{\E{-\frac{1}{2}(\widetilde{\rvec{x}}-m\,\widetilde{\rvec{p}})\bm{\cdot}\dyadic{\Sigma}_m^{-1}\bm{\cdot}(\widetilde{\rvec{x}}-m\,\widetilde{\rvec{p}})}}{\sqrt{(2\pi)^{M-1}\det\{\dyadic{\Sigma}_m\}}}\,,
		\label{eq:CLT_multinom}
	\end{align}
	where $\widetilde{\rvec{x}}=(k_1\,\,k_2\,\,\ldots\,\,k_{M-1})^\top$,
	\begin{equation}
		\dyadic{\Sigma}_m=m\!\left(\widetilde{\dyadic{P}}-\widetilde{\rvec{p}}\,\widetilde{\rvec{p}}^\top\right),\,\,\det\{\dyadic{\Sigma}_m\}=m^{M-1}p_1\,p_2\,\ldots\,p_M
	\end{equation}
	and the ($M-1$)-dimensional matrices are defined as
	\begin{equation}
		\widetilde{\dyadic{P}}\,=\begin{pmatrix}
			p_1 & 0 & \cdots & 0\\
			0 & p_2 & \cdots & 0\\
			\vdots & \vdots & \ddots & \vdots\\
			0 & 0 & \cdots & p_{M-1}
		\end{pmatrix},\,\,\widetilde{\rvec{p}}\,=\begin{pmatrix}
			p_1\\
			p_2\\
			\vdots\\
			p_{M-1}
		\end{pmatrix},\,\,\onevec\,=\begin{pmatrix}
			1\\
			1\\
			\vdots\\
			1
		\end{pmatrix}.
		\label{eq:P_and_p}
	\end{equation}
	
	Using this familiar approximation,
	\begin{align}
		\AVG{}{\E{-\frac{\gamma}{2}\sum_j(k_j-k'_j)^2}}
		\cong&\,\int(\D\,\widetilde{\rvec{x}})\,\dfrac{\E{-\frac{1}{2}(\widetilde{\rvec{x}}-m\,\widetilde{\rvec{p}})\bm{\cdot}\dyadic{\Sigma}_m^{-1}\bm{\cdot}(\widetilde{\rvec{x}}-m\,\widetilde{\rvec{p}})}}{\sqrt{(2\pi)^{M-1}\det\{\dyadic{\Sigma}_m\}}}\nonumber\\
		&\times\int(\D\,\widetilde{\rvec{x}}')\,\dfrac{\E{-\frac{1}{2}(\widetilde{\rvec{x}}'-n\,\widetilde{\rvec{p}})\bm{\cdot}\dyadic{\Sigma}_n^{-1}\bm{\cdot}(\widetilde{\rvec{x}}'-n\,\widetilde{\rvec{p}})}}{\sqrt{(2\pi)^{M-1}\det\{\dyadic{\Sigma}_n\}}}\,\E{-\frac{\gamma}{2}(\widetilde{\rvec{x}}'-\widetilde{\rvec{x}})^2}.
	\end{align}
	Using the standard result
	\begin{equation}
		\int(\D\rvec{x})\,\E{-\rvec{x}\bm{\cdot}\dyadic{A}\bm{\cdot}\rvec{x}+\rvec{b}\bm{\cdot}\rvec{x}}=\sqrt{\dfrac{\pi^{D}}{\det\{\dyadic{A}\}}}\,\E{\frac{1}{4}\,\rvec{b}\bm{\cdot}\dyadic{A}^{-1}\bm{\cdot}\rvec{b}}
		\label{eq:real_gauss}
	\end{equation}
	for real $D$-dimensional $\dyadic{A}>0$ and $\rvec{b}$, with tenacity, one~gets
	\begin{equation}
		S_{mn}=\dfrac{\E{-\frac{\gamma}{2}(m-n)^2\sum^M_{j=1}p^2_j}}{2^{M-1}\sqrt{\det\{\dyadic{\Sigma}_m\}\det\{\dyadic{\Sigma}_n\}\det\{\dyadic{C}\}}}\,\E{\frac{1}{4}\rvec{b}\bm{\cdot}\dyadic{C}^{-1}\bm{\cdot}\rvec{b}}
		\label{eq:gen_Smn}
	\end{equation}
	that holds for \emph{any} $\widetilde{\rvec{p}}$, where $\dyadic{B}=(\gamma/2)(\dyadic{1}+\onevec\onevec^\top)$,
	\begin{align}
		\dyadic{C}=&\begin{pmatrix}
			\frac{1}{2}\dyadic{\Sigma}_m^{-1}+\dyadic{B} & -\dyadic{B}\\
			-\dyadic{B} &\!\!\! \frac{1}{2}\dyadic{\Sigma}_n^{-1}+\dyadic{B}
		\end{pmatrix}\!,\,\rvec{b}=\gamma(m-n)\!\begin{pmatrix}
			p_M\onevec-\widetilde{\rvec{p}}\\
			\widetilde{\rvec{p}}-p_M\onevec
		\end{pmatrix}\!.
	\end{align}
	Some more acrobatics turn the determinant triple product~into
	\begin{align}
		\det\{\dyadic{\Sigma}_m\}\det\{\dyadic{\Sigma}_n\}\det\{\dyadic{C}\}
		=&\,\det\!\left\{\begin{pmatrix}
			\frac{1}{2}\dyadic{1}+\dyadic{\Sigma}_m\dyadic{B} & -\dyadic{\Sigma}_m\dyadic{B}\\
			-\dyadic{\Sigma}_n\dyadic{B} & \frac{1}{2}\dyadic{1}+\dyadic{\Sigma}_n\dyadic{B}
		\end{pmatrix}\right\}\nonumber\\
		=&\,\det\!\left\{\left(\frac{1}{2}\dyadic{1}+\dyadic{\Sigma}_m\dyadic{B}\right)\left(\frac{1}{2}\dyadic{1}+\dyadic{\Sigma}_n\dyadic{B}\right)-\dyadic{\Sigma}_m\dyadic{B}\,\dyadic{\Sigma}_n\dyadic{B}\right\}\nonumber\\
		=&\,\det\!\left\{\frac{1}{4}\dyadic{1}+\frac{1}{2}\left(\dyadic{\Sigma}_m+\dyadic{\Sigma}_n\right)\dyadic{B}\right\}\,,
	\end{align} 
	where we note that $\dyadic{\Sigma}_m\dyadic{B}$ commutes with $\dyadic{\Sigma}_n\dyadic{B}$ and that for 
	\begin{equation}
		\dyadic{\Gamma}=\begin{pmatrix}
			\dyadic{A} & \dyadic{B}\\
			\dyadic{C} & \dyadic{D}
		\end{pmatrix}\text{with an invertible $\dyadic{A}$},
	\end{equation}
	the identity $\det\{\dyadic{\Gamma}\}=\det\{\dyadic{A}\}\det\{\dyadic{D}-\dyadic{C}\dyadic{A}^{-1}\dyadic{B}\}$ holds.
	
	For a Hadamard~$\mathcal{U}$,
	\begin{align}
		&\dyadic{\Sigma}_m=\dfrac{m}{M}\left(\dyadic{1}-\frac{1}{M}\onevec\onevec^\top\right),\,\,\rvec{b}=\zerovec\nonumber\\
		&\dyadic{\Sigma}_m\dyadic{B}=\dfrac{\gamma\,m}{2M}\left(\dyadic{1}-\frac{1}{M}\onevec\onevec^\top\right)\left(\dyadic{1}+\onevec\onevec^\top\right)=\dfrac{\gamma\,m}{2M}\dyadic{1}\,,\nonumber\\
		&\det\!\left\{\frac{1}{4}\dyadic{1}+\frac{1}{2}\left(\dyadic{\Sigma}_m+\dyadic{\Sigma}_n\right)\dyadic{B}\right\}=\left[\frac{1+\frac{\gamma}{M}(m+n)}{4}\right]^{M-1},
	\end{align}
	leading to~\eqref{eq:S_Hadmn}. Based on numerical evidence, for finite~$M$, Eq.~\eqref{eq:S_Hadmn} is also rather accurate even for small~$m$ and~$n$.
	
	\subsection{Approximating a multinomial distribution}
	\label{app:multinomial}
	
	The $M$-mode multinomial distribution characterized by $n>0$, $\widetilde{\rvec{p}}=(p_1\,\,p_2\,\,\ldots\,\,p_{M-1})^\top$ and $p_M=1-p_1-p_2-\ldots-p_{M-1}$,
	\begin{equation}
		\PR(n,\widetilde{\rvec{p}})=\dfrac{n!}{k_1!\,k_2!\,\ldots\,k_M!}\,\delta_{k_1+\ldots+k_M,n}\,p_1^{k_1}p_2^{k_2}\ldots p_M^{k_M}\,,
	\end{equation}
	can be very well approximated using the Stirling formula $k!\cong k^k\E{-k}\sqrt{2\pi k}$, which works extremely well even for a very small~$k$. Together with the definitions $\nu_j=k_j/n$, $y=1-\sum^{M-1}_{j=1}\nu_j$ and the crucial recognition that there are \emph{only $M-1$ independent~$p_j$s and~$\nu_j$s},
	
	\begin{align}
		\log \PR(n,\widetilde{\rvec{p}})
		\cong&\,-\log\sqrt{(2\pi n)^{M-1}}-\sum^{M-1}_{j=1}\left(n\,\nu_j\log \nu_j+\dfrac{1}{2}\log \nu_j\right)\nonumber\\
		&+n\sum^{M-1}_{j=1}\nu_j\log p_j+n\,y\log p_M-\left(n\,y\log y+\dfrac{1}{2}\log y\right).
		\label{eq:logp}
	\end{align}
	
	A standard approach is to consider the situation in which $\nu_j\cong p_j$ for a sufficiently large~$n$. One may then expand the constituents of $\log \PR(n,\widetilde{\rvec{p}})$ to second order in $\nu_j-p_j$ for \mbox{$1\leq j\leq M-1$}, for which we acquire the first-order~derivatives 
	\begin{align}
		\partial_{\nu_k}\!\left(\sum^{M-1}_{j=1}\log\nu_j+\log y\right)=&\,\dfrac{1}{\nu_k}-\dfrac{1}{y}\,,\nonumber\\
		\partial_{\nu_k}\!\left(\sum^{M-1}_{j=1}\nu_j\log\nu_j+y\log y\right)=&\,\log\nu_k-\log y\,,
	\end{align}
	and also the second-order derivatives (Hessian elements)
	\begin{align}
		\partial_{\nu_l}\partial_{\nu_k}\!\left[\sum^{M-1}_{j=1}\log\nu_j+\log y\right]=&-\dfrac{\updelta_{k,l}}{\nu_k^2}-\dfrac{1}{y^2}\,,\nonumber\\
		\partial_{\nu_l}\partial_{\nu_k}\!\left[\sum^{M-1}_{j=1}\nu_j\log\nu_j+y\log y\right]=&\,\dfrac{\updelta_{k,l}}{\nu_k}+\dfrac{1}{y}\,.
	\end{align}
	With these, after keeping only terms of $\mathcal{O}(n)$ and $-(1/2)\sum^M_{j=1}\log p_j$ for consistent $\PR(n,\widetilde{\rvec{p}})$ normalization,
	\begin{align}
		&-\sum^{M-1}_{j=1}\left(n\nu_j\log \nu_j+\dfrac{1}{2}\log \nu_j\right)-\left(n\,y\log y+\dfrac{1}{2}\log y\right)\nonumber\\
		\cong&\,-\dfrac{1}{2}\sum^M_{j=1}\log p_j-n\sum^{M-1}_{j=1}\nu_j \log p_j-n\,y\log p_M-\dfrac{n^2}{2}(\widetilde{\rvec{\nu}}-\widetilde{\rvec{p}})\bm{\cdot}\dyadic{\Sigma}_n^{-1}\bm{\cdot}(\widetilde{\rvec{\nu}}-\widetilde{\rvec{p}})\,,
	\end{align}
	where
	\begin{equation}
		\dyadic{\Sigma}_n^{-1}= \frac{1}{n}\!\left(\widetilde{\dyadic{P}}^{-1}+\dfrac{\onevec\onevec^\top}{p_M}\right)=\left[n\!\left(\widetilde{\dyadic{P}}-\widetilde{\rvec{p}}\,\widetilde{\rvec{p}}^\top\right)\right]^{-1}
	\end{equation}
	and both $\widetilde{\dyadic{P}}$ and $\widetilde{\rvec{p}}$ are the ($M-1$)-dimensional objects as defined in \eqref{eq:P_and_p}. Hereafter, exponentiating the approximated right-hand side of~\eqref{eq:logp} nabs us~\eqref{eq:CLT_multinom}.

    \section{State-preparation noise on the ``cat'' code}
    \label{app:stateprep}

    \begin{figure}[hp]
		\centering
		\includegraphics[width=0.45\columnwidth]{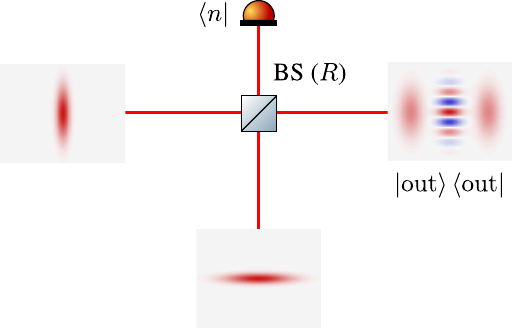}
		\caption{\label{fig:cat_gen} Schematic diagram for generation of squeezed-``cat'' logical states, following Fig.~1(b) in Ref.~\cite{Takase:2021generation}. The heralding of a particular photon number~$n$ after the merging of squeezed-vacuum states using a beam splitter results logical-state outputs for appropriate choices of the beam-splitter reflectance~$R$.}
    \end{figure}
    
    \begin{figure}[h!]
		\centering
		\includegraphics[width=0.60\columnwidth]{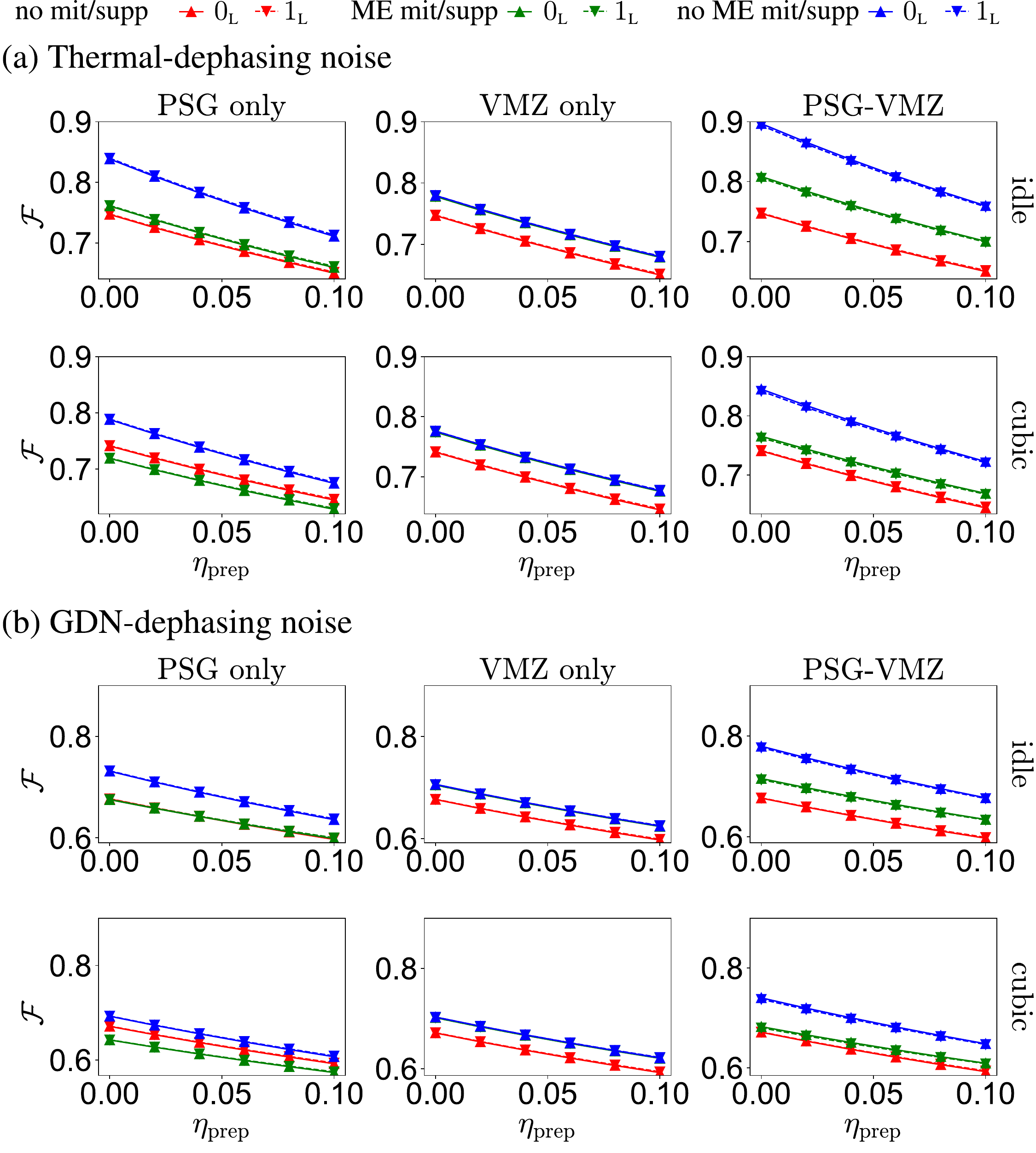}
		\caption{\label{fig:stateprep}Effects of photon loss during the state preparation of \texttt{cat(2,2)} logical codewords~$0_\textsc{l}$ and~$1_\textsc{l}$ on the mitigation and suppression of thermal-dephasing composite noise during idle operation and the cubic phase gate~$\E{\I\,0.02\,q^3}$. As expected, the combination of PSG and VMZ schemes works better at mitigating composite noise channels than individual schemes~themselves. State preparation errors arise from lossy detectors (loss rate~$\eta_{\rm{prep}}$) used in heralding the output using the scheme illustrated in~Fig.~\ref{fig:cat_gen}. Additionally, all PSG-mitigation and VMZ-suppression schemes are considered with or without measurement errors~(ME), with the measurement-noise specifications being the same as in~Fig.~\ref{fig:hybridPSGVMZ}. The additional PSGN and detector noise for these parameters were sufficient to nullify the advantage from PSG mitigation for the cubic phase gate but the advantage remains when it is used together with the VMZ. It also highlights the significance of detector imperfections in applying these techniques and, more importantly, the effectiveness in combating state-preparation and measurement imperfections when PSG and VMZ are \emph{combined} to mitigate composite noise~channels.}
    \end{figure}

    In general, the state preparation for different encodings is different, and hence a broad analysis of the impact of state preparation errors is beyond the scope of this article. However, here we show that the proposed mitigation and suppression schemes are robust under state preparation errors and state preparation losses for the two-component ``cat'' state codewords. 
    
    These are prepared by first mixing two equally but orthogonally squeezed vacuum states defined by kets $S(r)\left|0\right\rangle$ and $S(-r)\left|0\right\rangle$ with $r>0$, at a beam splitter with a reflectance
    \begin{equation}
    R=\frac{1}{1+\E{2r}}
        \label{eq:stateprep-R}
    \end{equation}
    and conditionally measuring the first mode in the Fock state $\ket{n}\bra{n}$ as shown in Fig.~\ref{fig:cat_gen}. In terms of eigenstates of the position operator, $q=\int\D x \ket{x} x \bra{x}$, the position wavefunction $\psi(x)=\inner{x}{\mathrm{out}}$ of the normalized output state $\ket{\mathrm{out}}\bra{\mathrm{out}}$ in the second mode is given by~\cite{Takase:2021generation}
    \begin{equation}
        \psi(x)=\sqrt{\frac{\left[\cosh\!\left(2r\right)\right]^{n+\frac{1}{2}}}{\left(n-\frac{1}{2}\right)!} }x^n \E{-\frac{\cosh\left(2r\right)}{2}x^2}.
        \label{eq:squeezed-output}
    \end{equation}
    This state is approximated to that of a squeezed~``cat'' of the form $S(r_{\textrm{eff}})\left|{\textrm{\texttt{cat(2,$\sqrt{n}$)}}}\right\rangle$ with the parity of the cat given by $(-1)^n$ and where
    \begin{equation}
        r_{\rm{eff}}=\frac{1}{2}\log\left(2\cosh \left(2r\right)\right).
        \label{eq:effective-squeezing}
    \end{equation}
    Subsequently, a deterministic anti-squeeze operation $S(-r_{\rm{eff}})$ leads to the desired ``cat'' states. For Fig.~\ref{fig:stateprep}, $5$dB-squeezed vacuum states are used with conditional measurements of $4$ and $5$ photons to generate even-parity $0_{\rm{L}}$ and odd-parity $1_{\rm{L}}$ states respectively, in \texttt{Mr~Mustard} with a sufficiently large cutoff dimension of $100$. Finally, a noiseless linear attenuator with a transmittance of~$0.8$ on the odd-parity state equalizes its coherent-state amplitudes with the even counterpart. The dimensions are truncated further down to $30$ after the state preparation. Absence of truncation artifacts were verified~numerically.

    The prepared states then experience composite noise channels, subject to our mitigation or suppression passages for the idle and cubic phase gate operations. The fidelities of the resulting states to those from the noiseless gates on ideal ``cat'' states in  Fig.~\ref{fig:stateprep} demonstrate robustness against two noise sources in state preparation: (1)~approximation of the output states prepared with the generalized photon subtraction method~\cite{Takase:2021generation} to the ideal states and (2)~detector losses during preparation. However, they also highlight that the additional noise from the PSGN layers in PSGs and detector noise in VMZ methods lower the advantage. 
     
  \section*{References}
	
	\bibliographystyle{unsrt}

\end{document}